\documentclass[12pt]{article} 
\pdfoutput=1

\usepackage[utf8]{inputenc}
\usepackage[T1]{fontenc} 
\usepackage{lmodern}
\usepackage[english]{babel}           
\usepackage{cancel}
\usepackage[pdftex]{graphicx}
\usepackage{epsfig}
\usepackage{graphicx}
\usepackage{comment}
\usepackage{latexsym}
\usepackage[hypertexnames=false]{hyperref}
\usepackage{amsmath}
\usepackage{mathrsfs}
\usepackage[usenames, dvipsnames]{color}
\usepackage{amsbsy}
\usepackage{amssymb}
\usepackage{amsthm}
\usepackage{amsfonts}
\usepackage{cite}
\usepackage{enumitem}
\usepackage{xcolor}
\usepackage{color}
\usepackage{mathtools}
\usepackage{caption} 
\usepackage{subcaption} 
\usepackage{euscript} 
\usepackage{float}
\usepackage{diagbox}
\usepackage[title]{appendix}
\usepackage{tabularx}
\usepackage{tikz}
\usepackage{pgfplots} 
\pgfplotsset{compat=1.16} 
\usepackage{tcolorbox}
\usepackage{soul}

\usetikzlibrary{decorations.pathmorphing,arrows.meta}

\usepackage{amsmath}
\allowdisplaybreaks
\usepackage{amsfonts}
\usepackage{amssymb}
\usepackage{bbm}
\usepackage{verbatim}
\usepackage{dsfont}
\usepackage{booktabs}
\usepackage{slashed}

\renewcommand\d{{\rm d}}

\newcommand{\C}{{\cal C}}

\newcommand{\be}{\begin{equation}}
\newcommand{\ee}{\end{equation}}
\newcommand{\dis}{\displaystyle}
\renewcommand{\thefootnote}{\fnsymbol{footnote}}

\newcommand{\Eq}[1]{Eq.~\eqref{#1}}
\newcommand{\Eqs}[1]{Eqs.~\eqref{#1}}

\newcommand{\Sect}[1]{Sect.~\ref{#1}}

\newcommand{\Appendix}[1]{Appendix~\ref{#1}}
\newcommand{\Fig}[1]{Fig.~\ref{#1}}

\newcommand{\R}{\mathbb{R}}
\newcommand{\Z}{\mathbb{Z}}
\renewcommand{\natural}{\mathbb{N}}

\renewcommand{\O}{{\cal O}}

\DeclareMathOperator\sign{sign}

\DeclareMathOperator{\arcsinh}{arcsinh}
\renewcommand{\Re}{{\rm Re}\,}

\DeclareMathOperator{\A}{Area}
\DeclareMathOperator{\hA}{\widehat\A}
\DeclareMathOperator{\tA}{\widetilde\A}

\newcommand{\ie}{{\it i.e.} }
\newcommand{\eg}{{\it e.g.} }
\newcommand{\via}{{\it via} }

\newcommand{\where}{\mbox{where}}
\newcommand{\with}{\mbox{with}}
\newcommand{\when}{\mbox{when}}
\renewcommand{\and}{\mbox{and}}

\newcommand{\esps}{\phantom{\!\!\!\overset{|}{a}}}
\newcommand{\esp}{\phantom{\!\!\overset{\displaystyle |}{|}}}
\newcommand{\espD}{\phantom{\!\!\underset{\displaystyle |}{\cdot}}}

\newcommand{\bm}{\boldmath} 

\newcommand{\red}{\color{red}}

\catcode`\@=11
\def\marginnote#1{}
\newcount\hour
\newcount\minute
\newtoks\amorpm
\hour=\time\divide\hour by60 \minute=\time{\multiply\hour by60
\global\advance\minute by-\hour}
\edef\standardtime{{\ifnum\hour<12 \global\amorpm={am}%
        \else\global\amorpm={pm}\advance\hour by-12 \fi
        \ifnum\hour=0 \hour=12 \fi
        \number\hour:\ifnum\minute<10 0\fi\number\minute\the\amorpm}}
\edef\militarytime{\number\hour:\ifnum\minute<10 0\fi\number\minute}
\def\draftlabel#1{{\@bsphack\if@filesw {\let\thepage\relax
   \xdef\@gtempa{\write\@auxout{\string
      \newlabel{#1}{{\@currentlabel}{\thepage}}}}}\@gtempa
   \if@nobreak \ifvmode\nobreak\fi\fi\fi\@esphack}
        \gdef\@eqnlabel{#1}}
\def\@eqnlabel{}
\def\@vacuum{}
\def\draftmarginnote#1{\marginpar{\raggedright\scriptsize\tt#1}}
\def\draft{\oddsidemargin -.2truein
        \def\@oddfoot{\sl preliminary draft \hfil
        \rm\thepage\hfil\sl\today\quad\militarytime}
        \let\@evenfoot\@oddfoot \overfullrule 3pt
        \let\label=\draftlabel
        \let\marginnote=\draftmarginnote
   \def\@eqnnum{(\theequation)\rlap{\kern\marginparsep\tt\@eqnlabel}%
\global\let\@eqnlabel\@vacuum}  }
\def\thebibliography#1{
\vskip 0.5cm \centerline{\bf \Large References}
\list{
[\arabic{enumi}]}{\settowidth\labelwidth{[#1]}
\leftmargin\labelwidth
\advance\leftmargin\labelsep
\usecounter{enumi}}
\def\newblock{\hskip .11em plus .33em minus .07em}
\sloppy\clubpenalty4000\widowpenalty4000
\sfcode`\.=1000\relax}

\renewcommand{\theequation}{\arabic{section}.\arabic{equation}}
\renewcommand{\section}{\setcounter{equation}{0}\@startsection
{section}{1}{0mm}{-\baselineskip}{0.5\baselineskip} {\normalfont\Large\bfseries}}
\renewcommand{\subsection}{\@startsection
{subsection}{2}{0mm}{-\baselineskip}{0.5\baselineskip} {\normalfont\large\bfseries}}
\renewcommand{\subsubsection}{\@startsection
{subsubsection}{3}{0mm}{-\baselineskip}{0.5\baselineskip}
{\normalfont\normalsize\slshape}}

\begin{document}


\begin{titlepage}
\begin{flushright}
CPHT-RR018.042023, May 2023
\vspace{0.0cm}
\end{flushright}
\begin{centering}
{\bm\bf \Large Bridging the static patches:\\ \vspace{0.2cm}de Sitter holography and entanglement }

\vspace{6mm}

 {\bf Victor Franken,$^1$\footnote{victor.franken@polytechnique.edu} Herv\'e Partouche,$^1$\footnote{herve.partouche@polytechnique.edu} Fran\c{c}ois Rondeau$^2$\footnote{rondeau.francois@ucy.ac.cy} \\ \vspace{0.1cm}and Nicolaos Toumbas$^2$\footnote{nick@ucy.ac.cy}}

 \vspace{3mm}

$^1$  {\it CPHT, CNRS, Ecole polytechnique, IP Paris, \\F-91128 Palaiseau, France}

$^2$ {\it Department of Physics, University of Cyprus, \\Nicosia 1678, Cyprus}

\end{centering}
\vspace{0.5cm}
$~$\\
\centerline{\bf\large Abstract}\vspace{0.2cm}
In the context of de Sitter static-patch holography, two prescriptions have been put forward for holographic entanglement entropy computations, the monolayer and bilayer proposals. In this paper, we reformulate both prescriptions in a covariant way and extend them to include quantum corrections. We argue that the bilayer proposal is self-consistent, while the monolayer proposal exhibits contradictory behavior. In fact, the bilayer proposal leads to a stronger holographic description, in which the full spacetime is encoded on two screens at the cosmological horizons. At the classical level, we find large degeneracies of minimal extremal homologous surfaces, localized at the horizons, which can be lifted by quantum corrections. The entanglement wedges of subregions of the screens exhibit non-trivial behaviors, hinting at the existence of interesting phase transitions and non-locality in the holographic theory. In particular, while each screen encodes its corresponding static patch, we show that the entanglement wedge of the screen with the larger quantum area extends and covers the causal diamond between the screens, with a phase transition occurring when the quantum areas of the screens become equal. We argue that the capacity of the screens to encode the  region between them is lost, when these are pushed further in the static patches of the observers and placed on stretched horizons.


\end{titlepage}
\newpage
\setcounter{footnote}{0}
\renewcommand{\thefootnote}{\arabic{footnote}}
 \setlength{\baselineskip}{.7cm} \setlength{\parskip}{.2cm}

\setcounter{section}{0}

\newpage
{\small \tableofcontents}
\newpage

\section{Introduction}
\label{sect:intro}

The area law for black hole entropy \cite{Bekenstein:1972tm, Hawking:1975vcx},
\begin{equation}
\label{eq:area_law}
   S = \frac{A}{4G\hbar},
\end{equation}
where $A$ is the area of the horizon and $G$ is the gravitational Newton's constant, has offered a beautiful thermodynamical description of black holes. This observation had immense consequences as it was the spark that ignited the holographic principle \cite{tHooft:1993dmi, Susskind:1994vu}. The holography conjecture was realized in the context of string theory in Anti de Sitter spaces (AdS) \cite{Maldacena:1997re, Gubser:1998bc, Witten:1998qj}, leading to many theoretical successes concerning the description of quantum field theory systems (and conformal field theories in particular) at strong coupling. At the same time, great puzzles of quantum gravity such as the black-hole information loss paradox and the description of physics at curvature singularities may find a natural resolution in terms of the physics of the dual holographic theory.  

Another significant step towards the understanding of quantum gravity has been the realization of a deep connection between geometry and quantum entanglement. 
In particular, recent developments in the understanding of entanglement entropy in quantum field theory and gravity led to a well defined prescription for the calculation of the fine grained, Von Neumann entropy of gravitational systems. This prescription, which was also inspired by \Eq{eq:area_law}, involves the extremization of a generalized entropy formula, with both geometrical area and semiclassical entropy contributions. See \eg \cite{Almheiri:2020cfm} for a review. In its original form, the prescription was proposed by Ryu and Takayanagi in the context of the AdS/CFT correspondence \cite{Ryu:2006bv}, so as to calculate the entanglement entropy between a subsystem of the dual quantum field theory and its complement. The subsystem in question is associated with a subregion of space, and the entanglement entropy is computed in terms of the area of an extremal homologous bulk surface. It was then generalized to include non-static cases \cite{Hubeny:2007xt} and quantum corrections \cite{Faulkner:2013ana, Engelhardt:2014gca}. Equivalent prescriptions have also been found in \cite{Wall:2012uf, Freedman:2016zud, Headrick:2022nbe}. 

Building on these ideas, it was realized that quantum entanglement plays a crucial role in the reconstruction of spacetime from the holographic data \cite{VanRaamsdonk:2009ar, VanRaamsdonk:2010pw}. Indeed, in the examples of eternal AdS black holes, it is the quantum entanglement between the degrees of freedom of the two dual CFT copies that bridges together the asymptotic causally disconnected regions of the bulk space \cite{Maldacena:2001kr}. Destroying quantum entanglement destroys the connectivity and smoothness of spacetime \cite{VanRaamsdonk:2016exw}. This idea was further developed under the name ``ER = EPR'' by Maldacena and Susskind \cite{Maldacena:2013xja}, conjecturing that any bipartite entangled state admits a geometrical description in terms of a bridge, or wormhole, connecting the two entangled parties. This is in analogy with the non-traversable wormhole connecting two distant black holes, which nonetheless are quantum mechanically entangled. More recently, it has been conjectured that the Ryu-Takayangi formula leads to the existence of a bulk region, called the entanglement wedge, which is always reconstructible from the considered subsystem of the boundary quantum theory~\cite{Wall:2012uf, Dong:2016eik, Cotler:2017erl}.

The generalized entropy formula has been applied \cite{ Penington:2019npb, Almheiri:2019psf, Almheiri:2019hni, Penington:2019kki, Almheiri:2019qdq} to compute the fine-grained entropy of Hawking radiation produced during the process of black-hole evaporation. It was found to follow the Page curve \cite{Page:1993wv}, as required by the conservation of information. Remarkably, there are contributions from ``islands'' in the interior of the black hole, implying a redundancy of the underlying degrees of freedom, as required by holography. The precise quantum state of the Hawking radiation encodes information about the interior of the black hole \cite{Almheiri:2020cfm}.

In parallel with the above developments concerning quantum gravity in Anti de Sitter space, the question of extending holography to cosmological spacetimes has been another active area of research. 
Indeed, \Eq{eq:area_law} has been shown to be verified by the cosmological horizon of de Sitter space \cite{Gibbons:1977mu}, obviously asking for a microscopic understanding. Unfortunately, a quantum gravity description of de Sitter space appears to be much more cumbersome than the case of AdS \cite{Banks:2000fe, Witten:2001kn, Dyson:2002nt, Dyson:2002pf, Goheer:2002vf, Banks:2002wr, Banks:2003cg}, and a string-theory description is still lacking. In fact, two parallel routes have been taken in order to realize holography for de Sitter space. The first, called ``the dS/CFT correspondence,'' utilizes the de Sitter symmetry group and describes quantum gravity in de Sitter space in terms of a Euclidean conformal field theory on the future spacelike boundary of the geometry \cite{Strominger:2001pn, Bousso:2001mw}. The time-reversed cosmological evolution can be obtained as an RG flow in the CFT \cite{Strominger:2001gp}.\footnote{For further attempts and developments along these lines, see \eg \cite{ Anninos:2011af, Anninos:2017hhn, Anninos:2011ui, Hikida:2022ltr, Banihashemi:2022jys,  Banihashemi:2022htw}.}
The second proposal by Susskind is to realize the hologram in terms of a dual theory living on the horizon of the static patch associated with a comoving, freely falling observer \cite{Susskind:2021omt}. This realization requires a Hamiltonian theory with discrete energy levels, as dictated by the finiteness of the Gibbons-Hawking entropy of the horizon \cite{Dyson:2002nt, Dyson:2002pf, Goheer:2002vf, Banks:2003cg, Banks:2022irh}, and it is in accordance with the holographic covariant-entropy conjecture of Bousso \cite{Bousso:1999xy, Bousso:2002ju}.\footnote{See also \cite{Alishahiha:2004md, Alishahiha:2005dj, Dong:2010pm, Dong:2018cuv, Arias:2019pzy, Arias:2019zug, Arenas-Henriquez:2022pyh, Emparan:2022ijy,Svesko:2022txo, Panella:2023lsi} for other approaches to holography in de Sitter space. For additional work discussing islands in de Sitter and other cosmological spaces, see \cite{Balasubramanian:2020coy, Hartman:2020khs, Balasubramanian:2020xqf, Geng:2021wcq, Sybesma:2020fxg, Aalsma:2021bit, Kames-King:2021etp, Levine:2022wos,Narayan:2015vda, Narayan:2017xca, Chapman:2021eyy,Diaz:2007mh, Caginalp:2019fyt, Chandrasekaran:2022cip,Kundu:2021nwp}.}

In this work, we will examine and build further on the second proposal, which entails static-patch holography. In global de Sitter space, static patches come in pairs, associated with comoving observers at two antipodal points of the spatial sphere. Each patch is enclosed by a cosmological, event horizon and can be thought of as an interior region. The surrounding space between the two patches is the exterior region. The two patches are causally disconnected, but the degrees of freedom associated with them are quantum mechanically entangled. Therefore, the dual holographic theory comprises two identical quantum mechanical systems, each living on a screen at the horizon surrounding a patch, which are entangled between them. In this holographic scenario, two distinct proposals have been put forward concerning the adaptation of the RT and HRT prescriptions \cite{Ryu:2006bv,Hubeny:2007xt} for holographic entropy computations, the monolayer and the bilayer proposals \cite{Susskind:2021esx, Shaghoulian:2021cef, Shaghoulian:2022fop}, with yet no sharp decisive argument hinting towards one of them. In the bilayer prescription, the entanglement entropy of a dual subsystem is computed \via an extremization problem in the two interior and in the exterior region, producing three extremal HRT surfaces. In the monolayer proposal on the other hand, the entropy receives contributions only from the exterior region between the cosmological horizons. 

We proceed to reformulate the bilayer and monolayer proposals in a covariant way and extend them to incorporate quantum corrections. 
In particular, we consider generic foliations of de Sitter space with Cauchy slices, including non-spherically symmetric ones, leading to general slicings of the  cosmological horizons in terms of holographic screens. The total area of the two screens divided by $4G\hbar$ is always constant, equal to twice the Gibbons-Hawking entropy, even for the cases of non-spherically symmetric bulk slicings. The bilayer and monolayer prescriptions yield completely different results for  the minimal extremal surfaces associated with various subsystems of the holographic dual theory on the screens,
as well as for the entanglement-wedge structures. At the classical level, we find large degeneracies of homologous minimal extremal surfaces in the exterior causal-diamond region, which can be lifted by the quantum corrections. Indeed, there are surfaces of minimal area, lying on the segments of the horizons joining the screens with the bifurcate horizon, which can be seen as solutions of an extremization problem in an enlarged domain. The area functional is supplemented with Lagrange multipliers that enforce the surfaces to lie inside the exterior causal-diamond region.  Indeed, the method of Lagrange multipliers can be generalized to impose inequality rather than equality constraints.\footnote{The issue is similar to what happens for the function $f(x)=-x^2$, for $x\in[-1,2]$. Besides the ``bulk'' maximum at $x=0$, two local minima at the boundaries $x=-1$ and $x=2$ exist. These three values are the extrema of the function $f$ supplemented with Lagrange multipliers. Among these extrema, the minimal one is reached for $x=2$.} In the bilayer case, the entanglement wedges of various subsystems can extend in the interior as well as in the exterior regions. On the other hand, in the monolayer case, the entanglement wedges restrict in the exterior region, leading to contradictory behavior and hinting towards the inconsistency of this proposal. Taking into account quantum corrections, we find in the context of the bilayer proposal that the screen with the larger quantum area gives rise to an entanglement wedge that covers not only its interior region but also the exterior causal-diamond region. An interesting phase transition occurs when the quantum areas of the screens become equal, hinting on the existence of non-local interactions between the degrees of freedom on the two screens. On the other hand, the entanglement wedge of a single screen system in the monolayer case confines to the screen itself, without extending in the bulks of the exterior and interior regions. We also revisit and generalize existing examples of arc subsystems in the case of three-dimensional de Sitter space \cite{Susskind:2021esx, Shaghoulian:2021cef, Shaghoulian:2022fop}, mainly obtaining the entanglement entropy.

In fact, the entanglement-wedge structure of the full two-screen system in the bilayer proposal is suggestive of an even stronger holography conjecture, where {\it the entire spacetime} (the two interiors together with the exterior region) can be encoded on the cosmological horizons associated with the two static patches. Quantum entanglement builds an effective bridge between the static patches, with the entanglement wedge of the two screens comprising complete bulk Cauchy slices, thus capturing the full set of bulk degrees of freedom. However, when we push the holographic screens in the interior regions, away from the cosmological horizons, the capability to encode the exterior region is lost and the bridge between the screens becomes ineffective. Only the interior regions are captured by the screens.

The plan of the paper is as follows. In \Sect{sect:static_patch_holography}, 
we review the arguments based on the Bousso entropy bound that support placing the pair of holographic screens on the cosmological horizons. We recall the original formulations of the monolayer and bilayer proposals for computing the entanglement entropies of subsystems of the screens. We reformulate both proposals in a covariant way, detail the entanglement-wedge structures and describe our proposal for evaluating quantum corrections to the entropies at the semiclassical level. 

In \Sect{sec:entropy_subsystems}, we argue that the entanglement-wedge structure leads to contradictions in the monolayer case. On the contrary, entanglement wedges motivate the fact that, in the bilayer case, the holographic theory should encode the entire de Sitter space rather than the static patches of the observers only. This is illustrated by applying both proposals for a single-screen subsystem and for the two-screen system. In both cases, the classical degeneracy of the areas of the minimal extremal surfaces must be lifted at the semiclassical level. We also study the situation where the screens are pushed on stretched horizons in the interiors of the static patches, and argue that there can be no exterior contributions to their entropy.

In \Sect{12arcs}, we restrict to the case of three-dimensional de Sitter space. When the subsystem of the screens consists of an arc circle on one screen, we show the existence of continuous families of possible HRT-like curves for computing the classical contributions to the entropy. These homologous curves lie on the boundaries of the interior or exterior causal diamonds 
along the horizons. They are minimal extrema of the length functional supplemented by Lagrange multipliers. We then consider the susbsystem of the screens comprising two slightly misaligned arcs, lying respectively on each screen. The HRT curve in the exterior region is not a pair of geodesics running through the bulk. As before, it belongs to a continuous family of curves on the boundary of the exterior region along the horizons. 

A summary of our results and some perspectives can be found in \Sect{sect:conclu}. \Appendix{Appendix:Lagrange_multipliers} and~\ref{Appendix:geodesics_dS} present detailed computations to search for homologous extremal surfaces (or curves in dS$_3$), which lie either on the boundary or in the bulk of the causal diamonds associated with Cauchy slices in the interior or exterior regions. \Appendix{ptran} describes a phase transition~\cite{Hartman:2020khs} that could occur in the two-arc subsystem in dS$_3$ if one omits the HRT curves lying on the boundary horizons.  


\section{Static-patch holography and the monolayer/bilayer proposals}
\label{sect:static_patch_holography}

A number of attempts have been made to apply the holographic principle to cosmological backgrounds, and in particular to de Sitter space. We begin our discussion in this paper with the static-patch holographic proposal for de Sitter space~\cite{Susskind:2021omt}, which states that the static patch associated with a comoving observer is described by a quantum system localized on its boundary, namely the cosmological horizon. In order to describe the proposal in more detail, we first review some basic geometrical properties of global de Sitter space and motivate the location of the dual system. We then recall the monolayer and bilayer versions of the static-patch holography  conjecture. We reformulate these proposals in a covariant way and describe how to incorporate quantum corrections.


\subsection{de Sitter geometry and setup}
\label{dSgeo}

\begin{figure}[!h]
    \centering
\begin{tikzpicture}
\begin{scope}[transparency group]
\begin{scope}[blend mode=multiply]
\path
       +(3,3)  coordinate (IItopright)
       +(-3,3) coordinate (IItopleft)
       +(3,-3) coordinate (IIbotright)
       +(-3,-3) coordinate(IIbotleft)
      
       ;
\draw (IItopleft) --
          node[midway, above, sloped]    {$\cal{J}^+$}
      (IItopright) --
          node[midway, above, sloped] {Antipode}
      (IIbotright) -- 
          node[midway, below, sloped] {$\cal{J}^-$}
      (IIbotleft) --
          node[midway, above , sloped] {Pode}
      (IItopleft) -- cycle;




\fill[fill=blue!20] (-3,3) -- (-3,-3) -- (3,3);

\fill[fill=red!20] (-3,-3) -- (-3,3) -- (3,-3);

\draw (IItopleft) -- (IIbotright)
              (IItopright) -- (IIbotleft) ;



\end{scope}
\end{scope}
\end{tikzpicture}
    \caption{\footnotesize Penrose diagram for de Sitter space. The diagonal lines are the past and future horizons for an observer on the north and south poles (pode and antipode). The regions $\mathcal{O}^+$ (purple or blue) and $\mathcal{O}^-$ (purple or red) intersect in the purple region, the northern causal patch. The southern causal patch is the blank region.}
    \label{fig:Penrose_dS}
\end{figure}
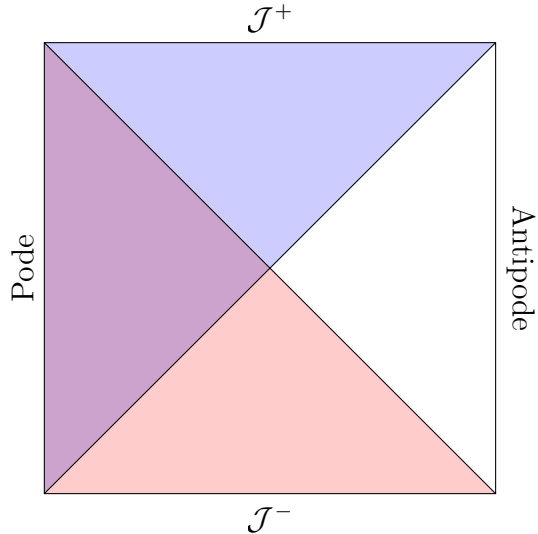

The metric of $(n+1)$-dimensional de Sitter space dS$_{n+1}$ in global coordinates is given by
\begin{equation}
\d s^2=-\d\tau^2+\cosh^2\tau\, \d\Omega^2_{n},
\end{equation}
where $\tau\!\in\!(-\infty,+\infty)$ is the global time and $\d\Omega_n^2\!=\!\d\theta_1^2+\sin^2\theta_1 \d\theta_2^2+\cdots+\sin^2\theta_1\cdots\sin^2\theta_{n-1}\d\theta^2_{n}$ is the metric on the unit $n$-dimensional sphere S$^n$. The radius of curvature has been set to unity for simplicity. A slice of constant global time corresponds to an $n$-dimensional sphere with scale factor $\cosh\tau$, which takes infinite values at $\tau=\pm\infty$ and its minimal value at $\tau=0$. 
By means of a time reparametrization
\begin{equation}
\cosh\tau=\frac{1}{\cos\sigma},
\label{tau-sigma}
\end{equation}
the metric admits the following conformal form
\begin{align}
\d s^2&=\frac{1}{\cos^2\sigma}\left(-\d\sigma^2+\d\Omega^2_{n}\right)\!,\nonumber \\
&=\frac{1}{\cos^2\sigma}\left(-\d\sigma^2+\d \theta_1^2+\sin^2\theta_1\,\d\Omega^2_{n-1}\right)\!,
\label{ds2}
\end{align}
where $\sigma\in(-\pi/2,\pi/2)$ is the conformal time and $\sign \tau=\sign \sigma$. Using the conformal coordinates $(\theta_1,\sigma)$, where $\theta_1\in[0,\pi]$ when $n\ge 2$, we can construct the Penrose diagram for de Sitter space, which is a square depicted in Figure~\ref{fig:Penrose_dS}. 
Each horizontal slice of constant $\sigma$ represents an S$^n$. Every point of the diagram represents an S$^{n-1}$, except for points on the left and right vertical sides, which are actual points, corresponding to the north and south poles of S$^{n}$. We will refer to these points as the pode and the antipode, respectively. The spacelike surfaces $\cal{J}^-$ and $\cal{J}^+$ at $\sigma=\mp \pi/2$ are the past and future null infinity. They correspond to the surfaces where null geodesics originate and terminate, respectively. 

Throughout this section, we will consider a comoving observer at the pode. The purple and blue shaded regions of Figure~\ref{fig:Penrose_dS}, called $\mathcal{O}^+$, are the parts of spacetime that the observer at the pode will ever be able to send a signal to. The purple and red shaded regions, called~$\mathcal{O}^-$, are the parts of spacetime that the observer at the pode will ever be able to receive a signal from. The intersection $\mathcal{O}^+\cap\mathcal{O}^-$ of these two regions, which is purple, is called the northern causal diamond, or the causal patch associated to the observer at the pode. Likewise, we can construct the southern causal diamond associated with an observer at the antipode, which is represented by the blank region in the Penrose diagram. These two patches are causally disconnected from each other, and they are bounded by cosmological horizons depicted by the diagonal lines in Figure~\ref{fig:Penrose_dS}.

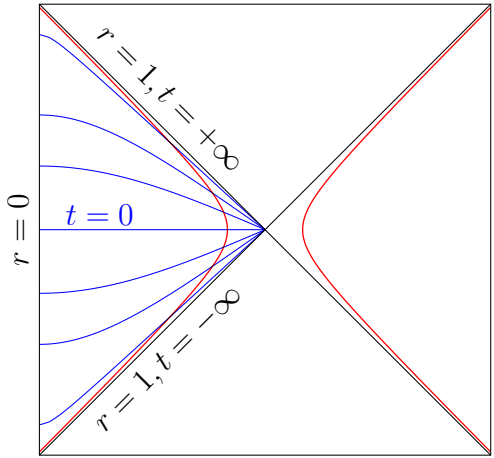
\begin{figure}[!h]
    \centering
\begin{tikzpicture}
\begin{scope}[transparency group]
\begin{scope}[blend mode=multiply]
\path
       +(3,3)  coordinate (IItopright)
       +(-3,3) coordinate (IItopleft)
       +(3,-3) coordinate (IIbotright)
       +(-3,-3) coordinate(IIbotleft)
      
       ;
\draw (IItopleft) --
          node[midway, above, sloped] {}
      (IItopright) --
          node[midway, above, sloped] {}
      (IIbotright) -- 
          node[midway, below, sloped] {}
      (IIbotleft) --
          node[midway, above , sloped] {$r=0$}
      (IItopleft) -- cycle;

\draw (IItopleft) -- node[midway, above, sloped]    {$r=1, t=+\infty$} (0,0)
              (0,0) -- node[midway, below, sloped]    {$r=1, t=-\infty$} (IIbotleft) ;

\draw (IItopright) -- (0,0)
              (0,0) -- (IIbotright) ;

\draw[line width= 0.5 pt,red] plot[variable=\t,samples=1000,domain=-80.4:80.4] ({0.5*sec(\t)},{0.5*tan(\t)});

\draw[line width= 0.5 pt,red]  plot[variable=\t,samples=1000,domain=-80.4:80.4] ({-0.5*sec(\t)},{-0.5*tan(\t)});

\draw[domain=-3:0, smooth, variable=\x, blue] plot ({\x}, {(3/1.7)*(3.14/180)*asin(cos(deg((3.14/6)*(\x+3)))*tanh(0.5))});

\draw[domain=-3:0, smooth, variable=\x, blue] plot ({\x}, {(3/1.7)*(3.14/180)*asin(cos(deg((3.14/6)*(\x+3)))*tanh(1))});

\draw[domain=-3:0, smooth, variable=\x, blue] plot ({\x}, {(3/1.7)*(3.14/180)*asin(cos(deg((3.14/6)*(\x+3)))*tanh(3))});

\draw[color=blue] (-3,0)--(0,0);
\draw (-2.8,0.2) node[color=blue,right] {$t=0$};

\draw[domain=-3:0, smooth, variable=\x, blue] plot ({\x}, {-(3/1.7)*(3.14/180)*asin(cos(deg((3.14/6)*(\x+3)))*tanh(0.5))});

\draw[domain=-3:0, smooth, variable=\x, blue] plot ({\x}, {-(3/1.7)*(3.14/180)*asin(cos(deg((3.14/6)*(\x+3)))*tanh(1))});

\draw[domain=-3:0, smooth, variable=\x, blue] plot ({\x}, {-(3/1.7)*(3.14/180)*asin(cos(deg((3.14/6)*(\x+3)))*tanh(3))});

\end{scope}
\end{scope}
\end{tikzpicture}
    \caption{\footnotesize Constant static time $t$ slices (blue curves) covering the causal patch associated with the pode. The stretched horizons are represented by the red hyperbolas.}
    \label{fig:Penrose_diag_glob_coord}
\end{figure}

The global and conformal coordinates cover the entire de Sitter spacetime. On the other hand, in these coordinate systems the cosmological horizons associated with the comoving observers at the pode and antipode are not manifest. For this purpose, one introduces the static coordinates, in which the metric of dS$_{n+1}$ takes the form
\begin{equation}
    \d s^2 = -(1-r^2) \d t^2 + \frac{\d r^2}{1-r^2}+ r^2 \d\Omega_{n-1}^2,
    \label{staticds}
\end{equation}
with $t\in(-\infty,+\infty)$ being the static time and $r\in[0,1]$. These static coordinates cover only the causal patch associated with the pode. The three remaining patches of the Penrose diagram can also be covered by independent sets of $(r,t)$ coordinates. Note that in the blue and red triangular regions of Figure~\ref{fig:Penrose_dS}, $r$ takes values from $1$ to infinity and becomes timelike, while $t$ is spacelike. As depicted in Figure~\ref{fig:Penrose_diag_glob_coord},
the cosmological horizon associated with the pode is located at $r=1$, while the pode is at $r=0$. The metric is manifestly static, independent of $t$, with $\partial_t$ being a Killing vector associated to the isometry $t\rightarrow t + \text{constant}$. In static coordinates, $\partial_t$ can only be used to define a consistent unitary Hamiltonian evolution in the northern causal patch. In particular, this patch can be foliated by constant $t$ slices, with the state of the system evolving unitarily from slice to slice. Notice that these surfaces end on the bifurcate horizon, which lies at the intersection of the diagonals of the Penrose diagram, and approach the past and future horizons as $t\to \mp \infty$ (see Figure~\ref{fig:Penrose_diag_glob_coord}).

Finally, one can also define a stretched horizon, which is a hyperbola $r=1-\varepsilon$, where $\varepsilon$ is a small positive constant. The latter is located at an infinitesimal proper distance from the horizon and serves as a regularized, timelike boundary for the static patch. This stretched horizon plays the role of a cutoff surface regulating the temperature, which from the point of view of a static accelerating observer (following the line $r=\rm constant$) becomes arbitrarily large as the horizon is approached. Indeed, from the point of view of the observer at the pode, the patch behaves like a thermal cavity with hot walls \cite{Susskind:2021omt}. As seen in Figure~\ref{fig:Penrose_diag_glob_coord}, the constant $t$ slices intersect the stretched horizon at distinct points, spanning the whole hyperbola as $t$ evolves from $-\infty$ to $+\infty$. 


\subsection{Location of the dual holographic system}

\begin{figure}[!h]
    \centering
\begin{tikzpicture}
\begin{scope}[transparency group]
\begin{scope}[blend mode=multiply]
\path
       +(3,3)  coordinate (IItopright)
       +(-3,3) coordinate (IItopleft)
       +(3,-3) coordinate (IIbotright)
       +(-3,-3) coordinate(IIbotleft)
      
       ;
\draw (IItopleft) -- (IItopright) -- (IIbotright) -- (IIbotleft) --(IItopleft) -- cycle;

\draw (-0.2,1.7) -- (0,1.9) -- (0.2,1.7);
\draw (-1.9,0.038+0.7) -- (-1.7,0.038+0.5) -- (-1.9,0.038+0.3);
\draw (1.9,0.85+0.7) -- (1.7,0.85+0.5) -- (1.9,0.85+0.3);
\draw (-0.2,-1.7) -- (0,-1.9) -- (0.2,-1.7);

\draw[domain=-3:-1.7, smooth, variable=\x, red] plot ({\x}, {sin(deg((\x/2-1)))+1.5});
\draw[domain=-1.7:1.7, smooth, variable=\x, black, dashed] plot ({\x}, {sin(deg((\x/2-1)))+1.5});
\draw[domain=1.7:3, smooth, variable=\x, red] plot ({\x}, {sin(deg((\x/2-1)))+1.5});

\node at (-2.4,0.7) [label = above:$\red L$]{};
\node at (2.4,1.6) [label = above:$\red L$]{};

\node at (-0.55,0.55) [circle, fill, inner sep=1.5 pt, label = below:$H_1$]{};
\node at (1,1) [circle, fill, inner sep=1.5 pt, label = below:$H_2$]{};

\node at (-1.7,0.5) [label=below:$\Sigma_{\rm left}$]{};
\node at (2.1,1.3) [label=below:$\Sigma_{\rm right}$]{};
\node at (0,1.5) [label=below:$\Sigma_{\rm ext}$]{};





\draw (IItopleft) -- (IIbotright)
              (IItopright) -- (IIbotleft) ;



\end{scope}
\end{scope}
\end{tikzpicture}
    \caption{\footnotesize A Cauchy slice $\Sigma$ can be divided into 3 parts by the cosmological horizons. Each part is bounded by $H_1$ or/and $H_2$. Bousso wedges are indicated for each of the four regions of the Penrose diagram. On the parts $\Sigma_{\rm left}$ and $\Sigma_{\rm right}$ of the slice lying in the left and right  static patches, sub-slices $L$ are depicted in red. The holographic screens are located on $H_1$ and $H_2$.}
    \label{Sigmas}
\end{figure}
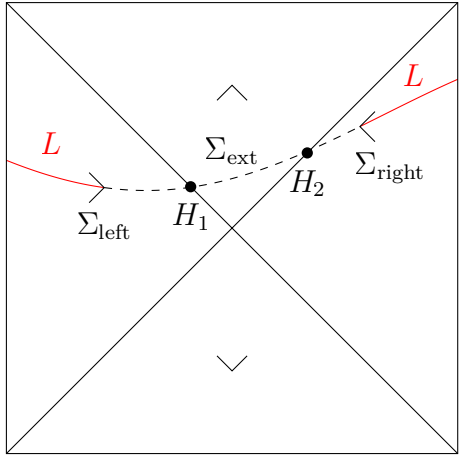

Let us consider for a moment an  ${\rm SO}(n)$-symmetric Cauchy slice $\Sigma$ (\eg a slice of constant global time~$\tau$). As shown in Figure~\ref{Sigmas}, 
%
%
it can be divided by the cosmological horizons into three parts.
The first one, $\Sigma_{\rm left}$, lies in the left static patch. It has the topology of a sphere cap, whose boundary is an S$^{n-1}$, denoted by $H_1$, on the horizon of the pode. Similarly, the second part, $\Sigma_{\rm right}$, extends in the right static patch and has the topology of a sphere cap, with boundary $H_2$ on the horizon of the antipode. The remaining part, $\Sigma_{\rm ext}$, is located in the region between the static patches. It has the topology of a barrel, with boundary $H_1\cup H_2$, and behaves as a wormhole bridge between the two sphere caps. We will refer to the left and right static patches as ``the two interior regions'' of de Sitter space (where the pode and antipode are living in) and the region between them as the ``exterior region.'' In the special case the Cauchy slice passes through the bifurcate horizon, it is split into two parts only, which lie in the two interior regions, respectively. 

Focusing on $\Sigma_{\rm left}$, we would like to investigate whether it is possible to describe any state on it in terms of a holographic dual quantum system on a holographic screen. To answer this question, we make use of the Bousso covariant entropy bound \cite{Bousso:1999xy, Bousso:2002ju}:

\noindent\textit{Let ${\rm Area}(A)$ be the area of an arbitrary codimension $2$ surface $A$. A light-sheet of $A$ is a codimension 1 surface generated by light-like rays which begin at $A$, extend orthogonally away from $A$ and are of non-positive expansion. Let $S(\Omega)$ be the coarse-grained entropy passing through any such light-sheet $\Omega$ of $A$. Then $S(\Omega)\leq {\rm Area}(A)/(4G\hbar)$.\footnote{At the quantum level, one must add to the area term in the right-hand side the entanglement entropy across the surface $A$, obtaining the quantum Bousso bound \cite{Strominger:2003br}.}}

\noindent When $A$ is codimension 2 also on the Penrose diagram, the light-sheets are represented on the diagram by the so-called Bousso wedges, with $A$ corresponding to the tip of the wedges. The Penrose diagram can be divided into distinct regions depending on the directions of the light-sheets. In the case of de Sitter space, there are four such regions, which turn out to coincide with the four regions delimited by the cosmological horizons, as shown in Figure~\ref{Sigmas}. The static-patch holographic proposal of \cite{Susskind:2021omt} then relies on a corollary of the covariant entropy bound, the ``spacelike projection theorem:''

\noindent\textit{Let $A$ be a closed-surface admitting at least one future directed light-sheet $\Omega$, whose only boundary is $A$. Let $S(L)$ be the coarse-grained entropy in a spatial region $L$ with boundary $A$ on the same side as $\Omega$. Then $S(L)\leq S(\Omega)\leq {\rm Area}(A)/(4G\hbar)$.} 

\noindent This corollary follows from the second law of thermodynamics and Bousso's covariant entropy bound.
In Figure~\ref{Sigmas}, two such spatial regions $L$ of $\Sigma_{\rm left}$ and $\Sigma_{\rm right}$ are depicted in red. The question asked is then: Where could we locate the holographic screen so as to encode the whole Cauchy slice $\Sigma_{\rm left}$ of the left static patch, maximizing the entropy bound? 
From the spacelike projection theorem, one can see that the whole $\Sigma_{\rm left}$ can indeed be encoded holographically, if we bring $A$ all the way to the horizon and place the screen there.
A similar construction applies for $\Sigma_{\rm right}$. Therefore, we end up with two causally disconnected holographic screens $H_1$ and $H_2$, which are respectively located on the left and right cosmological horizons. The dual holographic systems on $H_1$ and $H_2$ must be quantum mechanically entangled because the dual bulk regions in the left and right static patches are entangled when the quantum state on the full Cauchy slice $\Sigma$ is pure. Notice that a bulk foliation in terms of Cauchy slices produces an evolution of the screens along the two cosmological horizons. We end up with a holographic theory on the boundaries of the two causal patches, which are the two cosmological horizons.

The above conclusions have been reached in the particular case of an ${\rm SO}(n)$-symmetric Cauchy slice $\Sigma$ and a surface $A$ represented by a single point on the Penrose diagram. While these assumptions are useful for illustrating the ideas using Bousso wedges, they can be relaxed. Indeed, any (\ie non necessarily ${\rm SO}(n)$-symmetric) Cauchy slice $\Sigma$ can be considered and decomposed into 3 pieces, $\Sigma=\Sigma_{\rm left}\cup\Sigma_{\rm ext}\cup\Sigma_{\rm right}$. However, the boundaries $H_1$ or/and $H_2$ will not be necessarily spherical. For any codimension 2 closed surface $A$ lying on $\Sigma_{\rm left}$ ($\Sigma_{\rm right}$), the spacelike projection theorem applies. In order to encode holographically the whole slice $L=\Sigma_{\rm left}$ ($L=\Sigma_{\rm right}$), we have to place its boundary $A$ on the horizon of the pode (antipode). Hence, two screens $H_1$ and $H_2$ have to be placed on the horizons, respectively. 

When the Cauchy slice $\Sigma$ is ${\rm SO}(n)$-symmetric, the areas of the screens $H_1$ and $H_2$ remain constant, in Planck units. To see this, we use the fact that along the horizon of the northern causal patch, the coordinates $(\theta_1,\sigma)$ of the screen satisfy
\be
 \left\{\!\begin{array}{ll}
\dis \sigma+\theta_1={\pi\over 2}\, , &\mbox{if } \dis 0\le \sigma <{\pi\over 2}\, , \\
\dis \sigma-\theta_1=-{\pi\over 2}\, , &\mbox{if } \dis -{\pi\over 2}<\sigma \le 0\, .\esp
\end{array}\right.
\label{dH1}
\ee
Using \Eq{ds2}, the radius of $H_1$, which is a sphere S$^{n-1}$, is
\be
{\sin\theta_1\over \cos\sigma}=1.
\label{rad}
\ee
Similarly, the coordinates $(\theta_1,\sigma)$ along the horizon of the southern causal patch satisfy
\be
 \left\{\!\begin{array}{ll}
\dis \sigma-\theta_1=-{\pi\over 2}\, , &\mbox{if } \dis 0\le \sigma <{\pi\over 2}\, , \\
\dis \sigma+\theta_1={\pi\over 2}\, , &\mbox{if } \dis -{\pi\over 2}<\sigma \le 0\, .\esp
\end{array}\right.
\label{dH2}
\ee
Therefore, the radius of the screen $H_2$ associated with the southern static patch is also 1. The areas of the screens remain constant, irrespective of the ${\rm SO}(n)$ symmetric Cauchy slice~${\rm \Sigma}$.

The conclusion above generalizes to the non-${\rm SO}(n)$ symmetric cases due to the fact that the de Sitter cosmological horizons share similar properties with black hole event horizons.\footnote{This property becomes more manifest when we consider the static coordinates which cover the static patches. However, it does not hold for cosmological event/particle or apparent horizons associated with generic closed FRW cosmologies, a fact that highlights its importance for the de Sitter static-patch holographic proposal. Such cosmologies and their holographic descriptions will be studied in a forthcoming publication~\cite{FPRT}.} Let us consider a generic spacelike Cauchy slice $\Sigma$. 
Suppose first that $\Sigma$ and the bifurcate horizon do not overlap. Then $\Sigma$ necessarily intersects the cosmological horizons on two non overlapping surfaces (one on each horizon), which are closed and co-dimension $2$. Let $H_+$ be the surface of intersection with the positive slope horizon $\sigma=\theta_1-\pi/2$, and $H_-$ the surface of intersection with the negative slope horizon $\sigma=-\theta_1+\pi/2$. We can always choose to parameterize $H_+$ and $H_-$ with the remaining $n-1$ spherical angles $\theta_2, \dots, \theta_{n}$, and describe them as  parametric surfaces in dS$_{n+1}$ \via the equations $\sigma=\theta_1-\pi/2=f_+(\theta_2,\dots,\theta_n)$ and $\sigma=-\theta_1+\pi/2=f_-(\theta_2,\dots,\theta_n)$, respectively. 
Using \Eq{rad} and the fact that along $H_+$ and $H_-$ we have $\partial \sigma/\partial\theta_i=\pm\partial\theta_1/\partial\theta_i$, $i\in\{2,\dots, n\}$, one readily derives that the induced metrics on $H_+$ and $H_-$ are equal to the metric on the unit round sphere S$^{n-1}$. Therefore, the areas of both $H_+$ and $H_-$ are equal to the area of the unit sphere S$^{n-1}$. Since $H_+$ and $H_-$ do not overlap in this case, the functions $f_+(\theta_2,\dots,\theta_n)$ and $f_-(\theta_2,\dots,\theta_n)$ must be strictly positive or negative.\footnote{For example, if $f_+$ vanishes at some point, then this point lies on the bifurcate horizon, where $\sigma=0$ and $\theta_1=\pi/2$, and belongs to both $H_+$ and $H_-$.} If $f_+(\theta_2,\dots,\theta_n)<0$ ($f_+(\theta_2,\dots,\theta_n)>0$), $H_+$ lies below (above) the symmetric slice $\sigma=0$, and should be identified with the left pode (right antipode) screen $H_1$ ($H_2$). Since the points of $H_+$ must be spacelike separated from the points of $H_-$, $f_-$ must be also negative (positive). So $H_-$ should be identified with the screen $H_2$ ($H_1$) associated with the antipode (pode). 

Next suppose that $\Sigma$ and the bifurcate horizon intersect. Then, $H_+$ and $H_-$ overlap at the intersection of $\Sigma$ with the bifurcate horizon. In this case, we split $H_+$ into two parts $H_{++}$ and $H_{+-}$, satisfying at each of their points $f_+ \geq 0$ or $f_+ \leq 0$, respectively. Likewise, $H_-$ can be divided into $H_{-+}$ and $H_{--}$, satisfying $f_- \geq 0$ or $f_- \leq 0$, respectively. We take $H_1=H_{-+} \cup H_{+-}$ and $H_2=H_{--} \cup H_{++}$. 
Bousso's covariant entropy bound ensures that the area of the left (right) screen divided by $4G\hbar$ bounds the coarse-grained entropy on $\Sigma_{\rm left}$ ($\Sigma_{\rm right}$). In these cases as well, the sum of the areas of $H_1$ and $H_2$ is equal to twice the area of the unit sphere S$^{n-1}$.

The total area of the screens divided by $4G\hbar$ is a measure of the number of  underlying microscopic degrees of freedom needed to describe the dual bulk system.
The fact that the total area of the screens divided by $4G\hbar$ remains constant, equal to twice the Gibbons-Hawking entropy, as they evolve along the horizons, suggests that the two-screen system at any time has the capacity of describing the full de Sitter space. The precise microscopic nature of the theory living on the screens is not known. Only very general properties of it have been inferred and the problem of specifying it precisely will not be treated in this work either. See, however, \cite{Susskind:2021esx} for an interesting proposal concerning two-dimensional de Sitter~space.

In the special case where a bulk SO$(n)$-symmetric Cauchy slice $\Sigma$ contains the intersection of the two cosmological horizons (as in the case of the symmetric $\tau=0$ slice), the left and right screens coincide and become identical with the bifurcate horizon. This degenerate situation can be regulated by pushing $H_1$ and $H_2$ an infinitesimal proper distance away on the stretched horizons, at $r=1-\varepsilon$, as in \cite{Susskind:2021omt}. Then the holographic two-screen system lies in an entangled thermofield-double like pure state \cite{Goheer:2002vf, Susskind:2021omt}, in the same fashion as the two copies of the dual CFT for the eternal AdS black hole \cite{Maldacena:2001kr, VanRaamsdonk:2010pw}. Taking a partial trace over the degrees of freedom of the right (left) screen produces a mixed density matrix for the left (right) screen. As can be seen in Figure~\ref{fig:Penrose_diag_glob_coord}, spacelike slices in the left causal patch, which end on the bifurcate horizon, have the same causal diamond as the $t=0$ slice (namely, the static patch itself), for which the density matrix is thermal with respect to the Hamiltonian conjugate to $t$. (This statement holds if the whole of de Sitter space is in the Bunch-Davies vacuum.) Therefore, the corresponding Von Neumann entropy must be equal to the maximal, Gibbons-Hawking entropy, as it is the case for the $t=0$ slice.

As we will argue in \Sect{sec:entropy_subsystems}, the holographic entanglement entropy prescription implies that the pair of screens at any time is in a pure state, and the leading geometrical contribution to the entanglement entropy of a single-screen system is constant, despite the growing size of the barrel like region between the two static patches. The presence of the barrel is a manifestation of the ER = EPR relation due to the entanglement between the two single-screen systems. The situation is reminiscent of the growth of the Einstein-Rosen wormhole in the interior region of the eternal AdS black hole. In this latter example, the entanglement entropy of a dual CFT copy remains constant as time evolves, and so the growing size of the Einstein-Rosen bridge corresponds to the growing quantum complexity of the state \cite{Susskind:2018pmk}. 


\subsection{Monolayer and bilayer proposals}
\label{monobil}

Following the static-patch holography conjecture, two proposals have been made by  Susskind and Shaghoulian~\cite{Susskind:2021esx, Shaghoulian:2021cef, Shaghoulian:2022fop} to compute entanglement entropies in the holographic dual description, extending to de Sitter space the Ryu-Takayanagi and the generalized HRT prescriptions \cite{Ryu:2006bv,Hubeny:2007xt}. Indeed, the usual HRT prescription must be suitably adapted in order to encode the entanglement structure in the de Sitter case. Perturbatively in $G\hbar$, the prescriptions in~\cite{Susskind:2021esx, Shaghoulian:2021cef, Shaghoulian:2022fop} incorporate the leading contributions, which are of order $(G\hbar)^{-1}$. Sect.~4.5 of reference \cite{Shaghoulian:2021cef} provides interesting discussions on the need to extremize a generalized entropy in the presence of quantum corrections and examples of cases where quantum corrections will be important involving entangled pairs of particles, which separate in the interior and exterior regions, as well as evaporating black holes.\footnote{Explicit, detailed proposals concerning the extremization problems in the presence of quantum corrections, in the context of both the monolayer and bilayer proposals, will be provided in \Sect{quantumcorrections}.}  
Let us recall the two proposals:

\noindent $\bullet$ {\bf Monolayer proposal:} \textit{The entanglement entropy of a subregion $A$ of the two screens and its complement is $1/(4G\hbar)$ times the area of an extremal surface homologous\footnote{If there are multiple such extremal surfaces, we choose the one with the minimal area.} to $A$, and lying between the two sets of degrees
of freedom; in this case between the two cosmological horizons.}

\noindent $\bullet$ {\bf Bilayer proposal:} \textit{The entanglement entropy of a subregion $A$ of the two screens and its complement is $1/(4G\hbar)$ times the sum of the areas of the extremal surfaces homologous to $A$ in each of the three subregions of the bulk, that is, in the exterior and the two interior regions.}

\noindent Both proposals are natural modifications of the HRT prescription, where the bulk manifold admits a holographic dual on its boundary and the fine-grained entropy of subregions of the boundary theory is given in terms of an extremal area in the bulk. Indeed, by putting the holographic dual of de Sitter space on the two cosmological horizons, we must seek for extremal surfaces in bulk regions bounded by the horizons. Considering the whole de Sitter space would be inconsistent since it is not bounded by the horizons.\footnote{A complete spacelike Cauchy slice is topologically a sphere with no boundary.} On the other hand, the interior regions and the exterior region are the only regions bounded by the horizons. 
In \cite{Susskind:2021esx, Shaghoulian:2021cef, Shaghoulian:2022fop}, it was observed that applying an HRT-like prescription to the left (right)  interior region leads to a vanishing entropy for the single-screen system $H_1$ ($H_2$). On the other hand, applying an HRT-like prescription to the exterior region gives the expected result to leading order, namely the area of the screen divided by $4G\hbar$. This was the main motivation for the monolayer proposal. On the other hand, one could argue that since both the interiors and the exterior region are bounded by horizons, one should look for extremal surfaces in all three regions and add the contributions. This leads to the bilayer proposal of \cite{Shaghoulian:2021cef}, which is reminiscent of some AdS models \cite{Akal:2020wfl,Geng:2020fxl}. The bilayer prescription also yields the expected result for the entropy of a single-screen system at the horizon of de Sitter space~\cite{Shaghoulian:2021cef, Shaghoulian:2022fop}.

In the spirit of the HRT approach, we may specify in more detail the monolayer and bilayer proposals by formulating them in a covariant way and by specifying the entanglement wedges of the holographic dual subsystem of the screens:

\noindent($i$) Consider an arbitrary Cauchy slice $\Sigma$ of dS$_{n+1}$.

\noindent($ii$) Define the screens $H_1$ and $H_2$ as the intersections of $\Sigma$ with the cosmological horizons of the pode and the antipode, respectively. 

\noindent($iii$) Divide $\Sigma$ into 3 parts. The first one, $\Sigma_{\rm left}$, lies in the left causal patch and has boundary~$H_1$. The second one, $\Sigma_{\rm right}$, is in the right causal patch and has boundary $H_2$. The remaining one, $\Sigma_{\rm ext}$, lies in the exterior region and has boundary $H_1\cup H_2$. Any Cauchy slice~$\Sigma'$ of dS$_{n+1}$  passing through $H_1$ and $H_2$ is equivalent to $\Sigma$, in the following sense. Decomposing in a similar way $\Sigma'=\Sigma'_{\rm left}\cup \Sigma'_{\rm ext}\cup\Sigma'_{\rm right}$, the causal diamonds of $\Sigma'_{i}$ and $\Sigma_{i}$ in region $i$ are the same, for all $i\in\{\rm left, ext, right\}$.

\noindent($iv$) Consider any subregion $A$ of $H_1\cup H_2$. For the monolayer proposal, look for a surface~$\chi_{\rm ext}$ of minimal extremal area that is homologous to $A_{\rm ext}=A$ in the exterior region and is lying on a Cauchy slice $\Sigma'_{\rm ext}$. In the bilayer case, also look for a  surface $\chi_{\rm left}$ ($\chi_{\rm right}$) of minimal extremal area, homologous to $A_{\rm left}=A\cap H_1$ ($A_{\rm right}=A\cap H_2$) in the left (right) causal patch and lying on a Cauchy slice $\Sigma'_{\rm left}$ ($\Sigma'_{\rm right}$). 

\noindent($v$) At leading order in $G\hbar$, the Von Neumann entropy of the holographic dual subsystem on $A$, that is the entanglement entropy between $A$ and its complement in $H_1\cup H_2$, is (the sum of) the area(s) of the above surface(s) divided by $4G\hbar$.  

\noindent($vi$) For all $i\in\{\rm left, ext, right\}$, recall that the codimension 2 surface $\chi_i$ being homologous to the codimension 2 surface $A_i$ means that there exists a codimension 1 surface $\mathcal{C}_i$ such that $\partial \mathcal{C}_i =\chi_i \cup A_i$.\footnote{This implies in particular that $\chi_i$ is anchored on $A_i$, \ie $\partial \chi_i=\partial A_i$.\label{nb}} 
Among all possible choices of surfaces~$\mathcal{C}_i$, a particular one,  $\mathcal{C}'_i$, lies on~$\Sigma_i'$. The part in region $i$ of the ``entanglement wedge''  of the dual system living on $A$  is the causal diamond of~$\mathcal{C}'_i$.
In the monolayer proposal, the only causal diamond to be considered is that lying in the exterior region. 
For the bilayer proposal, the entanglement wedge is the union of those in each of the three regions.
Notice that we adopt the minimal choice concerning the entanglement wedge structure in the monolayer and bilayer cases, given the rules of quantum-corrected HRT surfaces. This choice implies in particular that the entanglement wedges in the monolayer case do not extend in the interior regions. Assuming entanglement wedge reconstruction, this conclusion is in tension with static patch holography.\footnote{It would be interesting to explore the possibility that the monolayer proposal of \cite{Susskind:2021esx} gets supplemented by a consistent set of rules concerning the structure of the entanglement wedges, allowing them to extend in the interior regions. However, it is difficult to see how such a consistent set of rules could be formulated, since in all other well understood examples in the literature, the entanglement wedge structure depends on the location of the extremal surfaces, and the monolayer proposal does not involve any extremal surfaces in the interior regions.}

\noindent Some remarks are in order: 

\noindent $\bullet$ In the original HRT formulation, one looks for a homologous surface that extremizes the area functional. If several extremal surfaces exist, one selects that of smallest area. As we will see, in the de Sitter case, homologous extremal surfaces may not exist in the causal diamond located in some region~$i$. However, whether extremal surfaces exist or not, non-extremal surfaces lying on the boundary of the causal diamond can be seen as solutions of an extremization problem in an enlarged domain that has no boundary. This point of view can be relevant in all path-integral computations in the semiclassical limit. In practice, one supplements the area functional with Lagrange multipliers that impose the surface to lie inside the diamond. We will see in explicit examples that this procedure produces extremal surfaces on the boundary of the causal diamond in region $i$. Considering the latter as well, $\chi_i$ is that of minimal area and lying on some Cauchy slice $\Sigma_i'$. In case of degeneracy at the classical level, quantum corrections can lift the ambiguity.


\noindent $\bullet$ Since the homologous surface $\chi_i$ lies on a Cauchy slice $\Sigma'_i$, it is spacelike and has a real area.\footnote{In limiting cases where $\Sigma'_i$ becomes null along the boundary of region $i$, we restrict to spacelike homologous surfaces $\chi_i$. \label{lim}} Timelike and lightlike extremal homologous surfaces may exist but should not be considered. Timelike extremal surfaces would not even make sense in the computation of the entropy since their areas are pure imaginary.\footnote{See however \cite{Doi:2023zaf} for recent discussions on timelike entanglement entropy.}

\noindent $\bullet$ Extremal homologous surfaces only partially lying in the causal diamond of $\Sigma_i$ may exist, but should not be considered either.


\subsection{Quantum corrections}
\label{quantumcorrections}

To incorporate quantum corrections, we must extremize a generalized entropy that includes both geometrical and semiclassical entropy contributions \cite{Almheiri:2020cfm}. In the case of the monolayer proposal, it is straightforward to define and associate a generalized entropy to a subregion $A$ of the two screens. Indeed, it is natural to add to the geometrical area term, associated with the homologous surface $\chi_{\rm ext}$ in the exterior region, the semiclassical entropy of the bulk fields (including gravitons) on the codimension 1 slice ${\cal {C}}'_{\rm ext}\subseteq {\rm \Sigma}'_{\rm ext}$, which is bounded by $\chi_{\rm ext} \cup A$:
\begin{equation}
\label{Sgenmono}
S_{\rm gen}(\chi_{\rm ext})=\frac{{\rm Area}(\chi_{\rm ext})}{4G\hbar}+ S_{\rm semicl}({\cal {C}}'_{\rm ext}).
\end{equation}
The semiclassical entropy term is positive and of order $(G\hbar)^0$. It can only vanish in some special cases for which ${\cal {C}}'_{\rm ext}$ shrinks to a higher-codimension surface on the screens.\footnote{Divergent contributions from the semiclassical entropy term, which obey an area law, can be absorbed in the leading geometrical term \via renormalization of the gravitational Newton's constant \cite{Almheiri:2020cfm}.}  To obtain the entropy of the dual subsystem $A$ in the monolayer case, we extremize the generalized entropy functional in \Eq{Sgenmono} with respect to $\chi_{\rm ext}$. If more than one extremal surface exist, we pick the one for which the generalized entropy is the smallest:  \begin{equation}
\label{Squantummono}
    S_{\rm mono}(A) = \min \text{ext}\, S_{\rm gen}(\chi_{\rm ext}) .
\end{equation}
Such a minimal extremal surface $\chi_{\rm ext}$ is called a ``quantum  extremal surface.'' 

For the bilayer proposal, the generalization to incorporate quantum corrections turns out to be more subtle. Since the geometrical contributions are additive (and positive), the computation of the entropy to leading order reduces to three separate extremization problems respectively in the left, exterior and right regions. At the quantum level, however, we expect these extremization problems to mix, reducing to a single joint extremization problem.\footnote{The necessity of a joint extremization can be argued for even at the classical level, when we consider classical perturbations of the pure de Sitter background, \eg Schwarzschild-de Sitter \cite{Shaghoulian:2021cef, Shaghoulian:2022fop}.} Indeed, the semiclassical entropy associated with the bulk quantum field system is not additive. It is constrained, however, by the subadditivity property of entanglement entropies \cite{VanRaamsdonk:2016exw}, which implies that the semiclassical entropy on the union of different bulk regions is {\it in general smaller} than the sum of the individual entropies. Therefore, in order to minimize the generalized entropy associated with a subregion $A$ of the screens, we add to the area terms of the  homologous surfaces $\chi_i$, where $i\in\{\rm left, ext, right\}$, the semiclassical entropy on the union of the corresponding bulk slices ${\cal {C}}'_{i}$, rather than the sum of the individual semiclassical entropies. This is reminiscent of the treatment of the island contribution to the entropy of Hawking radiation in the black hole case \cite{Almheiri:2020cfm}. Thus, in the context of the bilayer proposal, we associate to $A$ the following generalized entropy
\begin{equation}
\label{Sgenbil}
    S_{\rm gen}(\chi_{\rm left},\chi_{\rm ext},\chi_{\rm right})= \sum_{i\in\{\rm{left,ext,right}\}}{\frac{{\rm Area}(\chi_i)}{4G\hbar}} + S_{\rm semicl}\big(\bigcup_{j}{\cal {C}}'_{j} \big).
\end{equation}
To obtain the entropy of $A$ in the bilayer proposal, we extremize $S_{\rm gen}$ with respect to the three surfaces $\chi_i$ and select the combination of extremal surfaces for which the generalized entropy is the smallest:
\begin{equation}
\label{bilayer_with_corrections}
    S_{\rm bil}(A)=\min \text{ext}~S_{\rm gen}(\chi_{\rm left},\chi_{\rm ext},\chi_{\rm right}).
\end{equation}

The two proposals thus extended to incorporate quantum corrections give different predictions for the holographic entropies and the entanglement-wedge structures, which we can already anticipate without studying the details of particular systems. First, notice that since the Cauchy slices ${\cal {C}}'_{\rm ext}$ that are relevant for the computation of the semiclassical entropy in the monolayer proposal are not complete bulk Cauchy slices, the semiclassical entropy on them cannot vanish, except for special cases for which ${\cal {C}}'_{\rm ext}$ collapses to a subregion of the screens. In the latter cases, however, the geometrical area terms associated with non-trivial subsystems are not zero. As a result the monolayer entropies associated with dual subsystems cannot vanish at the quantum level. This applies in particular to the union of the two screen systems, implying 
\begin{equation}
    S_{\rm mono}(H_1 \cup H_2) > 0
\end{equation}
at the quantum level. In other words, the monolayer proposal implies that the state of the two-screen system cannot be pure. 

On the other hand, the union of the three slices ${\cal {C}}'_{i}$ in the interior and the exterior regions can amount to a complete bulk Cauchy slice, leading to the vanishing of the semiclassical entropy contribution (since the bulk state on a full Cauchy slice is taken to be pure). If the quantum extremal surfaces are of zero area, we would get a vanishing bilayer entropy. As we will see in the following section, this turns out to be the case for the full two-screen system, implying that   
\begin{equation}
    S_{\rm bil}(H_1\cup H_2)=0,
\end{equation} 
including quantum corrections. The bilayer proposal predicts that the two-screen system is in a pure state. One can see that computing the semiclassical entropies independently in the three regions and adding them as separate contributions to the generalized entropy would yield a larger entropy, justifying our choice. 

Below, we will explore some further consequences of the monolayer and bilayer proposals in order to highlight inconsistencies for the former and consistency checks for the latter. Of course, one would like to obtain direct evidence for a proposal from a replica bulk path integral approach in this cosmological context\footnote{This would involve Lorentzian Schwinger-Keldysh path integrals.} (see \eg  \cite{Penington:2019kki,Almheiri:2019qdq} for the black-hole cases). 


\section{Preference for the bilayer over the monolayer proposal}
\label{sec:entropy_subsystems}

The initial aim of ``static-patch holography'' is to describe arbitrary states of Cauchy slices of the pode and antipode static patches in terms of a dual theory on holographic screens \cite{Susskind:2021omt}. Arguments based on Bousso wedges are used to propose that the screens be pushed along these Cauchy slices up to the cosmological horizons, in order to encode the entire Cauchy slices. However, for any subsystem of the screens, the monolayer prescription states that the entanglement entropy is related to the area of extremal surfaces lying entirely in the exterior region, and not in the interior ones. As a result, all entanglement wedges are restricted to subregions of the exterior one. Since entanglement wedges are the bulk regions encoded by the dual subsystems on the screens, it seems that the monolayer proposal leads to a contradiction. 

On the contrary, the bilayer proposal implies that the entanglement wedge can extend in the interior regions as well as in the exterior region, depending on the situation. This very fact indicates that this prescription can lead to a new holography conjecture that extends the static-patch holographic proposal in the sense that the two holographic screens may encode the entire spacetime:

\noindent{\bf Holography conjecture for de Sitter space:}  \textit{The full de Sitter spacetime can be encoded holographically in terms of a theory on the two cosmological horizons. The states of the interior regions are encoded on the screens at the corresponding horizons. The exterior region between the cosmological horizons emerges from the entanglement between the screens at the horizons.}

\noindent Indeed, there are two sources of entanglement between a subsystem $A$ and its complement in $H_1\cup H_2$. On one hand, $A_{\rm left}$ ($A_{\rm right}$) can be entangled with the complement of $A_{\rm right}$ in $H_2$ ($A_{\rm left}$ in $H_1$), which is causally disconnected from it. Indeed, based on the ER = EPR paradigm \cite{VanRaamsdonk:2009ar, VanRaamsdonk:2010pw, Maldacena:2013xja}, the presence of this source of quantum entanglement is manifested by the presence of the bridge connecting the two causal patches, \ie the exterior region. On the other hand, $A_{\rm left}$ ($A_{\rm right}$) can be entangled with its own complement in $H_1$ ($H_2$), which is causally connected to it.

To support the above proposals, we reconsider in this section and study in great detail the derivations of the entanglement entropies associated with a single screen or the two-screen system $H_1\cup H_2$~\cite{Susskind:2021esx, Shaghoulian:2021cef, Shaghoulian:2022fop}. To this extend, we apply our prescription ($i$)--($vi$) and proposals in \Sect{quantumcorrections} to incorporate quantum corrections in order to determine the relevant homologous surfaces and entanglement wedges, in the monolayer and bilayer cases.  
In all instances, we choose an ${\rm SO}(n)$-symmetric initial Cauchy slice $\Sigma$, which therefore  intersects the horizons of the pode and antipode at spherical screens $H_1$ and $H_2$, respectively. The global time coordinate of $H_1$ is denoted by $\tau_1$ and that of $H_2$ by $\tau_2$ (conformal times $\sigma_1$ and $\sigma_2$, respectively). Since the screens lie on a Cauchy slice~$\Sigma$ (which can be lightlike in a limiting case), we have
\be
 \sigma_1\,\sigma_2\ge 0, \qquad \tau_1\,\tau_2\ge 0.
\label{si}
\ee
At the end of this section, we also argue that the bilayer prescription can be applicable in situations where the screens are pushed in the interior regions, while the monolayer prescription leads to contradictory results.

Before presenting our arguments, let us mention that there is further supporting evidence for the bilayer proposal in the literature. As discussed in \cite{Shaghoulian:2022fop}, this proposal seems to reproduce features of entanglement entropies in SYK models~\cite{Zhang:2020kia}. Indeed, it has been argued that a double-scaling limit of the SYK model yields a good toy model for the holographic dual theory living on the screens at the horizons, in a de Sitter version of two-dimensional JT gravity~\cite{Susskind:2021esx, Susskind:2022dfz, Susskind:2022bia, Rahman:2022jsf}. Therefore, it would be interesting to compute the entropies of subsystems in this double-scaling limit of the SYK model and compare the results with those obtained by applying the bilayer proposal. 

\subsection{Entropy of a single screen}
\label{1s}

We begin our discussion with the single-screen subsystem $A=H_1$, which is entangled with its complement $H_2$. 
The monolayer and bilayer proposals yield the same results for the leading geometrical contributions to the Von Neumann entropies of $H_1$ and $H_2$. These contributions, which are of order $(G\hbar)^{-1}$, are both equal to the Gibbons-Hawking entropy for any $\tau_1$ and $\tau_2$ \cite{Susskind:2021esx, Shaghoulian:2021cef, Shaghoulian:2022fop}. However, we will see that the two proposals yield completely different entanglement-wedge structures, as well as different results for the entropies and the extremal surfaces at the quantum level. Furthermore, the bilayer prescription implies an interesting phase transition when $\tau_1$ and $\tau_2$ become equal.

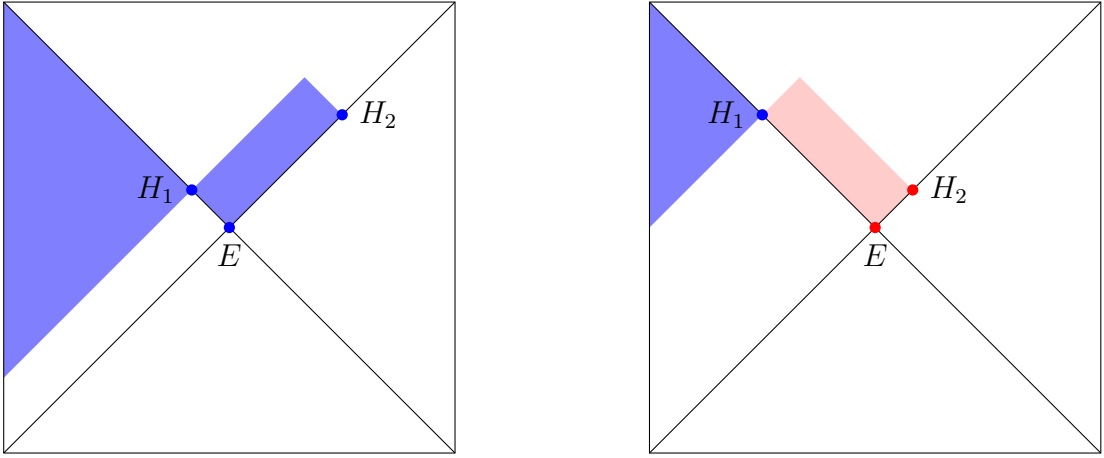
\begin{figure}[!h]
\begin{subfigure}[t]{0.48\linewidth}
\centering
\begin{tikzpicture}

\path
       +(3,3)  coordinate (IItopright)
       +(-3,3) coordinate (IItopleft)
       +(3,-3) coordinate (IIbotright)
       +(-3,-3) coordinate(IIbotleft)
      
       ;
       
\fill[fill=blue!50] (-1/2,1/2) -- (1,2) -- (3/2,3/2) -- (0,0) -- cycle;

\fill[fill=blue!50] (-3,3) -- (-1/2,1/2) -- (-3,-2) --  cycle;
       
\draw (IItopleft) --
      (IItopright) --
      (IIbotright) -- 
      (IIbotleft) --
      (IItopleft) -- cycle;
      
\draw (IItopleft) -- (IIbotright)
              (IItopright) -- (IIbotleft) ;
              

\node at (-1/2,1/2) [circle,fill,inner sep=1.5pt, blue, label = left:$H_1$]{};
\node at (3/2,3/2) [circle,fill,inner sep=1.5pt, blue, label = right:$H_2$]{};
\node at (0,0) [circle,fill,inner sep=1.5pt, blue, label = below:$E$]{};

\end{tikzpicture}
\caption{\footnotesize When $|\sigma_1|<|\sigma_2|$, the causal diamond of all Cauchy slices $\Sigma_{\rm ext}'$ is the blue rectangle. The minimal surface $\chi_{\rm ext}$ is $H_2$ and the entanglement wedge in the exterior region is the blue rectangle. \label{PL}}
\end{subfigure}
\quad \,
\begin{subfigure}[t]{0.48\linewidth}
\centering
\begin{tikzpicture}

\path
       +(3,3)  coordinate (IItopright)
       +(-3,3) coordinate (IItopleft)
       +(3,-3) coordinate (IIbotright)
       +(-3,-3) coordinate(IIbotleft)
      
       ;
       
\fill[fill=red!20] (-3/2,3/2) -- (-1,2) -- (1/2,1/2) -- (0,0) -- cycle;

\fill[fill=blue!50] (-3,3) -- (-3/2,3/2) -- (-3,0) --  cycle;
     
\draw (IItopleft) --
      (IItopright) --
      (IIbotright) -- 
      (IIbotleft) --
      (IItopleft) -- cycle;
      
\draw (IItopleft) -- (IIbotright)
              (IItopright) -- (IIbotleft) ;

\node at (-3/2,3/2) [circle,fill,inner sep=1.5pt, blue, label = left:$H_1$]{};
\node at (1/2,1/2) [circle,fill,inner sep=1.5pt, red, label = right:$H_2$]{};
\node at (0,0) [circle,fill,inner sep=1.5pt, red, label = below:$E$]{};

\end{tikzpicture}
\caption{\footnotesize When $|\sigma_1|>|\sigma_2|$, the minimal surface $\chi_{\rm ext}$ is $H_1$ and the entanglement wedge in the exterior region reduces to $H_1$.\label{PR}}
\end{subfigure}
    \caption{\footnotesize Causal diamonds (in blue or red) and entanglement wedges (in blue) for the single-screen subsystem $H_1$, for the bilayer proposal. The blue triangle is the common causal ``diamond'' of all Cauchy slices $\Sigma_{\rm left}'$.  It is also the entanglement wedge of $H_1$ in the left interior region. $E$ is the bifurcate horizon.}
    \label{fig:Penrose_dS_ext}
\end{figure}

 Consider any Cauchy slice $\Sigma'$ containing $H_1$ and $H_2$. Focusing on the left static patch, which is relevant for the bilayer proposal, $\Sigma_{\rm left}'$ and $\Sigma_{\rm left}$ have a common ``causal diamond,'' which  corresponds to the blue triangular domain in Figure~\ref{fig:Penrose_dS_ext}.
%
%
The Cauchy slice $\Sigma_{\rm left}'$ has the topology of a spherical cap bounded by $A_{\rm left}=H_1$. Therefore, the minimal extremal surface homologous to $A_{\rm left}$ is the empty set, $\chi_{\rm left}=\varnothing$. Indeed, it ``lies'' on $\Sigma_{\rm left}'$ (as it is a subset of the slice). Moreover, it is homologous to~$H_1$ since $\partial \mathcal{C}'_{\rm left}=\varnothing\cup H_1$ is trivially satisfied for $\mathcal{C}'_{\rm left}=\Sigma_{\rm left}'$. Finally, being of vanishing area, it is  trivially of minimal extremal area.\footnote{It is an absolute minimum of the area functional. It can be viewed both as a ``minimin'' and a ``maximin'' surface in the causal diamond of all possible Cauchy slices $\Sigma_{\rm left}'$. Indeed, the empty set is the homologous ``surface of minimal area'' on any $\Sigma_{\rm left}'$. It is therefore both the minimal and the maximal-area surface in the set that contains it.} The geometrical contribution from the pode interior region to the Von Neumann entropy of the screen $H_1$ is thus vanishing.  Moreover, the entanglement wedge of $H_1$ in this region is the causal diamond of $\mathcal{C}'_{\rm left}=\Sigma_{\rm left}'$, \ie the entire blue triangular domain of the left static patch. 

The geometrical contribution to the entanglement entropy arising from the antipode interior region is also null. Indeed, since $A_{\rm right}=\varnothing$, we can take $\chi_{\rm right}=\varnothing$ as the  minimal extremal homologous surface, which indeed ``lies'' on the Cauchy slice $\Sigma'_{\rm right}$.  Moreover, $\mathcal{C}'_{\rm right}=\varnothing$, as it satisfies trivially the condition $\partial \mathcal{C}'_{\rm right}=\chi_{\rm right}\cup A_{\rm right}$. The causal diamond of the empty set $\mathcal{C}'_{\rm right}$ being the empty set, there is no entanglement wedge in the right static patch. 

Since there are no geometrical contributions to the entropy of $H_1$ from the interior regions in the context of the bilayer prescription, 
the monolayer and bilayer proposals yield the same classical contribution to the entropy, from the exterior region. Consider all possible choices of Cauchy slices $\Sigma'_{\rm ext}$ in the exterior region, which by definition have boundaries $H_1\cup H_2$. They span a common causal diamond, which is the colored rectangle in Figure~\ref{fig:Penrose_dS_ext}.
Since $A_{\rm ext}=H_1$ has no boundary, the homologous surface $\chi_{\rm ext}$ lying on some Cauchy slice $\Sigma'_{\rm ext}$ must be closed (see Footnote~\ref{nb}). We will look for minimal extremal surfaces that are ${\rm SO}(n)$-symmetric.\footnote{ In other words, we assume that the ${\rm SO}(n)$ symmetry is not spontaneously broken when extremizing the area functional.} All ${\rm SO}(n)$-symmetric surfaces are spheres represented by points on the Penrose diagram and their areas depend on their positions on the diagram. In Appendix~\ref{A1}, we rederive the fact that in the exterior causal diamond, the only S$^{n-1}$ of extremal area\footnote{What is meant by this is that the variation of the area of the sphere located in the diamond vanishes as its position varies infinitesimally in the diamond, $\delta\!\A=0$. } is the bifurcate horizon~\cite{Shaghoulian:2021cef}, denoted $E$ in Figure~\ref{fig:Penrose_dS_ext}. Notice that it is a sphere of minimal area in the diamond.\footnote{On the contrary, with respect to global de Sitter space, $E$ is a saddle point of the area functional. Indeed, it is a minimax surface. That is, one first finds the maximal area sphere for each Cauchy slice connecting the pode and the antipode and then takes the minimal area one within this set. The fact that $E$ is a minimax rather than a maximin of the area functional was one of the motivations for the bilayer and monolayer proposals \cite{Shaghoulian:2021cef}.}
It also lies on the particular Cauchy slice $\Sigma'_{\rm ext}$ represented on the Penrose diagram as the lightlike line segments from $H_1$ to $E$ and from $E$ to $H_2$. This slice is the part of the boundary of the causal diamond that lies on the horizons. 
However, all spheres along this slice have the same area as $E$ (see \Eq{rad}). Therefore, they are also minimal-area surfaces in the causal diamond, even though they do not extremize the area functional (except $E$). In Appendix~\ref{A2}, we show that these extra spheres of minimal areas that are lying on the boundary of the causal diamond at the horizons are solutions of an extremization problem. In this problem, the area function depends on the position of the sphere as well as extra variables that include Lagrange multipliers. The key point is that in this formulation, the domain of definition of the area function has no boundary, allowing all minima and maxima to be seen as extremal points. Notice that restricting to the exterior causal diamond, the spheres described above become maximin surfaces \cite{Wall:2012uf,Akers:2021fut}.\footnote{Since all Cauchy slices $\Sigma'_{\rm ext}$ are bounded by $H_1\cup H_2$, the minimal-area spheres on them are $H_1$ and $H_2$ or degenerate spheres in the special cases where $\Sigma'_{\rm ext}$ has parts along the horizons.}
Since they have degenerate areas equal to that of the screen $H_1$, they all yield the same geometrical contributions to the monolayer and bilayer entropies $S_{\rm mono}$ and $S_{\rm bil}$, equal to the area of $H_1$ divided by $4G\hbar$. We conclude that 
\begin{equation}
    S_{\rm mono}(H_1) =\frac{{\rm Area}(H_1)}{4G\hbar} +{\mathcal{O}}(G\hbar)^{0}, \qquad S_{\rm bil}(H_1) =\frac{{\rm Area}(H_1)}{4G\hbar} +{\mathcal{O}}(G\hbar)^{0}.
\end{equation}
To leading order, the entropy of $H_1$ is independent of $\tau_1$, $\tau_2$ and is equal to the Gibbons-Hawking entropy of the horizon \cite{Gibbons:1977mu}, irrespective of the choice among the minimal area spheres. The same conclusion generalizes for the antipode-screen subsystem $A=H_2$: To leading order, the entropy of $H_2$ is equal to the Gibbons-Hawking entropy (in both the monolayer and bilayer proposals). 

However, choosing one or the other of the minimal-area spheres leads to different entanglement wedges in the exterior region. Indeed, letting $\chi_{\rm ext}$ be a minimal sphere on the left lightlike segment between $H_1$ and $E$, the only possible choice for $\mathcal{C}_{\rm ext}'$ is the smaller segment between $H_1$ and $\chi_{\rm ext}$ on the Penrose diagram (because the boundary of $\mathcal{C}_{\rm ext}'$ must be $H_1 \cup \chi_{\rm ext}$ and $H_1$ and $\chi_{\rm ext}$ are lightlike separated). As a result, the corresponding entanglement wedge in the exterior is the lightlike segment between $H_1$ and this minimal sphere. The entanglement wedge confines on the left horizon. In particular, if $\chi_{\rm ext}=H_1$, the entanglement wedge shrinks to $H_1$ and does not extend in the exterior region. On the other hand, if $\chi_{\rm ext}$ is a minimal area sphere on the right horizon between $E$ and $H_2$, $\mathcal{C}_{\rm ext}'$ can be chosen to be any Cauchy slice bounded by $H_1\cup \chi_{\rm ext}$.\footnote{Indeed, such a codimension one surface $\C'_{\rm ext}$ can be completed into a Cauchy slice $\Sigma_{\rm ext}'$ by adding to it the surface represented on the Penrose diagram by the line segment between $\chi_{\rm ext}$ and $H_2$.} All these codimension one surfaces have the same causal diamond. The corresponding entanglement wedge is the common causal diamond of the various $\mathcal{C}_{\rm ext}'$, which is a smaller rectangle inside the colored rectangle of Figure~\ref{fig:Penrose_dS_ext}. In the special case $\chi_{\rm ext}=H_2$, the entanglement wedge assumes its largest possible extent, covering the whole colored rectangular region. This would imply that the full causal diamond in the exterior region can be reconstructed holographically in terms of degrees of freedom on $H_1$.    

Hence it is important to figure out which minimal area sphere is actually the correct one.\footnote{Let us remark that only the screens themselves are marginally stable maximin surfaces according to the criteria of \cite{Akers:2021fut}. It would be interesting to understand the applicability of these criteria in this time-dependent, cosmological context. As we will show in this section, quantum effects seem to always favor one of the two screens for being the correct extremal surface.}
The classical degeneracy can be lifted at the quantum level.\footnote{The degeneracy can also be lifted in the presence of classical perturbations of the geometry. Such cases will be examined as part of a future work \cite{FPRT}.} To incorporate quantum corrections in the bilayer and monolayer proposals, we apply the prescriptions of \Sect{quantumcorrections}. Here, we shall consider the first order correction to the entropy, which can be obtained by adding to the leading geometrical contributions the semiclassical entropy on a Cauchy slice of the entanglement wedge~\cite{Faulkner:2013ana}. We expect the actual quantum minimal extremal surface, which is the minimal extremal surface for the generalized entropy, to be close to the classical minimal one for which the first-order corrected entropy is the smallest.\footnote{By doing so, we neglect the backreaction of the semiclassical corrections on the classical surface.} The bilayer and monolayer proposals give different results since they yield different entanglement wedges. 

Let $\chi_{\rm ext}$ be one of the minimal area spheres on any of the two horizons, and denote $\bar{\sigma}$ its conformal time. In the bilayer proposal, we apply \Eq{Sgenbil} and add to the geometrical contribution the semiclassical entropy on $\mathcal{C}_{\rm left}'\cup \mathcal{C}_{\rm ext}'\cup\mathcal{C}_{\rm right}'=\Sigma_{\rm left}'\cup \mathcal{C}_{\rm ext}'$. Indeed, the full entanglement wedge contains the blue triangular region of the left interior region (see Figure~\ref{fig:Penrose_dS_ext}) as well as a component in the exterior one. The data on $\Sigma_{\rm left}'\cup \mathcal{C}_{\rm ext}'$ can be unitarily related to the data on the constant time slice running between the pode and $\chi_{\rm ext}$, which we denote by $\Sigma_1(\bar{\sigma})$. Hence the semiclassical entropy on $\Sigma_{\rm left}'\cup \mathcal{C}_{\rm ext}'$ is equal to the semiclassical entropy on $\Sigma_1(\bar{\sigma})$. Since the bulk field system on the full constant time slice $\Sigma_{\rm full}(\bar{\sigma})$ is in a pure state, the semiclassical entropy on $\Sigma_1(\bar{\sigma})$ is equal to the semiclassical entropy on $\Sigma_2(\bar{\sigma})$, which runs between $\chi_{\rm ext}$ and the antipode. Equivalently, this is the entanglement entropy $S_{\rm ent}(\chi_{\rm ext})$ through the sphere $\chi_{\rm ext}$:
\begin{equation}
    S_{\rm semicl}\big(\Sigma_{\rm left}'\cup \mathcal{C}_{\rm ext}'\big)=S_{\rm semicl}\big(\Sigma_1(\bar{\sigma})\big)=S_{\rm ent}(\chi_{\rm ext}).
\end{equation}
The first-order corrected entropy is then
\begin{align}
    S_{\rm gen}(\chi_{\rm ext}) &=\frac{{\rm Area}(\chi_{\rm left})+{\rm Area}(\chi_{\rm ext})+{\rm Area}(\chi_{\rm right})}{4G\hbar}+S_{\rm semicl}\big(\Sigma_{\rm left}'\cup \mathcal{C}_{\rm ext}'\big)\nonumber\\
    &=\frac{{\rm Area}(\chi_{\rm ext})}{4G\hbar}+S_{\rm ent}(\chi_{\rm ext}),
\end{align}
which is the ``quantum area'' of the union of the sphere $\chi_{\rm ext}$ and the empty sets $\chi_{\rm left}$ and $\chi_{\rm right}$, divided by $4G\hbar$~\cite{Strominger:2003br}. The dominant minimal area sphere $\chi_{\rm ext}$ is the one for which the total quantum area is the smallest. This is the minimal area sphere for which $S_{\rm ent}(\chi_{\rm ext})$ is minimal (since the classical areas are degenerate).

Let us examine how $S_{\rm gen}(\chi_{\rm ext})$ behaves as a function of $\bar\sigma\in [-\pi/2, \pi/2]$. During the contracting phase of de Sitter space, particles can leave the region between the cosmological horizons and enter the static patches without being able to escape. Hence, entangled particles produced in pairs during the contracting cosmological evolution can separate into different patches, leading to a growth of $S_{\rm gen}(\chi_{\rm ext})$ associated with $\Sigma_1(\bar{\sigma})$ (or $S_{\rm ent}(\chi_{\rm ext})$ since the geometrical term is constant) as $\bar{\sigma}$ increases from $-\pi/2$ to $0$. On the other hand, during the expanding phase, particles can exit from the static patches to the exterior region, but not the other way. So, eventually entangled particles can reunite in the region between the cosmic horizons, leading to a decrease of $S_{\rm gen}(\chi_{\rm ext})$ associated with $\Sigma_1(\bar{\sigma})$.

Therefore, $S_{\rm ent}(\chi_{\rm ext})$ is expected to increase as $\bar\sigma$ increases from $-\pi/2$ to $0$, reaching its maximal value when $\chi_{\rm ext}$ coincides with the bifurcate horizon $E$ at $\bar\sigma=0$, and then it should decrease as $\bar\sigma$ increases from $0$ to $\pi/2$. This is irrespective of whether $\chi_{\rm ext}$ is on the left or the right horizon. Thus, along a future horizon ($\bar \sigma\geq 0$), the total quantum area  is expected to obey the quantum Bousso bound of \cite{Strominger:2003br}, in accordance with the quantum focusing conjecture \cite{Bousso:2015mna}. Indeed, since the rate $\d S_{\rm gen}/\d\lambda$, where $\lambda$ is an affine parameter along a null line, cannot increase as $\lambda$ increases, if this rate is established to be negative at a point on the horizon, the quantum focusing conjecture ensures that $S_{\rm gen}$ will decrease monotonically as $\lambda$ increases. Along a past horizon, the quantum Bousso bound is expected to be obeyed in the past direction.

We can demonstrate explicitly that these expectations hold for the case of two-dimensional de Sitter space, taking for the matter theory a conformal field theory of central charge $c$. The CFT is assumed to be in its vacuum. The metric of dS$_2$ (\ie $n=1$) is given by
\begin{equation}
    \d s^2 = \frac{1}{\cos^2\sigma}\!\left(-\d\sigma^2 + \d\theta_1^2\right)\!,
\end{equation}
where $\theta_1 \sim \theta_1 +2\pi$.\footnote{In the two-dimensional JT models of \cite{Sybesma:2020fxg}, the value of the quantum corrected dilaton along the horizons, which determines the geometrical contributions, remains constant.} We take $\theta_1$ to lie in the interval $(-\pi, \pi]$ and identify the ends. The metric is conformal to that on a flat (Lorentzian) cylinder with the spatial direction compactified on a circle of unit radius. The Penrose diagram can be taken to be a square, where each point represents two actual spacetime points with coordinates $(\theta_1,\sigma)$ and $(-\theta_1,\sigma)$. The pode is at $\theta_1=0$. Since the metric is conformal to that on the cylinder (and the CFT is in its vacuum), the semiclassical entropy of an arc of comoving size $\Delta \theta_1\in(0,2\pi)$, at constant time~$\sigma$, is given by
\begin{equation}
    S_{\rm semicl}= \frac{c}{3}\ln\!\left[\frac{2\sin{\frac{\Delta \theta_1}{2}}}{\epsilon_{{\rm uv},\,\theta}}\right]\!,
\end{equation}
where $\epsilon_{{\rm uv},\,\theta}$ is a UV cutoff in the angular comoving coordinate \cite{Calabrese:2004eu,Chen:2020tes}. For a CFT, the Von Neumann entropy associated with an arc at constant time is independent of the conformal factor of the metric (which in this case depends only on time). Notice that the Von Neumann entropy for the complementary arc, which has comoving size $2\pi - \Delta \theta_1$, is the same.

Now let $\chi_{\rm ext}$ correspond to a pair of points on the negative (or positive) slope horizon of the Penrose diagram. Then, $\Sigma_1(\bar{\sigma})$ is an arc of comoving size $\theta_1-(-\theta_1)=2(\pi/2 - \bar{\sigma})$ (or $2(\pi/2 + \bar{\sigma})$). The comoving size of the arc changes with $\bar \sigma$. In both cases, we obtain
\begin{equation}
    S_{\rm ent}(\chi_{\rm ext})=\frac{c}{3}\ln\left[\frac{2\cos{\bar{\sigma}}}{\epsilon_{{\rm uv},\,\theta}}\right]\!.
\end{equation}
As expected, the entanglement entropy through $\chi_{\rm ext}$ grows as $\bar \sigma$ increases from $-\pi/2$ to $0$, reaches a maximal value at $\bar\sigma=0$, and then it decreases monotonically as $\bar \sigma$ increases from $0$ to $\pi/2$.

The considerations above show that in the context of the bilayer prescription, quantum corrections lift the classical degeneracy, favoring the minimal area sphere for which $|\bar\sigma|$ is the biggest. This is the sphere with the smallest quantum area. Let us assume for instance that $|\sigma_1|<|\sigma_2|$. In this case, this sphere is the second screen $H_2$. Therefore, the entanglement wedge of $H_1$ covers the left blue triangular region and the blue causal-diamond region in the exterior, as shown in Figure~\ref{PL}. It is easy to generalize the discussion for the second single-screen system $A=H_2$. The relevant $\chi_{\rm ext}$ is now the screen $H_2$ itself. The entanglement wedge does not extend in the exterior region, covering the causal triangular region in the right patch only. We conclude that among the minimal surfaces $H_1$ and $H_2$, the one with the greatest quantum area\footnote{Recall that the quantum area does not depend on the single-screen system we consider, since $S_{\rm semicl}\big(\Sigma_1(\bar{\sigma})\big)=S_{\rm semicl}\big(\Sigma_2(\bar{\sigma})\big)$.} can be viewed as a single-screen system that encodes the causal-diamond region in the exterior. It is the one that is closest to $E$ on the Penrose diagram. However, the Von Neumann entropies of the two single-screen subsystems $H_1$ and $H_2$ are always equal, even at the semiclassical level.\footnote{We will see in the next subsection that this should remain true at the exact quantum level.} Indeed, both Von Neumann entropies are equal to the smaller of the two quantum areas of the minimal surfaces $H_1$ and $H_2$, divided by $4G\hbar$.

An interesting phase transition occurs at $\sigma_1 = \sigma_2$. At equal screen conformal times, the quantum areas of the minimal surfaces $H_1$ and $H_2$ are equal and the degeneracy between them is not lifted by the first-order quantum corrections. When $|\sigma_1|$ becomes bigger than~$|\sigma_2|$, the entanglement wedge of the single-screen subsystem $H_1$ now confines in the left static patch (the blue triangle in Figure~\ref{PR}), while the entanglement wedge of the single-screen subsystem $H_2$ extends and covers the causal-diamond region in the exterior. This phase transition implies the presence of non-local interactions between the holographic degrees of freedom of the two screens, as was argued to be the case in \cite{Shaghoulian:2021cef}. This also follows from the fact that, as we will see in the following subsection, the two-screen system describes the exterior.  

The conclusions are completely different in the context of the monolayer prescription. Since the entanglement wedge extends only in the exterior region, the generalized entropy \Eq{Sgenmono} for the single-screen subsystem $A=H_1$ becomes
\be
    S_{\rm gen}(\chi_{\rm ext}) =\frac{{\rm Area}(\chi_{\rm ext})}{4G\hbar}+S_{\rm semicl}\big(\mathcal{C}_{\rm ext}'\big).
\ee
The semiclassical contribution $S_{\rm semicl}$ being always positive, it is minimized for the minimal-area sphere $\chi_{\rm ext}=H_1$, as it vanishes in this case. This has two consequences. First, the Von Neumann entropy of a single-screen subsystem remains uncorrected, equal to the classical area of the screen divided by $4G\hbar$. Second, the entanglement wedge for a single-screen subsystem is minimal, covering only the screen itself.  


\subsection{Entropy of both screens} 

\begin{figure}[!h]
\centering
\begin{tikzpicture}

\path
       +(3,3)  coordinate (IItopright)
       +(-3,3) coordinate (IItopleft)
       +(3,-3) coordinate (IIbotright)
       +(-3,-3) coordinate(IIbotleft)
      
       ;
       
\fill[fill=blue!50] (-1/2,1/2) -- (1,2) -- (3/2,3/2) -- (0,0) -- cycle;

\fill[fill=blue!50] (-3,3) -- (-1/2,1/2) -- (-3,-2) --  cycle;

\fill[fill=blue!50] (3/2,3/2) -- (3,3) -- (3,0) --  cycle;
       
\draw (IItopleft) --
      (IItopright) --
      (IIbotright) -- 
      (IIbotleft) --
      (IItopleft) -- cycle;
      
\draw (IItopleft) -- (IIbotright)
              (IItopright) -- (IIbotleft) ;

\node at (-1/2,1/2) [circle,fill,inner sep=1.5pt, blue, label = left:$H_1$]{};
\node at (3/2,3/2) [circle,fill,inner sep=1.5pt, blue, label = right:$H_2$]{};
\node at (0,0) [circle,fill,inner sep=1.5pt, blue, label = below:$E$]{};

\end{tikzpicture}
\caption{\footnotesize The three blue regions are the causal diamonds of arbitrary Cauchy slices $\Sigma_{\rm left}'$, $\Sigma_{\rm ext}'$, $\Sigma_{\rm right}'$. In the bilayer case, they are also the entanglement wedges in the left interior, exterior and right interior regions for the two-screen system $H_1\cup H_2$.}
    \label{fig:Penrose_dS_ext2}
\end{figure}
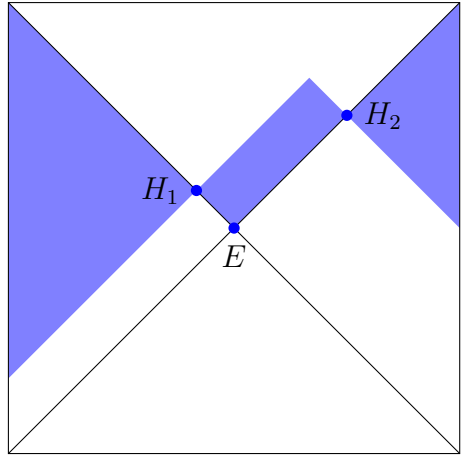

We can also consider the full two-screen system $A=H_1 \cup H_2$. In the bilayer case, the relevant causal diamonds are depicted in blue in Figure~\ref{fig:Penrose_dS_ext2}. 
%
%
The minimal extremal surface homologous to $A_{\rm left}=H_1$ is $\chi_{\rm left}'=\varnothing$, for the same reasons as given for the single-screen subsystem $H_1$. Similarly, \mbox{$\chi_{\rm right}=\varnothing$} is the minimal extremal surface homologous to $A_{\rm right}=H_2$, in the right causal patch. Finally, the same arguments apply to the exterior region, for which we have \mbox{$A_{\rm ext}=H_1\cup H_2$} and $\chi_{\rm ext}=\varnothing$. Indeed, the empty surface  is a subset of any Cauchy slice $\Sigma_{\rm ext}'$ and the homology constraint $\partial \mathcal{C}'_{\rm ext}=\varnothing\cup A_{\rm ext}$ is trivially satisfied for $\mathcal{C}'_{\rm ext}=\Sigma_{\rm ext}'$.  The empty surface is also of minimal extremal area.
Therefore, at any screen global times $\tau_1$, $\tau_2$, the leading geometrical contributions to the Von Neumann entropy of the dual holographic two-screen system are vanishing, in both the monolayer and the bilayer proposals: 
\begin{equation}
     S_{\rm mono}(H_1\cup H_2) =0 +{\mathcal{O}}(G\hbar)^{0}, \qquad S_{\rm bil}(H_1\cup H_2) =0 +{\mathcal{O}}(G\hbar)^{0}.
\label{eq:entropy_H1_H2}
\end{equation}

In any region $i\in\{{\rm left}, {\rm ext},{\rm right}\}$, we have $\mathcal{C}_i'=\Sigma'_i$. Hence, the entanglement wedge in region $i$ is nothing but the causal diamond of $\Sigma_i$. This very fact is consistent with the bilayer proposal. Indeed, since the entanglement wedges in the interior regions contain respectively any entire Cauchy slices $\Sigma'_{\rm left}$ and $\Sigma'_{\rm right}$, it is consistent to conjecture that the screens $H_1$ and~$H_2$ encode all information on them, as was motivated by the Bousso wedges. As a result, the vanishing of the entropy must be exact at the quantum level, when the two screens coincide at $\tau_1=\tau_2=0$. The reason is that in this case, there is no barrel region between them and so the union of $\Sigma'_{\rm left}$ and $\Sigma'_{\rm right}$ is the entire Cauchy slice $\Sigma'$. Since the full de Sitter state is assumed to be in a pure state, the entropy of the two-screen system at $\tau_1=\tau_2=0$ must vanish to all orders in~$G\hbar$. 

Furthermore, this result should persist regardless of the existence of the barrel region, when the screens are considered at arbitrary~$\tau_1$ and $\tau_2$. This statement is motivated by the entanglement wedge of $H_1\cup H_2$, which is the union of the causal diamonds of all three regions\footnote{As a consistency check, the entanglement wedge of $H_1 \cup H_2$ contains the union of the entanglement wedges associated with the two single-screen subsystems.} and thus contains entirely any Cauchy slice $\Sigma'$.\footnote{In particular, as the screens evolve with time, each spacetime point will have been inside the entanglement wedge at least once.} The state on $\Sigma'$ can be unitarily evolved from the state on a Cauchy slice of de Sitter space that passes through the bifurcate horizon. Since the state on this slice is pure, then, at any times $\tau_1$, $\tau_2$, the two-screen system $H_1\cup H_2$ must be in a pure state, with an exactly vanishing entropy.




A corollary of the above conclusion is that the bilayer entanglement entropy of the single-screen subsystem $H_1$ equals that of the single-screen subsystem $H_2$ at the exact level in $G\hbar$ and any times $\tau_1$, $\tau_2$, 
and not only at the leading and next to leading order, 
as argued in \Sect{1s}. 

Another argument for the purity of the two-screen system is the following.  
When the screens coincide at the bifurcation point,\footnote{Such a Cauchy slice must be included in a foliation of de Sitter space with $SO(n)$ symmetric slices.} we know that the state is pure. The evolution of the screens in time along the horizons should be described by the action of a (possibly unitary) linear operator on the closed quantum-mechanical system living on the screens. Such a linear operator cannot transform a pure state into a mixed state.

In the monolayer case, the entanglement wedge of $H_1\cup H_2$ is only the causal diamond of~$\Sigma_{\rm ext}$. As said at the beginning of this section, this seems incompatible with the conjecture motivated by the Bousso wedges, namely that $H_1$ and $H_2$ encode all the information respectively contained on Cauchy slices $\Sigma'_{\rm left}$ and $\Sigma'_{\rm right}$. Moreover, there is no obvious reason for explaining the fact that $S_{\rm mono}(H_1\cup H_2)$ should vanish, even at the semiclassical level, since the Cauchy slice $\Sigma'$, which is in a pure state, is not entirely contained in the entanglement wedge. 


\subsection{Pushing the screens on stretched horizons}

Let us now consider situations where the screens are pushed away from the cosmological horizons, and placed on stretched horizons in the interior regions. Applying the rules of the bilayer and monolayer proposals blindly, without any modification, leads to paradoxical behavior.  Indeed, we would end up in situations where the entanglement wedges continue to extend in the exterior region between the stretched horizons, indicating that the exterior region continues to be reconstructible by the two-screen system, while the number of degrees of freedom on the two screens becomes arbitrarily small. In fact, the areas of the screens decrease as these are pushed further and further in the interior regions, becoming vanishingly small as they approach the pode and antipode, respectively. 

In the context of the monolayer proposal, this paradoxical behavior is more acute. Indeed, we have seen that the entanglement wedge of a single-screen subsystem at the horizon confines to the screen itself. So, in the monolayer proposal, no part of the exterior region can be reconstructed by a single-screen system, when the two screens are at the cosmological horizons. Now, let us consider the case $\sigma_1=\sigma_2=\bar\sigma$ and push the spherical screens respectively in the left and right interior regions, along the constant time slice ${\rm \Sigma}_{\rm full}(\bar\sigma)$. Let us further place them at asymmetric points of the Penrose diagram, so that their areas are no longer degenerate, say ${\rm Area}(H_1)>{\rm Area}(H_2)$. Then, the minimal area sphere $\chi_{\rm ext}$ in the exterior causal diamond is the screen $H_2$ with the smallest area, for both single-screen subsystems. Suddenly, the entanglement wedge of $H_1$ becomes non-trivial and covers the entire causal-diamond region in the exterior. This is despite the fact that the number of degrees of freedom on both screens can become arbitrarily small.    

To remedy this behavior in the context of the bilayer proposal, we will argue below that we can replace the holographic theory on the screens on stretched horizons with a cutoff version of the initial one, where the degrees of freedom describing the region between the stretched horizons (that includes the region between the cosmological horizons) have been integrated out. Thus, no contributions to the entropies of subsystems of the screens can arise  from the exterior region between the stretched horizons, and the entanglement wedges are restricted to lie in the interior regions. While this cutoff version is applicable in the bilayer case, it is incompatible with the monolayer proposal, since no contributions arise from the interior regions in the latter case. 

Indeed, recall that in static coordinates, the left static patch behaves like a thermal cavity with varying proper temperature, given by
\begin{equation}
    T=\frac{1}{2\pi\sqrt{1-r^2}},
\end{equation}
in units of the de Sitter radius of curvature (and likewise for the right static patch). The temperature is precisely $1/2\pi$ at the pode, and diverges at the cosmological horizon at $r=1$. As we remarked in \Sect{dSgeo}, we can regulate this behavior by introducing a cutoff surface along a hyperbola $r=1-\varepsilon$, for suitably small positive $\varepsilon$, which we identify with the stretched horizon \cite{Susskind:2021omt}. We may tune $\varepsilon$ so that the temperature at the stretched horizon does not exceed the string or the Planck scale. The proper distance of the stretched horizon from the cosmological horizon is of order $\sqrt{\epsilon}$.

In the framework of the bilayer proposal, let us now push the left screen in the interior region, and place it on the stretched horizon. As the screen evolves along the hyperbola $r=1-\varepsilon$, its area remains constant, but now this is smaller by a factor $(1-\varepsilon)^{n-1}$ (see \Eq{staticds}). Evidently a number of holographic degrees of freedom have been integrated out (modifying the effective theory on the screen). We associate these with the degrees of freedom giving rise to the strip bulk region between the stretched and the left cosmological horizon, as well as degrees of freedom encoding part of the exterior region between the two cosmological horizons. With this strip bulk region effectively removed, the screen can no longer be deformed in the exterior region between the cosmological horizons, and so there can be no contributions to the entropy of the left screen or subsystems of it from the exterior region. 
Rather, the entanglement entropy of the left-screen subsystem with the rest arises from thermal, semiclassical contributions from the interior region, which can be computed on a constant $t$ slice between the pode and the screen.\footnote{The thermal semiclassical entropy on a constant time slice $t$ diverges as the cutoff $\varepsilon \to 0$ and obeys an area law. Upon adding the geometrical area term at the bifurcate horizon, the divergence at $\varepsilon =0$ can be regulated \via a renormalization of the gravitational Newton's constant $G$.} This semiclassical bulk computation (in the static $(r,t)$ coordinates) of entanglement entropy can be trusted for sufficiently large~$\varepsilon$, so that the cutoff scale is smaller than the Planck-energy scale. As the screen is pushed further in the interior, more and more holographic degrees of freedom are integrated out, and less and less part of the static patch is covered holographically by the left-screen subsystem. 

Since the screen at the stretched horizon lies on a constant $t$ surface, this can be obtained from the screen at the symmetric slice $t=0$ \via the unitary evolution operator $e^{i\mathcal{H}t}$, where~$\mathcal{H}$ is the Hamiltonian conjugate to the static time $t$. So the entropy of the screen subsystem remains constant as this evolves along the stretched horizon.       

We may also regulate both static patches and place both screens on stretched horizons. We may take the screens to lie on constant $t_{\rm L}$ and $t_{\rm R}=-t_{\rm L}$ surfaces, where $t_{\rm L}$ and $t_{\rm R}$ are the static times in the left and right patches, respectively. The screens can be unitarily evolved from the pair of screens at the symmetric $\sigma=0$ slice, \via the Hamiltonian $\mathcal{H}_{\rm L}+\mathcal{H}_{\rm R}$. Notice however that the state of the two-screen system is no longer pure, since the degrees of freedom giving rise to the part of the symmetric $\sigma=0$ slice between the screens have been integrated out. These UV degrees of freedom are responsible  for the emergence of the exterior region between the two cosmological horizons. This region, together with the strip regions between the stretched and the cosmological horizons in the interiors, are no longer encodable in the regulated holographic two-screen system.


\section{Entropy of subsystems of the screens in \bm dS$_3$}
\label{12arcs}

The monolayer and bilayer proposals can also be used in the case of subsystems partially covering the screen(s) $H_1$ and/or $H_2$. Examples can be considered for de Sitter space in three dimensions ($n=2$), for which explicit computations can be done. The initial Cauchy slice $\Sigma$ is chosen to be ${\rm SO}(2)$-symmetric, which implies the screens to be circles. In the following, we reexamine the computations of the geometrical contributions to the entanglement entropies for subsystems of $H_1\cup H_2$ consisting in one or two arcs~\cite{Shaghoulian:2021cef, Shaghoulian:2022fop}. This leads to extra arguments in favour of the bilayer over the monolayer proposal. 

\subsection{Entropy of an arc of a screen in \bm dS$_3$}
\label{1s'}

The metric~(\ref{ds2}) for dS$_3$ can be written as
\begin{equation}\label{eq:dS3_metric}
\d s^2=\frac{1}{\cos^2\sigma}\left(-\d\sigma^2+\d\theta_1^2+\sin^2\theta_1 \,\d\theta_2^2\right)\!,
\end{equation}
where $\theta_2\in[0,2\pi)$. The left screen $H_1$ and the right screen $H_2$ are at conformal times $\sigma_1$ and $\sigma_2$ satisfying \Eq{si}. They are circles parametrized by $\theta_2$ and of circumference $2\pi$ (see \Eq{rad}). As a subsystem $A$ of $H_1\cup H_2$ we take the arc of length $\Theta$ on the screen $H_1$, parametrized by $\theta_2\in[0,\Theta]$, for a given $\Theta\in(0, 2\pi)$. Following the proposals of \Sect{monobil}, the ``surfaces'' to be considered in this case are the minimal extremal curves $\chi_i$ homologous to $A_i$ and lying on Cauchy slices $\Sigma_i'$, where $i\in\{{\rm left}, {\rm ext}, {\rm right}\}$ ($i=\rm ext$) in the bilayer (monolayer) case. The proper lengths (length) of these curves (this curve), divided by $4G\hbar$, is the geometrical contribution to the entanglement entropy of the arc subsystem with its complement in $H_1\cup H_2$. 

The intersection of the arc $A$ and $H_2$, which is relevant for the bilayer proposal, is the empty set, $A_{\rm right}=\varnothing$. Therefore, as in the case of the full single-screen subsystem $H_1$ in \Sect{1s}, the classical contribution to the entropy arising from the antipode interior region vanishes and the entanglement wedge in this region is the empty set. 

For both proposals, to figure out the contribution of a curve $\chi_{\rm ext}$ in the exterior region, let us look for geodesics anchored on the endpoints of the arc. This can be done by noticing that given two distinct points in dS$_3$, there are two geodesic curves starting and ending at these points.\footnote{Their union is similar to the great circle passing through two distinct points on S$^3$.} As seen in \cite{Shaghoulian:2021cef} and discussed in more detail in Appendix~\ref{sec_embedding_one_arc},  specializing to the endpoints of the arc, we find that:

\noindent $\bullet$ The first geodesic lies exclusively on the horizon of the pode and satisfies at each of its points $|\sigma|\ge|\sigma_1|$. Its length $L$ is equal to the smaller between $\Theta$ and $2\pi-\Theta$, which are respectively the lengths of the arc and its complementary arc on $H_1$. 

\noindent $\bullet$ The second geodesic extends on both horizons and $\sigma$ changes sign along the curve. (The latter crosses at 2 points the bifurcate horizon $E$, which is a circle.) Its length $L'$ is the greater between $\Theta$ and $2\pi-\Theta$.

\noindent  Notice that even though the lengths of the geodesic curves are those of the arc and its complement on the circle $H_1$, they do not identify to them since the conformal time $\sigma$ equals~$\sigma_1$ only at the anchorage endpoints.


Both geodesics lie on the boundary of the exterior region. However, since the first satisfies $|\sigma|\ge |\sigma_1|$, it does not lie on any Cauchy slice $\Sigma'_{\rm ext}$. (Recall that these slices span the colored rectangle in the exterior region shown in Figure~\ref{fig:Penrose_dS_ext}.) Moreover, the second geodesic does not lie either on such a slice $\Sigma'_{\rm ext}$, since $\sigma$ changes sign along it. The fact that there is no curve of extremal length in the causal diamond in the exterior region may not be an issue. Indeed, we still have to consider non-extremal curves on the boundary of the diamond that can be seen as solutions of a suitable length-extremization problem. In Appendix~\ref{dege curves}, we consider the class of curves homotopic in the two horizons to the geodesic of length $L$, and similarly the class of curves homotopic in the two horizons to the geodesic of length~$L'$. One of these classes contains the arc of angle $\Theta$ and all curves in this class are homologous to the arc in the exterior region. The other class contains the complementary arc of angle $2\pi-\Theta$ in~$H_1$ and all curves in this class are homologous to the complementary arc in the exterior region.\footnote{In particular, the two complementary arcs are not homologous in the exterior region.} We are interested in the class of curves homologous to the arc of angle~$\Theta$. We have shown that their lengths are greater or equal to~$\Theta$, which is the value reached when $\theta_2$ varies monotonically along the curves. Let us consider the curves of length $\Theta$ that are also lying entirely on the boundary of the exterior causal diamond.\footnote{In Appendix~\ref{dege curves}, they are among those denoted $\chi$ or $\check \chi$ when $\Theta\in(0,\pi]$ and  $\chi'$ or $\check \chi'$  when $\Theta\in(\pi,2\pi)$, with  $\tilde\theta_2(\lambda)$ monotonic.} They are valid candidate curves $\chi_{\rm ext}$, first because they are located on the limiting Cauchy slice $\Sigma_{\rm ext}'$ represented on the Penrose diagram of Figure~\ref{fig:Penrose_dS_ext} as the line segments from  $H_1$ to $E$ and from $E$ to  $H_2$, and secondly because they are curves of ``minimal extremal lengths'' in a certain sense. In fact, by reasoning as in Appendix~\ref{nu(lambda)},\footnote{ Appendix~\ref{nu(lambda)} treats the case of the causal diamond in the pode interior region rather than in the exterior one.} one can extremize a length functional defined for curves attached to the arc with unrestricted locations, but supplemented with Lagrange multipliers enforcing the extremal curves to lie in the exterior causal diamond only on-shell. All extrema have lengths $\Theta+2\pi w$ or $2\pi-\Theta+2\pi w$, where $w\in\natural$ counts the number of times the curve winds around $\Sigma_{\rm ext}'$, which has the topology of a cylinder. Those of length~$\Theta$ are the only ones homologous to the arc in the exterior region. Since they have the same length, they are indeed ``minimal extremal curves,'' which are precisely the candidate curves $\chi_{\rm ext}$ mentioned above.  As a result, the monolayer proposal yields for the Von Neumann entropy of the arc 
\begin{equation}
    S_{\rm mono}(\Theta) = \frac{\Theta}{4G\hbar}+{\mathcal{O}}(G\hbar)^{0}, \quad \Theta\in[0,2\pi],
    \label{smt}
\end{equation}
where we include the limiting cases $\Theta\to 0$ and $\Theta\to 2\pi$. Indeed, this formula is compatible with a vanishing entropy for the empty subsystem ($\Theta=0$). It also reproduces the expected result for $\Theta = 2\pi$, since the entropy of the full screen $H_1$ should be given by its circumference divided by $4G\hbar$. 

However, lifting in the monolayer case the degeneracy between all candidate curves $\chi_{\rm ext}$ is important, as each choice leads to a different entanglement wedge in the exterior causal diamond. For this purpose, we add to the leading classical result the semiclassical contribution, $S_{\rm semicl}(\mathcal{C}_{\rm ext}')$, which is nonnegative. Since the arc is the only choice of curve $\chi_{\rm ext}$ for which $\mathcal{C}_{\rm ext}'$ shrinks to the arc itself and leads to a vanishing semiclassical contribution, it is the right choice. As a result, the entanglement wedge of the arc subsystem in the monolayer proposal reduces to the arc itself. It does not cover any region in the bulk of the exterior or of the interior regions. 

To apply the bilayer proposal, we need to add the contribution of a minimal extremal curve $\chi_{\rm left}$ homologous in the left static patch to $A_{\rm left}$, which is the arc itself, and lying on a Cauchy slice $\Sigma_{\rm left}'$. Among the two geodesics described above, only that of length $L=\inf(\Theta,2\pi-\Theta)$ satisfies both constraints.  It is actually homologous to the arc as well as to the complementary arc on $H_1$.\footnote{Indeed, both arcs are homologous in the interior region of the pode, since all Cauchy slices $\Sigma_{\rm left}'$ are caps with boundary the screen $H_1$, which is the union of the two complementary arcs.} Moreover, the geodesic lies for instance on the Cauchy slice $\Sigma_{\rm left}'$ represented in Figure~\ref{fig:Penrose_dS_ext} as the line segment between $H_1$ and the corner $(\theta_1,\sigma)=(0,\sign(\sigma_1)\pi/2)$. On the contrary, the second geodesic does not lie on any $\Sigma_{\rm left}'$ since it also extends on the horizon of the antipode. Let us now consider the class of curves homotopic on the two horizons to the geodesic of length $L$. Except the geodesic itself, none of the curves of the class is extremal. As shown in Appendix~\ref{dege curves}, their lengths are bounded from below by $L=\inf(\Theta,2\pi-\Theta)$, which is the value reached when $\theta_2$ varies monotonically along the curves. Among those of length~$L$, let us restrict to those that are extending on the boundary of the causal diamond in the pode interior region.\footnote{They are among those denoted $\chi$ in Appendix~\ref{dege curves}.}  These curves are valid candidate curves $\chi_{\rm left}$. Indeed, they are lying on the Cauchy slice $\Sigma_{\rm left}'$ mentioned above, namely between $H_1$ and the corner $(\theta_1,\sigma)=(0,\sign(\sigma_1)\pi/2)$ of Figure~\ref{fig:Penrose_dS_ext}. Moreover, they are curves of ``minimal extremal lengths'' in a certain sense. As shown in Appendix~\ref{nu(lambda)}, one can extremize a length functional defined for curves attached to the arc with unrestricted locations, but supplemented with Lagrange multipliers that impose the extremal curves to lie in the left interior causal diamond only on-shell. The extrema have lengths $\Theta+2\pi w$ or $2\pi-\Theta+2\pi w$, where $w\in\natural$ counts the number of times the curve winds around $\Sigma_{\rm left}'\backslash\{\mbox{the pode}\}$, which has the topology of a cap minus a point, \ie a cylinder. The shortest ones have length $L=\inf(\Theta,2\pi-\Theta)$ and they are homologous to the arc in the left static patch.  They are therefore ``minimal extremal curves,'' which match precisely with the candidate curves $\chi_{\rm left}$ mentioned above. 
Whatever the correct choice for the curve $\chi_{\rm left}$, we can nevertheless conclude that the entanglement entropy of the arc in the bilayer case is 
\begin{align}
    S_{\rm bil}(\Theta) &= \frac{\Theta}{4G\hbar}+\left\{\!\begin{array}{ll}
       \dis \frac{\Theta}{4G\hbar}  \, , &\mbox{if }0\le \Theta \leq \pi\espD\\
       \dis  \frac{2\pi-\Theta}{4G\hbar} \, , &\mbox{if } \pi<\Theta \le2 \pi
    \end{array}\right.+{\mathcal{O}}(G\hbar)^{0} \nonumber\\
    &= \left\{\!\begin{array}{ll}
       \dis \frac{\Theta}{2G\hbar} +{\mathcal{O}}(G\hbar)^{0} \, , &\mbox{if }0\le \Theta \leq \pi\espD\\
       \dis  \frac{\pi}{2G\hbar}+{\mathcal{O}}(G\hbar)^{0} \, , &\mbox{if } \pi<\Theta \le2 \pi
    \end{array}\right. \!\!,
\end{align}
where we include the limiting cases $\Theta\to 0$ and $\Theta\to 2\pi$. 
In the bilayer case, the entropy takes the correct values for $\Theta=0$ and $\Theta = 2\pi$. The leading contribution to the holographic entanglement entropy obeys a ``volume law'' (as opposed to an area law), suggesting that the holographic theory on the screens is non-local~\cite{Shaghoulian:2021cef}. 

To figure out the entanglement wedge of the arc in the bilayer case, it is necessary to determine the true curves $\chi_{\rm left}$ and $\chi_{\rm ext}$, obtained by minimizing the generalized entropy. This is easy to carry out when $0<\Theta<\pi$, since taking the two curves $\chi_{\rm left}$ and $\chi_{\rm ext}$ to be the arc itself implies that the surface $\mathcal{C}'_{\rm left}\cup \mathcal{C}'_{\rm ext}$ shrinks to the arc. In that case, the semiclassical contribution to the entanglement entropy is minimized, as it vanishes. The entanglement wedge is thus the same as in the monolayer case, and it reduces to the arc itself. An arc of angle $0<\Theta<\pi$ corresponds to a subsystem that cannot encode any part of ${\rm dS}_3$ larger than the arc itself.   

On the contrary, the situation is more involved and becomes non-trivial when \mbox{$\pi<\Theta<2\pi$}. Indeed, the curve $\chi_{\rm left}$ is now homotopic in the Cauchy slice $\Sigma_{\rm left}'$ to the complementary arc of length $L=2\pi-\Theta$ in $H_1$. As a result, $\mathcal{C}'_{\rm left}$ covers a non-trivial portion of $\Sigma_{\rm left}'$, which is~$\Sigma_{\rm left}'$ except for the closed part bounded by the union of $\chi_{\rm left}$ and the complementary arc. The entanglement wedge in the left interior region is thus always a non-trivial part of the full causal diamond of $\Sigma_{\rm left}$, \ie some non-trivial bulk region of the static patch of the pode. This shows that lifting the classical degeneracy of the length of the candidate curves $\chi_{\rm left}$ and $\chi_{\rm ext}$ requires a non-trivial extremization of the semiclassical contribution $S_{\rm semicl}(\mathcal{C}'_{\rm left}\cup\mathcal{C}'_{\rm ext})$. This goes beyond the scope of the present work. We can however make some interesting remarks. If the actual preferred surface $\chi_{\rm ext}$ confines on the part of $\Sigma_{\rm ext}'$ represented in Figure~\ref{fig:Penrose_dS_ext} as the line segment between $H_1$ and $E$,\footnote{In this case, $\chi_{\rm ext}$ is a curve denoted $\chi'$ in Appendix~\ref{dege curves}.} then the entanglement wedge in the exterior restricts to a subregion of this part of $\Sigma_{\rm ext}'$. In particular, it does not cover any bulk region of the exterior. On the contrary, if the actual, preferred surface $\chi_{\rm ext}$ extends on both parts of $\Sigma_{\rm ext}'$ between $H_1$ and $E$ and between $E$ and $H_2$,\footnote{In this case, $\chi_{\rm ext}$ is a curve denoted $\check\chi'$ in Appendix~\ref{dege curves}.} then the entanglement wedge in the exterior region contains a non-trivial part of the causal diamond of $\Sigma_{\rm ext}$. Notice that the entanglement wedge of the full single-screen subsystem $H_1$ when $\tau_2>\tau_1$ (see~\Sect{1s}) is not recovered by taking the limit $\Theta\to 2\pi$ of the arc subsystem on $H_1$.\footnote{The reason is very general. For instance, the causal diamond of a spacelike line segment in two-dimensional Minkowski space is larger than the union of the causal diamonds of two half line segments, obtained by removing a single point of the initial line segment.}


\subsection{Entropy of two arcs on both screens in \bm dS$_3$}
\label{misal}

In this subsection, we motivate the fact that, if true, the HRT-like prescription stated in \Sect{monobil} deserves a clear derivation from first principles. To see the issues, we consider the subsystem $A$ of $H_1\cup H_2$ consisting of two arcs on both screens. The first arc is on $H_1$ and is parametrized by $\theta_2\in[0,\Theta]$, where $\Theta\in(0,2\pi)$ as in the previous subsection. The second arc is on $H_2$ and is parametrized by $\theta_2\in[\alpha,\Theta+\alpha']$. For the sake of simplicity, the angles $\alpha$,~$\alpha'$, which may not be infinitesimal, are taken small enough in absolute values. 
The screens $H_1$ and~$H_2$ are at global (conformal) times $\tau_1$ and $\tau_2$ ($\sigma_1$ and  $\sigma_2$) satisfying \Eq{si}. 

To compute the geometrical contributions to the Von Neumann entropy, we need to find minimal extremal curves. The analysis builds on the discussion of the previous subsection. Let us first examine the curves that can potentially be recognized as $\chi_{\rm ext}$:

\noindent $(i)$ Let $\Sigma_{\rm ext}'$ be the Cauchy slice shown in  Figure~\ref{fig:Penrose_dS_ext2} comprising the line segments from $H_1$ to~$E$ and from $E$ to $H_2$. Consider any pair of curves respectively homotopic in $\Sigma_{\rm ext}'$ to the arcs in $H_1$ and~$H_2$, with $\theta_2$ varying monotonically along them. Their union is homologous in the exterior region to the two-arc subsystem and their total length is
\be
L_1=\Theta+(\Theta+\alpha'-\alpha). 
\ee

\noindent $(ii)$ The union of the two arcs is homologous in the exterior region to the union of their complementary arcs in the respective screens. This is the case since the union of the four arcs is $H_1\cup H_2$, which is the boundary of $\mathcal{C}_{\rm ext}'=\Sigma_{\rm ext}'$. As a result, consider any pair of curves respectively homotopic in $\Sigma_{\rm ext}'$ to the complementary arcs in $H_1$ and $H_2$, with $\theta_2$ varying monotonically along them. Their union is homologous in the exterior region to the two-arc subsystem\footnote{On the contrary, the union of a curve homotopic to one of the two arcs and a curve homotopic to the complement of the other arc on its screen is not homologous in the exterior region to the two-arc subsystem.} and their total length is
\be
L_2=(2\pi-\Theta)+\big(2\pi-(\Theta+\alpha'-\alpha)\big).
\ee 
Let us denote by $L_{\rm arcs}$ the smallest of the lengths of the curves $(i)$ and $(ii)$, 
\be
L_{\rm arcs}=\inf\!\big(2\Theta+\alpha'-\alpha\,,4\pi-2\Theta-\alpha'+\alpha\big).
\ee

\noindent $(iii)$  Another possibility consists of a geodesic connecting in the exterior region the endpoints at $\theta_2=0$ in $H_1$ and $\theta_2=\alpha$ in $H_2$, together with the geodesic connecting the endpoints at $\theta_2=\Theta$ in $H_1$ and $\theta_2=\Theta+\alpha'$ in $H_2$. This type of geodesics is described in Appendices~\ref{arm} and~\ref{appendix_non_sym}.\footnote{This is implemented in several steps. In Appendix~\ref{arm}, we restrict to the case where the screens are at equal global times $\tau_0\equiv \tau_1=\tau_2$, and take $\alpha=\alpha'=0$. Appendix~\ref{armsem} shows that in the full dS$_3$, there are two geodesics connecting the endpoints at $\theta_2=0$ on the screens. One of them lies exclusively in the exterior region and is referred to as an ``arm along the barrel.'' On the contrary, the second geodesic must be omitted, as it extends in the exterior and both interior regions. In Appendices~\ref{armsem} and~\ref{two arcs:eq_of_motion}, we show in two different ways that the arms are spacelike geodesics when the global time satisfies $\sinh|\tau_0|\le 1$. In Appendix~\ref{appendix_non_sym}, we show that the generic case where $\tau_1$, $\tau_2$ and $\alpha$ are arbitrary reduces to the above  particular one.} We refer to them as ``arms along the barrel,'' which are spacelike and have total length 
\be
    L_{\rm arms} = 2\arcsin\sqrt{\sinh\tau_1\sinh\tau_2+\sin^2{\alpha\over 2}}+2\arcsin\sqrt{\sinh\tau_1\sinh\tau_2+\sin^2{\alpha'\over 2}},
    \ee
when the conditions
 \be
 \sinh\tau_1\sinh\tau_2\le \cos^2{\alpha\over 2}\quad \and\quad \sinh\tau_1\sinh\tau_2\le \cos^2{\alpha'\over 2}
 \label{c1}
\ee
are satisfied. These arms may or may not cross each other once. 
When they do not cross each other, which is the case when $|\alpha|$, $|\alpha'|$ are small enough, their union extends on some Cauchy slice $\Sigma_{\rm ext}'$ and is homologous in the exterior region to the two-arc subsystem.\footnote{On the contrary, when the arms~($iii$) cross each other, the homology constraint is not satisfied and their union is not a candidate curve $\chi_{\rm ext}$.}

\noindent $(iv)$ One may also consider a pair of geodesics similar to those defined in case~($iii$), up to the exchange of the boundary endpoints in $H_2$. They are spacelike and their total length~is
\be
   L_{\rm crossed} = 2\arcsin\sqrt{\sinh\tau_1\sinh\tau_2+\sin^2{\Theta+\alpha'\over 2}}+2\arcsin\sqrt{\sinh\tau_1\sinh\tau_2+\sin^2{\Theta-\alpha\over 2}},
   \ee
when  
\be
\sinh\tau_1\sinh\tau_2\le \cos^2{\Theta+\alpha'\over 2}\quad \and\quad \sinh\tau_1\sinh\tau_2\le \cos^2{\Theta-\alpha\over 2}.
\label{c2}
\ee
These curves are also ``arms along the barrel.'' They cross each other once when the arms in case~($iii$) do not cross, and {\it vice versa}.
When they cross each other, which is the case when $|\alpha|$,~$|\alpha'|$ are small enough, their union extends on some Cauchy slice $\Sigma_{\rm ext}'$ and is homologous in the exterior region to the two-arc subsystem.\footnote{On the contrary, when the arms~($iv$) do not cross each other, the homology constraint is not satisfied. In fact, the two pairs of arms~($iii$) and~($iv$) are always homologous in the exterior region. } However, since on the one hand condition~(\ref{c1}) is satisfied when condition~(\ref{c2}) is met, and on the other hand $L_{\rm crossed}$ is larger than $L_{\rm arms}$, the arms~($iv$) are not of minimal extremal length.

\noindent $(v)$ Let us consider the Cauchy slice $\Sigma_{\rm ext}'$ defined in case~($i$). In $\Sigma_{\rm ext}'$, consider any curve connecting the endpoint $\theta_2=0$ in $H_1$ to the endpoint $\theta_2=\Theta+\alpha'$ in $H_2$, as well as any curve connecting the endpoint $\theta_2=\Theta$ in $H_1$ to the endpoint $\theta_2=\alpha$ in $H_2$. Reasoning as in Appendix~\ref{nu(lambda)}, when $\theta_2$ varies monotonically along them, their union is expected to be an extremum of the length functional supplemented with Lagrange multipliers. In that case, for small enough~\mbox{$|\alpha|$, $|\alpha'|$}, their total length is either 
\begin{align}
(\Theta+\alpha'+2\pi w_1)&+(\Theta-\alpha+2\pi w_2),\nonumber\\
\big(2\pi-(\Theta+\alpha')+2\pi w_1\big)&+(\Theta-\alpha+2\pi w_2),\nonumber\\
(\Theta+\alpha'+2\pi w_1)&+\big(2\pi-(\Theta-\alpha)+2\pi w_2\big),\nonumber\\
\mbox{or}\quad \big(2\pi-(\Theta+\alpha')+2\pi w_1\big)&+\big(2\pi-(\Theta-\alpha)+2\pi w_2\big),
\label{sss}
\end{align}
where $w_1,w_2\in\natural$ are their respective winding numbers around $\Sigma_{\rm ext}'$. The shortest pairs have $w_1=w_2=0$ and correspond either to the first or the last case. Their lengths are thus 
\be
L_5=\inf\!\big(2\Theta+\alpha'-\alpha\, , 4\pi-2\Theta-\alpha'+\alpha\big),
\ee
which is equal to $L_{\rm arcs}$. This is not surprising. Indeed, in both cases, like for the arms~($iv$), the two curves cross each other, here an odd number of times, and their union is homologous in the exterior region to the two-arc subsystem.\footnote{On the contrary, when $w_1=w_2=0$, the second and third cases in \Eq{sss} correspond to pairs of curves that are not homologous in the exterior region to the two-arc subsystem.} By reinterpreting each crossing as curves touching each other rather than intersecting, one sees that the union of the curves has already been taken into account in case~$(i)$ or~($ii$). 

\noindent $(vi)$ The last possibility to consider is that of any pair of curves in $\Sigma_{\rm ext}'$ defined as in case~($v$), up to the exchange of the boundary endpoints in $H_2$. The first curve connects the endpoint $\theta_2=0$ in $H_1$ to the endpoint  $\theta_2=\alpha$ in $H_2$, while the second  curve connects the endpoint $\theta_2=\Theta$ in $H_1$ to the endpoint  $\theta_2=\Theta+\alpha'$ in $H_2$. When $\theta_2$ varies monotonically along them, their union is expected to be an extremum of the length functional supplemented with Lagrange multipliers. 
Moreover, their total length is  
\be
|\alpha+2\pi w_1|+|\alpha'+2\pi w_2|,
\ee
where, in the present convention,  $w_1,w_2\in\Z$ are their winding numbers (of any signs) around~$\Sigma_{\rm ext}'$. When $|\alpha|, |\alpha'|\!<\pi$, the shortest pairs of curves have vanishing winding numbers. Their total lengths are 
\be
L_6=|\alpha|+|\alpha'|
\ee
and they are homologous to the two-arc subsystem. Notice that in the particular case where $\alpha=0$ ($\alpha'=0$), the set of curves connecting the points $\theta_2=0$ ($\theta_2=\Theta$) in $H_1$ and $H_2$ reduces to a unique lightlike curve.\footnote{This is due to the fact that they are on $\Sigma_{\rm ext}'$, which lies on the horizons, and that $\theta_2$ is constant on all their lengths ($\theta_2$ varies monotonically between the same initial and final values).} Hence, when $\alpha\alpha'=0$, the pairs of curves of length $L_6$ should be excluded, as they are not spacelike. Instead, curves with non-trivial winding numbers but still satisfying the homology constraint should be considered.

We are ready to discuss the leading contributions to the entanglement entropies of the two-arc subsystem in the monolayer and bilayer cases. Let us consider first the generic situation where $\alpha$ and $\alpha'$ are nonvanishing. Since $L_{\rm arms}$ increases when $|\tau_1|$ or $|\tau_2|$ increases, it is minimal when $\tau_1\tau_2=0$, which implies\footnote{When $\tau_1\tau_2=0$, the geodesics~($iii$) coincide with the curves of minimal lengths of case~($vi$).}
\be
L_{\rm arms}\ge \left.L_{\rm arms}\right|_{\tau_1\tau_2=0}=|\alpha|+|\alpha'|\quad \when \quad |\alpha|,|\alpha'|\le \pi.
\ee
As a result, at fixed angle $\Theta$ and global times $\tau_1$, $\tau_2$, the length $L_{6}$ is lower than $L_{\rm arms}$ and~$L_{\rm arcs}$, when $|\alpha|,|\alpha'|$ are small enough.\footnote{Of course, other cases exist. For instance, taking  $\Theta$ small and $\alpha$, $\alpha'$ big enough leads to a length $L_{\rm arcs}$ that is lower than $L_5$ and $L_6$.} The candidate curves $\chi_{\rm ext}$ are those described in case~($vi$) and the entropy of the two-arc subsystem satisfies in the monolayer case
\be
S_{\rm mono}={|\alpha|+|\alpha'|\over 4G\hbar}+\O(G\hbar)^0.
\label{sm}
\ee
In the bilayer proposal, we have to add to $S_{\rm mono}$ the contributions from the two interior regions. Since $A_{\rm left}$ is the arc of angle $\Theta$ in $H_1$, the candidate curves $\chi_{\rm left}$ are exactly those described in the case of the single-arc subsystem of angle $\Theta$, treated in \Sect{1s'}. Let $\Sigma_{\rm left}'$ be the  Cauchy slice represented in Figure~\ref{fig:Penrose_dS_ext} as the line segment between~$H_1$ and the corner $(\theta_1,\sigma)=(0,\sign(\sigma_1)\pi/2)$. 
$\chi_{\rm left}$ can be any curve homotopic in~$\Sigma_{\rm left}'$ to the geodesic of length $L=\inf(\Theta,2\pi-\Theta)$. Its length is also $L$. Similarly,  when $|\alpha|$, $|\alpha'|$ are small enough, $A_{\rm right}$ is an arc of length $\Theta+\alpha'-\alpha$ in $H_2$ and any curve $\chi_{\rm right}$ has length $\inf\!\big(\Theta+\alpha'-\alpha,2\pi-(\Theta+\alpha'-\alpha)\big)$. Since the sum of the lengths of the curves $\chi_{\rm left}$ and $\chi_{\rm right}$ is $L_{\rm arcs}$, we obtain
\begin{equation}
    S_{\rm bil} = S_{\rm mono} +\frac{L_{\rm arcs}}{4G\hbar}+\O(G\hbar)^0.
    \label{sb}
\end{equation}
In order to lift the classical degeneracy between all candidate curves in the monolayer and bilayer cases, one has to minimize the respective generalized entropies, which goes beyond the scope of our work.  However, since the curve $\chi_{\rm ext}$ on the limiting Cauchy slice $\Sigma_{\rm ext}'$ lies on both horizons, the entanglement wedge covers a non-trivial part of the exterior causal diamond. In the bilayer case, extremizing the generalized entropy is necessary to know whether the entanglement wedge can cover a non-trivial bulk part of the interior causal diamonds. 

Before we proceed to discuss the limiting case  of two aligned arcs, for which $\alpha=0$ and/or $\alpha'=0$, let us check whether the above results for the entropy of the two-arc subsystem, which are valid for sufficiently small but not vanishing $|\alpha|$ and $|\alpha'|$, are compatible with the strong subadditivity of entanglement entropies. We will discuss only the leading geometrical contributions to the monolayer and bilayer entropies given in \Eqs{sm} and~(\ref{sb}), respectively. For this purpose, let us denote the arc of length $\Theta$ in $H_1$ by $A_1$ and its complement in $H_1$ by ${\bar A}_1$. Likewise, we denote the arc of length $\Theta+\alpha'-\alpha$ in $H_2$ by $A_2$ and its complement in $H_2$ by ${\bar A}_2$. Consider now the following three non-overlapping subsystems: $\bar{A}_1\cup \bar{A}_2$, $A_1$ and $A_2$. Strong subadditivity requires that \cite{VanRaamsdonk:2016exw}
\begin{equation}
S(\bar{A}_1\cup \bar{A}_2 \cup A_1)+S(A_1 \cup A_2) \geq S(A_1)+S(\bar{A}_1\cup \bar{A}_2 \cup A_1 \cup A_2).
\end{equation}
Since the union of the four arcs is $H_1\cup H_2$, $S_{\rm bil}(\bar{A}_1\cup \bar{A}_2 \cup A_1 \cup A_2)$ vanishes to all orders in~$G\hbar$, while $S_{\rm mono}(\bar{A}_1\cup \bar{A}_2 \cup A_1 \cup A_2)$ vanishes to leading order. Moreover, $S_{\rm bil}(\bar{A}_1\cup \bar{A}_2 \cup A_1)=S_{\rm bil}(A_2)$ to all orders, while  
$S_{\rm mono}(\bar{A}_1\cup \bar{A}_2 \cup A_1)=S_{\rm mono}(A_2)$ at the level of the leading geometrical contributions. So, to leading order, the above inequality reduces to the Araki-Lieb inequality \cite{Araki:1970ba}
\begin{equation}
\label{ineq}
S(A_2)+S(A_1 \cup A_2)-S(A_1) \geq 0 ,
\end{equation}
in both the monolayer and the bilayer proposals. It suffices to show the inequality for the monolayer case, since the leading interior contributions to $S_{\rm bil}(A_1 \cup A_2)$ are equal to the leading interior contributions to the sum of the individual entropies, $S_{\rm bil}(A_1)+S_{\rm bil}(A_2)$, satisfying the inequality separately. Thanks to \Eq{smt} and multiplying with $4G\hbar$, \Eq{ineq} reduces  to
\begin{equation}
    \alpha'-\alpha+|\alpha'|+|\alpha|\geq 0
\end{equation}
in the monolayer case, which indeed holds.

Despite this consistency check, the smallness of the monolayer entropy in the cases of very small but not vanishing $|\alpha|$ and $|\alpha'|$ requires very special correlations for the degrees of freedom on the four arcs on the two screens, which seem to be difficult to reproduce from the bulk path integral point of view. For example, to get an almost vanishing  entropy for the two arc system $A_1\cup A_2$, the total state on the screens needs to assume a form close to a factorized one, $\Psi=\Psi_1(A_1,A_2)\Psi_2(\bar{A}_1,\bar {A}_2)$, so that when we take a partial trace over the degrees of freedom of the complementary arc subsystem $\bar{A}_1\cup \bar{A}_2$ we get an almost pure density matrix. 
This quasi-factorization should hold irrespective of the location of $A_1$ on the left screen and the value of $\Theta$. On the other hand, the bilayer entropy receives finite, large contributions from the interior regions, suggesting that the necessary correlations restrict to the fraction of the degrees of freedom encoding the exterior region. As we remarked in \Sect{1s}, these must be interacting in order to reproduce bulk interactions in the region between the cosmological horizons. These interactions may establish and maintain the correlations for this subset of the degrees of freedom, as the screens evolve along the horizons. So it seems that the bilayer proposal is more consistent and natural from this point of view.

Let us now consider to the instances where $\alpha=0$ or/and $\alpha'=0$. The curves of total length $L_6$ must be excluded and thus both $S_{\rm mono}$ and $S_{\rm bil}$ take values larger than those expected from \Eqs{sm} and~(\ref{sb}). In other words, the monolayer and bilayer entropies are not continuous functions of $\alpha$ and $\alpha'$, jumping at these exceptional values of $\alpha$ and $\alpha'$. This strange behavior may be the sign that, for some reason, the curves of lengths $L_6$ should also be excluded in the generic cases, when $\alpha\alpha'\neq 0$. In our list~($i$)--($vi$) of all possible curves~$\chi_{\rm ext}$, some of them cannot be infinitely differentiable. As explained in Appendix~\ref{dege curves}, in cases~($i$) and~$(ii$), the curves homotopic to an arc in $H_1$ ($H_2$) and extending on the part of $\Sigma'_{\rm ext}$ corresponding in Figure~\ref{fig:Penrose_dS_ext2} to the line segment between $E$ and $H_2$ ($H_1$) cannot be analytic in their affine parameters, at least at the points where they cross the bifurcate horizon. Hence, they are not infinitely differentiable. Similarly, all curves (of any winding numbers) described in cases~($v$) and~$(vi$) are not infinitely differentiable.  If in the HRT-like set of rules of \Sect{monobil} one has to restrict to perfectly smooth curves of minimal extremal lengths, then $\chi_{\rm ext}$ should be chosen only from the sets of curves~($i$)--($iv$). For arbitrary $\alpha$,~$\alpha'$, this would require comparing $L_{\rm arcs}$, $L_{\rm arms}$ and $L_{\rm crossed}$, whose relative magnitudes change as time evolves, thus giving rise to phase transitions. In Appendix~\ref{ptran}, we describe the phase transition that occurs in the case of perfectly identical and aligned arcs in $H_1$ and~$H_2$, \ie $\alpha=\alpha'=0$, when the screens are at equal global times $\tau_0\equiv \tau_1=\tau_2$~\cite{Shaghoulian:2021cef}.  

In order to avoid the very weird behavior of the entropies when $\alpha$ or/and  $\alpha'$ vanishe(s), another possibility may be to allow the minimal extremal curves to be lightlike in limiting cases. Indeed the fact that the results are compatible with entropy inequalities seems to support this. On the other hand, such a possibility may lead to problems when perturbing the de Sitter geometry since lightlike curves may become timelike. 

In fact, the dimensionless parameter $G\hbar/L_6$ is always small in the semiclassical limit, except when $L_6$ vanishes or becomes of the order of the Planck length. When it is not small, higher order corrections and actually exact results in $G\hbar/L_6$ for the entropy must be derived. As a result, a classical case for which $\alpha\alpha'=0$ may simply not make sense in a theory of quantum gravity.

 To clarify the issue, it would be very interesting to derive the set of HRT-like rules of \Sect{monobil} from first principles, by studying the replica path integral in this time-dependent cosmological context.

Another interesting remark is that the curves of length $L_6$ are not maximin curves, even when we restrict to the exterior causal-diamond region. In fact they are minimin spacelike curves. That is, we first find the minimal curve on each Cauchy slice ${\rm \Sigma}_{\rm ext}'$, and among the minima we take the one of smallest length. This shows that the HRT curves may not be necessarily maximin in the de Sitter cases. 


\section{Summary and outlook}
\label{sect:conclu}

In this work, we have reconsidered the monolayer and bilayer holographic proposals  describing the non-overlapping static patches of two freely falling observers in de Sitter space \cite{Susskind:2021esx, Shaghoulian:2021cef, Shaghoulian:2022fop}. We formulate these proposals in a covariant way by considering arbitrary foliations of spacetime in terms of Cauchy slices. Two screens are respectively located at the intersections of the slices with the horizon of each observer. As a result, every slice is divided into three parts. Two parts have the topologies of sphere caps, extending respectively in the two static patches, or ``interior regions.'' The third part has the topology of a barrel and lies in the region between the cosmological horizons, or the ``exterior region.'' 

In the monolayer case, the classical contribution to the 
entanglement entropy of a subsystem of the screens with its complement 
is given by one quarter of the area in Planck units of a minimal extremal surface homologous to the subsystem and lying on a 
barrel-like Cauchy slice between the two screens. The entanglement wedge, which is expected to be the bulk region that can be reconstructed from the holographic subsystem, is the causal diamond of the part of this barrel bounded by the screen subsystem and the extremal surface. As a result, the monolayer proposal seems to lead to a contradiction, as the screens are supposed to encode the static patches of the observers rather than the exterior region.   

In the bilayer case, the classical fine-grained entropy admits additional contributions from two extra minimal extremal surfaces, which are homologous to the screen subsystem and lie respectively on a cap-like Cauchy slice in each interior region. The entanglement wedge now covers two extra domains in the static patches, which are the causal diamonds of the parts of the caps bounded by the screen subsystem and the two extremal surfaces. For the system comprising both screens on a given Cauchy slice of de Sitter space, the entanglement wedge contains the full slice. Hence, rather than possibly providing a holographic description of the static patches of the two observers only, the bilayer proposal suggests that the screens encode the entire de Sitter space, capturing the full set of bulk degrees of freedom. Assuming that the whole spacetime is in a pure state, the entanglement entropy of the union of the holographic screens vanishes exactly in $G\hbar$. Consistently, the bilayer rule of thumb for computing the classical contribution to the Von Neumann entropy yields a vanishing result, since all three minimal extremal homologous surfaces are the empty set. 

We also considered the two-screen system in situations where the screens are pushed in the interior regions and placed on stretched horizons. The monolayer proposal leads to contradictory behavior in such situations, since the entanglement wedge of a single-screen system may extend in the exterior, even in limits where the number of degrees of freedom of the screens becomes very small. In the context of the bilayer proposal, such a situation can be understood in terms of integrating out the degrees of freedom. We argue that the degrees of freedom giving rise to the region between the stretched horizons 
are integrated out. As a result, no contributions to the entropy arise from the exterior region between the screens. The entanglement wedge of the two-screen system on the stretched horizons confines in the interior regions. 

We have also considered various subsystems of the two screens on the cosmological horizons. Even though the Cauchy slices of dS$_{n+1}$ can be generic, in all explicit examples we analyze, we restrict to foliations of dS$_{n+1}$ with Cauchy slices that are SO($n$) symmetric. Hence, it would be interesting to generalize our analyses by relaxing this assumption.\footnote{This is trivially allowed for the system comprising both screens described in the previous paragraph.} In the remainder of this concluding section, we restrict to the bilayer proposal for holographic entropy computations in de Sitter space. 

The simplest subsystem of the screens to analyse consists of only one of the two screens. The holographic recipe for  entanglement entropy reproduces the Gibbons-Hawking result, namely one quarter of the area of the screen in Planck units. However, the analysis reveals several subtle issues. Indeed, 
any sphere S$^{n-1}$ lying on the boundary of the exterior causal diamond 
along the horizons is a candidate minimal extremal homologous surface. This classical degeneracy can be lifted at the semiclassical level, by demanding the classical minimal surface to minimize a generalized entropy. The generalized entropy turns out to be equal to one quarter of the ``quantum area'' of the surface in Planck units. This implies that the minimal extremal homologous surface in the exterior region is the screen that lies the farthest from the bifurcate horizon on the Penrose diagram. Therefore, two phases can be identified: 

\noindent $\bullet$  When the minimal extremal surface coincides with the screen subsystem under study, the entanglement wedge reduces to the causal diamond of the cap-like Cauchy slices bounded by the screen subsystem. The entanglement wedge of the single-screen subsystem extends only in the corresponding interior region. 

\noindent $\bullet$  When the minimal extremal surface coincides with the second, complementary screen, the entanglement wedge acquires an additional component, which is the causal diamond of the barrel-like Cauchy slices between the two screens. This extra component lies in the exterior region. In other words, the screen with maximal quantum area encodes the causal diamond in the exterior region. This is the screen that lies closest to the bifurcate horizon on the Penrose diagram. 

\noindent The extremization at the semiclassical level we have performed is however approximate, since it amounted to minimizing the semiclassical entropy contribution to the quantum area associated with the classically degenerate minimal extremal surfaces. The fully correct result (at the semiclassical level) would instead be obtained by minimizing the generalized entropy among all surfaces, which are homologous to the screen subsystem and lie on a barrel-like Cauchy slice.  

One may think that seeking extremal surfaces lying on some cap-like or barrel-like Cauchy slice may be achieved by first extremizing the area functional and then eliminating the solutions that are not extending entirely in the causal diamond. We have seen that this is not the case. The reason is that relevant surfaces (partially) lying on the boundary of a diamond are generically not extremal, in the sense that the variation of their area does not vanish for all infinitesimal deformations of the surface. In order not to miss them, one considers the area functional supplemented by Lagrange multipliers, enforcing the homologous surfaces to lie in the diamond only ``on-shell,'' \ie when the upgraded functional is extremal. This requires an implementation of the Lagrangian multipliers in such a way as to impose inequality rather than equality constraints. In addition to the single-screen subsystems, we have seen that this approach is essential for computing the entropy of a subsystem comprising an arc of a circular screen in~dS$_3$. Indeed, none of the geodesic curves homologous to the arc in dS$_3$ lies on a barrel-like Cauchy slice. However, there is a continuous family of curves homologous to the arc that are minimal extrema of the upgraded area functional. These curves are lying on the boundary of the causal diamond of the barrel-like Cauchy slices, on the horizons. A~similar family of curves exists on the boundary of the causal diamond of the cap-like Cauchy slices bounded by the screen, along the horizon. To lift the classical degeneracy in both families, one has to minimize at the semiclassical level the generalized entropy associated with a pair of such curves: 

\noindent $\bullet$ When the arc angle is smaller or equal to $\pi$, both minimal extremal curves are identical to the arc itself, and so the entanglement wedge reduces to the arc. As a result, the arc subsystem does not encode anything of the bulk. 

\noindent $\bullet$ When the arc angle is greater than $\pi$, the curve in the interior region (actually on its boundary) is not the arc, since its length is equal to the angle of the complementary arc in the screen. The extremization of the generalized entropy associated with the pair of curves in this case is a difficult task. It would be very interesting to perform the computation and determine the precise curves and the entanglement wedge. In any case, the latter extends in the interior region and possibly in the exterior region. 

The final subsystem we have considered is a pair of arcs lying respectively on each screen in dS$_3$. In contrast with the previous example, 
there is now a competition between the geodesic curves connecting the endpoints of the arcs through the bulk of the exterior causal diamond 
and curves lying on the boundary of this diamond. When the arcs are of almost equal length and slightly misaligned, the dominant curves lie on the boundary, with lengths equal to the magnitude of the offset. Notice that these curves are minimin from the point of view of the exterior causal diamond, indicating that in time-dependent cosmological settings, not all minimal extremal HRT-surfaces are maximin surfaces. They are also not infinitely differentiable. Moreover, when the offset vanishes exactly, these boundary curves (or one of the curves) become(s) lightlike, and should not be allowed for (a) minimal extremal curves (curve). A weird discontinuity seems to arise in that the allowable minimal extremal classical curves are not the previous ones anymore and are of  finite large length, which can be an artifact of the perturbative expansion. It seems that one should include quantum corrections to remove this discontinuous behavior for the entanglement entropy, which may yield a quantum minimal extremal curve leading to a small entropy. In any case, a clear derivation of the bilayer proposal for computing the leading Von Neumann entropies is definitely desirable, in order to clarify further the rules of the proposal. 

If the bilayer proposal may be relevant for describing holographically de Sitter spacetime and computing entropies, a question is to understand how perturbations of the geometry affect qualitatively the entanglement entropies and entanglement wedges of subsystems of the screens. For instance, the ubiquitous classical degeneracy of minimal extremal homologous surfaces on the boundaries of causal diamonds in the interior or exterior regions can be easily lifted~\cite{FPRT}.   

Anyhow, the consistency of HRT-like rules is only an indication of the validity of the whole picture. Clearly, a major breakthrough would be to understand the nature of a possible holographic quantum theory dual to de Sitter space. In the literature, various proposals have been put forward~\cite{Banks:2000fe,Witten:2001kn,Banks:2006rx, Susskind:2021dfc, Susskind:2021esx, Susskind:2022dfz, Susskind:2022bia, Rahman:2022jsf}. In contrast to the Anti de Sitter case, the holographic theory is expected to be non-local and exotic, which makes it less familiar than local quantum field theories~\cite{Goheer:2002vf, Susskind:2021omt,Shaghoulian:2021cef,Susskind:2021esx,Susskind:2022dfz}.


\section*{Acknowledgements}
F.R. and N.T. would like to acknowledge hospitality by the Ecole Polytechnique, while H.P. and V.F. would like to thank the University of Cyprus for hospitality, where early stages of this work have been done. This work is partially supported by the Cyprus Research and Innovation Foundation grant EXCELLENCE/0421/0362.


\begin{appendices} 
\numberwithin{equation}{section}

\section{Area extremization in a causal diamond}
\label{Appendix:Lagrange_multipliers}

At fixed $(\theta_1,\sigma)$, the metric of dS$_{n+1}$ given in \Eq{ds2} is explicitly  ${\rm SO}(n)$ symmetric, as it reduces to the metric of an $(n-1)$-dimensional sphere. The area of this S$^{n-1}$ is 
\begin{equation}
\A=v_{n-1}\left(\frac{\sin\theta_1}{\cos\sigma}\right)^{n-1},
\end{equation}
where $v_{n-1}$ is the volume of the sphere of radius 1. Our first aim in this appendix will be to show that this function has a unique extremum, which is a minimum, when the domain of definition of the function is the causal diamond of the Cauchy slices $\Sigma'_{\rm ext}$. In a second step, we will show that extra minima and maxima exist on the boundary of the diamond, and can be seen as solutions of an extremization problem. 
As seen in \Eq{si}, the conformal times at which the screens $H_1$ and $H_2$ are considered have same signs. 
In the following, for simplicity, we restrict our discussion to positive $\sigma_1$ and $\sigma_2$, in the generic case where 
\be
0<\sigma_1<{\pi\over 2}, \qquad 0<\sigma_2<{\pi\over 2}.
\ee


\subsection{Extremal points of the area function}
\label{A1}

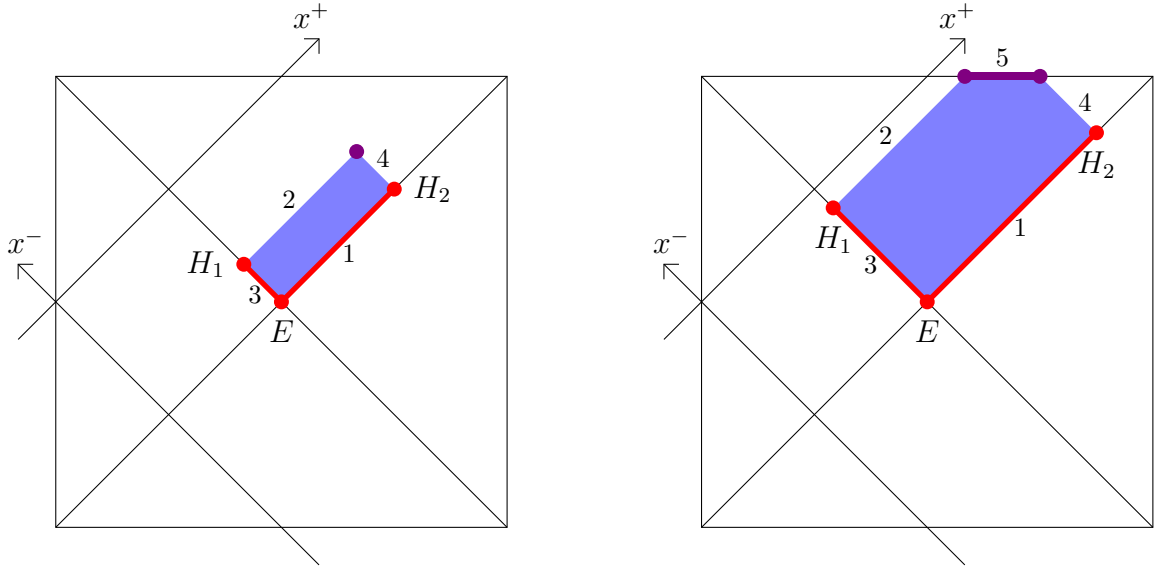
\begin{figure}[!h]
\begin{subfigure}[t]{0.48\linewidth}
\centering
\begin{tikzpicture}

\path
       +(3,3)  coordinate (IItopright)
       +(-3,3) coordinate (IItopleft)
       +(3,-3) coordinate (IIbotright)
       +(-3,-3) coordinate(IIbotleft)
      
       ;
       
\fill[fill=blue!50] (-1/2,1/2) -- node[pos=0.4, above] {{\footnotesize $2$}}(1,2) -- node[pos=0.7, above] {{\footnotesize $4$}}(3/2,3/2) -- node[pos=0.4, below] {{\footnotesize $1$}}(0,0) -- node[pos=0.7, below] {{\footnotesize $3$}}(-1/2,1/2);

\draw (IItopleft) --
      (IItopright) --
      (IIbotright) -- 
      (IIbotleft) --
      (IItopleft) -- cycle;
      
\draw (IItopleft) -- (IIbotright)
              (IItopright) -- (IIbotleft) ;

\draw (-3.5,0.5) -- (1/2,-3.5);
\draw (-3.5,-0.5) -- (1/2,3.5);
\draw (-3.5,0.3) -- (-3.5,0.5) -- node[midway, above, sloped] {$x^-$} (-3.3,0.5) ;
\draw (0.5,3.3) -- (0.5,3.5) -- node[midway, above, sloped] {$x^+$} (0.3,3.5) ;
              

\fill[fill=red] (-1/2,1/2+0.05) -- (0,0.05) -- (0,-0.05) -- (-1/2,1/2-0.05) -- cycle; 
\fill[fill=red] (3/2,3/2+0.05) -- (0,0.05) -- (0,-0.05) -- (3/2,3/2-0.05) -- cycle;

\node at (-1/2,1/2) [circle,fill,inner sep=2pt, red, label = left:$H_1$]{};
\node at (3/2,3/2) [circle,fill,inner sep=2pt, red, label = right:$H_2$]{};
\node at (0,0) [circle,fill,inner sep=2pt, red, label = below:$E$]{};
\node at (1,2) [circle,fill,inner sep=2pt, violet]{};

\end{tikzpicture}
\caption{\footnotesize When $0<\sigma_1+\sigma_2\le \pi/2$, the diamond has 4 boundary line segments. \label{dL}}
\end{subfigure}
\quad \,
\begin{subfigure}[t]{0.48\linewidth}
\centering
\begin{tikzpicture}

\path
       +(3,3)  coordinate (IItopright)
       +(-3,3) coordinate (IItopleft)
       +(3,-3) coordinate (IIbotright)
       +(-3,-3) coordinate(IIbotleft)
      
       ;
       
\fill[fill=blue!50] (-5/4,5/4) -- node[pos=0.4, above] {{\footnotesize $2$}} (1/2,3) -- node[midway, above] {{\footnotesize $5$}} (3/2,3) -- node[pos=0.8, above] {{\footnotesize $4$}}(9/4,9/4) -- node[pos=0.45, below] {{\footnotesize $1$}}(0,0) -- node[pos=0.5, below] {{\footnotesize $3$}}(-3/2,3/2);

\draw (IItopleft) --
      (IItopright) --
      (IIbotright) -- 
      (IIbotleft) --
      (IItopleft) -- cycle;
      
\draw (IItopleft) -- (IIbotright)
              (IItopright) -- (IIbotleft) ;

\draw (-3.5,0.5) -- (1/2,-3.5);
\draw (-3.5,-0.5) -- (1/2,3.5);
\draw (-3.5,0.3) -- (-3.5,0.5) -- node[midway, above, sloped] {$x^-$} (-3.3,0.5) ;
\draw (0.5,3.3) -- (0.5,3.5) -- node[midway, above, sloped] {$x^+$} (0.3,3.5) ;

\fill[fill=red] (-5/4,5/4+0.05) -- (0,0.05) -- (0,-0.05) -- (-5/4,5/4-0.05) -- cycle; 
\fill[fill=red] (9/4,9/4+0.05) -- (0,0.05) -- (0,-0.05) -- (9/4,9/4-0.05) -- cycle;
\fill[fill=violet] (1/2,3+0.05) -- (3/2,3+0.05) -- (3/2,3-0.05) -- (1/2,3-0.05) -- cycle;

\node at (-5/4,5/4) [circle,fill,inner sep=2pt, red, label = below:$H_1$]{};
\node at (9/4,9/4) [circle,fill,inner sep=2pt, red, label = below:$H_2$]{};
\node at (0,0) [circle,fill,inner sep=2pt, red, label = below:$E$]{};
\node at (1/2,3) [circle,fill,inner sep=2pt, violet]{};
\node at (3/2,3) [circle,fill,inner sep=2pt, violet]{};

\end{tikzpicture}
\caption{\footnotesize When $\pi/2<\sigma_1+\sigma_2<\pi$,  the diamond has 5 boundary line segments.\label{dR}}
\end{subfigure}
    \caption{\footnotesize Minima (in red) and maxima (in purple) of the area of the sphere S$^{n-1}$ as its position varies in the Causal diamond of all Cauchy slices $\Sigma_{\rm ext}'$.}
    \label{fig:homologous_regions}
\end{figure}

To describe the causal diamond, it is convenient to define null coordinates  
\be
x^+=\sigma+\theta_1,\qquad x^-=\sigma-\theta_1.
\label{x+-}
\ee
When $0<\sigma_1+\sigma_2\le \pi/2$, the diamond corresponds to the domain
\be
\left\{\!\!\begin{array}{l}
\dis \phantom{-}{\pi\over 2}\le x^+\le 2\sigma_2+{\pi\over 2}\\
\dis -{\pi\over 2}\le x^-\le 2\sigma_1-{\pi\over 2}\esp
\end{array}\right..
\label{diam}
\ee
Its boundary is composed of 4 segments labelled as $1,\dots,4$ in Figure~\ref{dL}. 
%
%
On the contrary, when $\pi/2<\sigma_1+\sigma_2<\pi$, the domain is further restricted to $\sigma\le \pi/2$, \ie
\be
x^+ + x^-\le \pi .
\label{extra}
\ee
Its boundary amounts to 5 segments labelled as $1,\dots,5$ on  Figure~\ref{dR}. 

In both cases, the derivatives of the area, 
\begin{align}
\frac{\partial \!\A}{\partial x^+}&=\phantom{-}v_{n-1}\,\frac{n-1}{2}\,\cos (x^-)\,\frac{(\sin\theta_1)^{n-2}}{(\cos\sigma)^n},\nonumber\\
\frac{\partial \!\A}{\partial x^-}&=-v_{n-1}\,\frac{n-1}{2}\,\cos (x^+)\,\frac{(\sin\theta_1)^{n-2}}{(\cos\sigma)^n},
\label{eq:deriv}
\end{align}
vanish simultaneously in the diamond only at $(x^+,x^-)=(\pi/2,-\pi/2)$ \ie $(\theta_1,\sigma)=(\pi/2,0)$, which corresponds to the bifurcate horizon, $E$. Hence, the S$^{n-1}$ at this location is the only one with extremal area, in the sense that it satisfies $\delta\!\A=0$ as the position of the sphere varies infinitesimally in the diamond. This area is minimal, since it increases when $\sigma$ increases from $E$. However, we know that all spheres located on the boundaries~1 and~3 of the diamond  are degenerate. They are therefore other minima, even though they are not extrema, if we mean by this that $\delta\!\A\neq 0$ when the sphere moves towards  the bulk of the diamond. On the contrary, none of the spheres lying on the remaining boundaries of the diamond is a minimum or a maximum, unless it is  located at the upper tip of the diamond in Figure~\ref{dL}, or along the boundary~5 in Figure~\ref{dR}. The reason for this is that the area increases as we move upward along the boundaries~2 and~4. The upper tip (boundary~5) thus corresponds to a maximum  (degenerate maxima) of the area, which again is  not an extremum (are not extrema). 


\subsection{Minima \& maxima as solutions of an extremization problem}
\label{A2}

In the following, we show how all minima and maxima of the area present in the exterior causal diamond can be recovered by solving an extremization problem.

\noindent $\bullet$ Let us  first consider the case where $0<\sigma_1+\sigma_2\le \pi/2$. The inequalities~(\ref{diam}) can be imposed by supplementing the area function with terms proportional to Lagrange multipliers~$\nu_I$, $I\in\{1,2,3,4\}$,
\begin{align}
\hA(x^+,x^-,\nu_I,a_I)=\A(x^+,x^-)&+\nu_1\left(x^-+\frac{\pi}{2}-a_1^2\right)+\nu_2\left(2\sigma_1-\frac{\pi}{2}-x^--a_2^2\right)\nonumber\\
&+\nu_3\left(x^+-\frac{\pi}{2}-a_3^3\right)+\nu_4\left(2\sigma_2+\frac{\pi}{2}-x^+-a_4^2\right)\!.
\label{hA}
\end{align}
Beside the multipliers, $a_I$ are extra variables whose squares are the ``positive distances from the boundary segment $I$ in the Penrose diagram.''\footnote{One may think that the distances $a_1^2$ and $a_2^2$ (and similarly $a_3^2$ and $a_4^2$) should be related from the onset. However, it is important to recover their relation only ``on-shell'' since otherwise $\nu_1$ and $\nu_2$ ($\nu_3$ and $\nu_4$) would simply add in \Eq{hA} and we would loose half of the constraints.} 
The key point is that $\hA$ is defined in a domain without boundary, as $x_\pm$, $\nu_I$, $a_I$ are all spanning $\R$. To find its extrema, \ie the points in $\R^{10}$ satisfying $\delta\!\hA=0$, we first vary the function $\hA$ with respect to $x_\pm$, which yields\!
\begin{subequations}
\label{x}
 \begin{align}
\nu_4-\nu_3&= \frac{\partial{\A}}{\partial x^+},\label{x+}\\
\nu_2-\nu_1&= \frac{\partial{\A}}{\partial x^-}.\label{x-}
\end{align}
\end{subequations} 
Varying with respect to the Lagrange multipliers leads to the equations 
\begin{subequations}
\label{n}
 \begin{align}
\label{n1}
a_1^2&= x^-+\frac{\pi}{2},\\
a_2^2&=2\sigma_1-\frac{\pi}{2}-x^-,\label{n2}\\
a_3^2&=x^+-\frac{\pi}{2},\label{n3}\\
a_4^2&=2\sigma_2+\frac{\pi}{2}-x^+,\label{n4}
\end{align}
\end{subequations} 
which imply $\hA=\A$ when satisfied.
Finally, the variations with respect to $a_I$ give
\begin{subequations}
\label{a}
\begin{align}
\nu_1a_1&=0,\label{a1}\\
\nu_2a_2&=0,\label{a2}\\
\nu_3a_3&=0,\label{a3}\\
\nu_4a_4&=0.\label{a4}
\end{align}
\end{subequations}
To find all solutions of the system of 10 equations, we organize our discussion from the location of $(x^+,x^-)\in\R^2$:
 
- When $(x^+,x^-)$ is not on the boundary of the diamond, \Eqs{n} impose it to lie in the bulk of the diamond and determine the values of $a^2_I> 0$, $I\in\{1,2,3,4\}$. As a result, $\nu_I=0$ from \Eqs{a}. However, \Eqs{x} are not satisfied, as seen below \Eq{eq:deriv}. 

- Take now $(x^+,x^-)$ in the bulk of the boundary segment 1, \ie not at its endpoints. We have $x^-+\pi/2=0$ and thus $a_1=0$ from \Eq{n1}. This implies that \Eq{a1} is satisfied. \Eqs{n2}--(\ref{n4}) determine $a^2_{2,3,4}> 0$, which imposes $\nu_{2,3,4}=0$ from \Eqs{a2}--(\ref{a4}). In that case, \Eq{x+} is solved for any $x^+$ and \Eq{x-} fixes $\nu_1$. We thus have found degenerate extrema of $\hA$. 

- For $(x^+,x^-)$ in the bulk of the boundary segment 3, a similar analysis can be carried out,  with the same conclusion.

- When $(x^+,x^-)$ is in the bulk of the boundary segment 2, we have $x^-=2\sigma_1-\pi/2$. \Eqs{n3}, (\ref{n4}) determine $a^2_{3,4}> 0$, which implies $\nu_{3,4}=0$ from  \Eqs{a3}, (\ref{a4}). However, since $\cos x^-=\sin(2\sigma_1)\neq 0$, \Eq{x+} is not satisfied. 

- Similarly, when $(x^+,x^-)$ is in the bulk of the boundary segment 4, \Eq{x-} cannot be satisfied. 

- If $(x^+,x^-)$ is at the upper tip of the diamond, \ie  the intersection of the boundary segments 2 and 4, we have $2\sigma_1-\pi/2-x^-=0$, $2\sigma_2+\pi/2-x^+=0$ and thus $a_{2,4}=0$ from \Eqs{n2}, (\ref{n4}). \Eqs{a2}, (\ref{a4}) are thus satisfied. We also have $a^2_1=2 \sigma_1> 0$, $a^2_3=2 \sigma_2> 0$ from \Eqs{n1}, (\ref{n3}), which implies $\nu_{1,3}=0$ from \Eqs{a1}, (\ref{a3}). \Eqs{x+},(\ref{x-}) then determine $\nu_{4,2}$. We thus have found an extremum of $\hA$. 

- Similarly, when $(x^+,x^-)$ is at another tip $H_1$, $H_2$ or $E$ of the diamond, one finds an extremum of $\hA$. 

To summarize, we have shown that the minima and the maximum of the function $\A$ in the diamond can be seen as extrema of $\hA$ in a domain of definition of higher dimension but without boundary. Conversely, all extrema of $\hA$ correspond to minima or the maximum of $\A$ in the diamond. However, it should be noticed that the description in terms of $\hA$ introduces an extra degeneracy, since any on-shell value of $a_I^2>0$ admits two roots, $a_I=\pm\sqrt{a_I^2}$. 
In Figure~\ref{dL}, the maximum is depicted as the purple dot, while the degenerate minima correspond to the red segments.


\noindent $\bullet$ In the case where  $\pi/2<\sigma_1+\sigma_2< \pi$, we introduce extra variables $\nu_5$, $a_5$ in $\R$ to impose the condition~(\ref{extra}) and define
\be
\tA(x^+,x^-,\nu_I,\nu_5,a_I,a_5)=\hA(x^+,x^-,\nu_I,a_I)+\nu_5\big(\pi-x^+-x^--a_5^2 \big).
\ee
Solving the equation $\delta\!\tA=0$ in $\R^{12}$ amounts to satisfying 
\begin{subequations}
\label{x'}
 \begin{align}
\nu_5+\nu_4-\nu_3&= \frac{\partial{\A}}{\partial x^+},\label{x+'}\\
\nu_5+\nu_2-\nu_1&= \frac{\partial{\A}}{\partial x^-},\label{x-'}
\end{align}
\end{subequations} 
\Eqs{n} and 
\be
a_5^2=\pi-x^+-x^-,
\label{n5}
\ee
along with \Eqs{a} and   
\be
\nu_5\,a_5=0.
\label{a5}
\ee
When \Eqs{n} and~(\ref{n5}) are fulfilled, we have $\tA=\A$. As before, we organize our discussion from the location of $(x^+,x^-)\in\R^2$:

- When $(x^+,x^-)$ is away from the line $\sigma=\pi/2$ (see Figure~\ref{dR}), \Eq{n5} imposes it to be below this line and determines $a_5^2> 0$. Therefore,  $\nu_5=0$ from \Eq{a5} and we have $\tA=\hA$. We can then apply all steps of the analysis below \Eq{a4} to the full rectangular diamond defined in \Eq{diam}, where now $\pi/2<\sigma_1+\sigma_2<\pi$, and simply omit the solution located above the line $\sigma=\pi/2$, namely the upper tip of the rectangular diamond. We have thus found degenerate extrema of $\tA$ located along the boudaries 1 and 2 of the diamond depicted in Figure \ref{dR}.

- Take  now $(x^+,x^-)$ along the line $\sigma=\pi/2$, but not at the boundary endpoints of the segment 5. \Eqs{n} impose it to lie in the bulk of segment 5 and determine $a_I^2>0$, $I\in\{1,2,3,4\}$. Hence, $\nu_I=0$ from \Eq{a}. We also have $x^++x^-=\pi$ and thus $a_5=0$ from \Eq{n5}, while \Eq{a5} is satisfied. Finally, \Eqs{x+'} and~(\ref{x-'}) are equivalent, giving
\be
\nu_5=\lim_{\sigma\to {\pi\over 2}^-}v_{n-1}\,\frac{n-1}{2}\,\frac{(\sin\theta_1)^{n-1}}{(\cos\sigma)^n}=+\infty.
\ee 
Indeed, since homologous surfaces at infinite distances (and thus infinite sizes) should be allowed in limiting cases, other variables such as Lagrange multipliers should also be allowed to be infinite. We have thus found degenerate extrema of $\tA$. 

- If $(x^+,x^-)$ is at the intersection of the boundary segments 2 and 5, we have $2\sigma_1-\pi/2-x^-=0$, $\pi-x^+-x^-=0$, which determine $x^+$, $x^-$, and thus $a_{2,5}=0$ from \Eqs{n2},~(\ref{n5}). \Eqs{a2},~(\ref{a5}) are thus satisfied. 
\Eqs{n1}, (\ref{n3}), (\ref{n4}) also fix $a^2_{1,3,4}> 0$, which implies $\nu_{1,3,4}=0$ from \Eqs{a1}, (\ref{a3}), (\ref{a4}). \Eqs{x+'}, (\ref{x-'}) then determine $\nu_5=+\infty$, $\nu_2=0$. We thus have found an extremum of $\tA$.  

- Similarly, $(x^+,x^-)$ at the intersection of the boundary segments 4 and 5 leads to an extremum of $\tA$. 

As a conclusion, all minima and maxima of the function $\A$ are recovered as extrema of the function $\tA$.


\section{Geodesics in three-dimensional de Sitter}
\label{Appendix:geodesics_dS}

In this Appendix, we consider in the case of dS$_3$ ($n=2$) an SO(2) symmetric Cauchy slice $\Sigma$, which implies the screens $H_1$ and $H_2$ to be circles. We choose subsystems $A$ of the two-screen system $H_1\cup H_2$ that are consisting in one arc in $H_1$, or two arcs respectively in $H_1$ and~$H_2$. Our goal is to compute the lengths of extremal curves anchored to $A$. These results are relevant for computing the geometrical contributions to the entanglement entropies between the subsystems $A$ and their complements in $H_1\cup H_2$. 


\subsection{Geodesic anchored on an arc of a screen}
\label{sec_embedding_one_arc}

Let us consider the subsystem $A$ of the screens consisting in the arc described below \Eq{eq:dS3_metric}. It is an arc lying on the circular screen $H_1$ at conformal time $\sigma_1$, with $\theta_2$ restricted to the range $0\le \theta_2\le \Theta$, for a given $\Theta\in(0,2\pi)$. The second screen $H_2$ is at conformal time $\sigma_2$.  In the following, we determine the lengths and properties of the geodesics anchored at the endpoints of the arc. 
The analysis is carried out using embedding coordinates~\cite{Shaghoulian:2021cef}. 

The  dS$_3$ manifold of unit radius is defined as the hypersurface
\begin{equation}
    \eta_{\mu\nu}X^\mu X^\nu=1,
    \label{hyperboloid}
\end{equation}
embedded in the four-dimensional Minkowski spacetime of coordinates $X^\mu$, $\mu\in\{0,\dots,3\}$,  and  metric $\eta_{\mu\nu}$. Using the definition~(\ref{tau-sigma}) for the conformal time $\sigma$ and the following conventions for the relations between embedding and global coordinates,
\begin{align}
&X^0=\sinh \tau, \quad X^a=\omega^a\cosh\tau, ~~a\in\{1,2,3\},\nonumber\\
\where\quad &\omega^1=-\cos\theta_1,\quad  \omega^2=\sin\theta_1\cos\theta_2, \quad \omega^3=\sin\theta_1\sin\theta_2,
\end{align}
we find that the pode and the antipode horizons defined in \Eqs{dH1} and~(\ref{dH2}) correspond to the loci 
\be
\left\{\!\begin{array}{lll}\mbox{pode horizon:} &X^1\!\!\!\!\!&=-|X^0| \\ 
\mbox{antipode horizon:} &X^1\!\!\!\!\!&=|X^0|\esps
\end{array}\right.\!\!,\quad \with\quad (X^2)^2+(X^3)^2=1.
\label{H12}
\ee  
Moreover, the arc we consider at some given time $X^0=T_1$ is parametrized as  
\be
\big(T_1,-|T_1|,\cos\theta_2,\sin\theta_2\big), \quad \theta_2\in[0,\Theta],
\label{h1}
\ee
where the relations between $T_1$, the global time $\tau_1$ and the conformal time $\sigma_1$ are given by
\be
T_1=\sinh\tau_1=\tan \sigma_1. 
\label{Ts}
\ee 

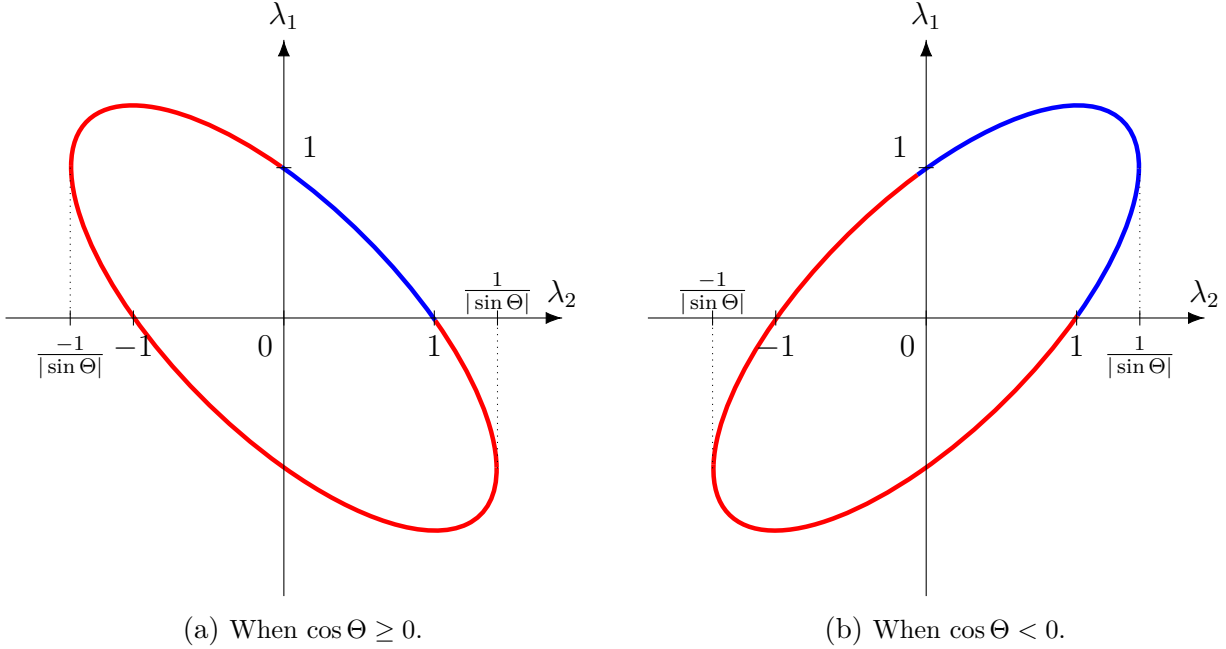
\begin{figure}[!h]
\centering
\begin{subfigure}[t]{0.48\linewidth}
\begin{tikzpicture}



\tikzset{elliparc/.style args={#1:#2:#3}{%
insert path={++(#1:#3) arc (#1:#2:#3)}}}

\draw[rotate=45, red, line width=0.6mm] (0,0) [elliparc=23:112:1.52cm and 3.7cm];
\draw[rotate=45, red, line width=0.6mm] (0,0) [elliparc=-67:-23:1.52cm and 3.7cm];
\draw[rotate=45, blue, line width=0.6mm] (0,0) [elliparc=-23:23:1.52cm and 3.7cm];
\draw[rotate=45, red, line width=0.6mm] (0,0) [elliparc=112:360-67:1.52cm and 3.7cm];


\draw[dotted] (-2.84,2) -- (-2.84,0);
\draw[dotted] (2.84,-2) -- (2.84,0);

\draw (-3.7, 0) -- (3.7, 0) node[above]{$\lambda_2$};
\draw (0, -3.7) -- (0, 3.7) node[above]{$\lambda_1$};


\draw (-2.84,0.1) -- ++ (0,-0.2) node[below, scale=1.0] {$\frac{-1}{|\sin\Theta|}$};
\draw (2.84,0.1) -- ++ (0,-0.2) node[above, scale=1.0] {$\frac{1}{|\sin\Theta|}$};
\draw (-2,0.1) -- ++ (0,-0.2) node[below, scale=1.0] {$-1$};
\draw (2,0.1) -- ++ (0,-0.2) node[below, scale=1.0] {$1$};
\draw (0,0.1) -- ++ (0,-0.2) node[below left, scale=1.0] {$0$};
\draw (-0.1,2) -- ++ (0.2,0) node[above right, scale=1.0] {$1$};


\draw[-{Latex[scale=1.3]}] (3.5,0) -- (3.7,0); 
\draw[-{Latex[scale=1.3]}] (0,3.5) -- (0,3.7);

\end{tikzpicture}
\caption{\footnotesize When $\cos\Theta \geq 0$.\label{ell+}}
\end{subfigure}
\quad \,
\begin{subfigure}[t]{0.48\linewidth}
\begin{tikzpicture}




\tikzset{elliparc/.style args={#1:#2:#3}{%
insert path={++(#1:#3) arc (#1:#2:#3)}}}

\draw[rotate=-45, red, line width=0.6mm] (0,0) [elliparc=160:246:1.52cm and 3.7cm];
\draw[rotate=-45, red, line width=0.6mm] (0,0) [elliparc=246:383:1.52cm and 3.7cm];
\draw[rotate=-45, blue, line width=0.6mm] (0,0) [elliparc=23:67:1.52cm and 3.7cm];
\draw[rotate=-45, blue, line width=0.6mm] (0,0) [elliparc=67:160:1.52cm and 3.7cm];


\draw[dotted] (-2.84,-2) -- (-2.84,0);
\draw[dotted] (2.84,2) -- (2.84,0);

\draw (-3.7, 0) -- (3.7, 0) node[above]{$\lambda_2$};
\draw (0, -3.7) -- (0, 3.7) node[above]{$\lambda_1$};


\draw (-2.84,0.1) -- ++ (0,-0.2) node[above, scale=1.0] {$\frac{-1}{|\sin\Theta|}$};
\draw (2.84,0.1) -- ++ (0,-0.2) node[below, scale=1.0] {$\frac{1}{|\sin\Theta|}$};
\draw (-2,0.1) -- ++ (0,-0.2) node[below, scale=1.0] {$-1$};
\draw (2,0.1) -- ++ (0,-0.2) node[below, scale=1.0] {$1$};
\draw (0,0.1) -- ++ (0,-0.2) node[below left, scale=1.0] {$0$};
\draw (0.1,2) -- ++ (-0.2,0) node[above left, scale=1.0] {$1$};


\draw[-{Latex[scale=1.3]}] (3.5,0) -- (3.7,0); 
\draw[-{Latex[scale=1.3]}] (0,3.5) -- (0,3.7);

\end{tikzpicture}
\caption{\footnotesize When $\cos\Theta < 0$.\label{ell-}}
\end{subfigure}
    \caption{\footnotesize Projection in the plane $(\lambda_2,\lambda_1)$ of the geodesics anchored on the endpoints of an arc of angle $\Theta$ on the circular screen $H_1$.}
    \label{ell}
\end{figure}

The geodesics joining two distinct points of dS$_3$ can be found by intersecting the hyperboloid \eqref{hyperboloid} and the plane that passes through the two points and the origin.\footnote{This is similar to the case of a sphere S$^3$, where the geodesics joining two points can be found by intersecting the sphere and the plane passing through the points and the origin, creating a great circle.} For the two endpoints of the arc, the plane we need to consider is the set of points 
\be
\Big\{\lambda_1\big(T_1,-|T_1|,1,0\big)+\lambda_2\big(T_1,-|T_1|,\cos\Theta,\sin\Theta\big), ~\where~\lambda_1,\lambda_2\in\R\Big\}. 
\label{pl}
\ee
Indeed, the endpoints are reached for $(\lambda_2,\lambda_1)=(0,1)$ and $(1,0)$, while $(\lambda_2,\lambda_1)=(0,0)$ corresponds to the origin. Moreover, the points in \Eq{pl} satisfy 
\be
X^1=-|X^0|\sign(\lambda_1+\lambda_2).
\ee
Hence, those also on dS$_3$ are exclusively located on the horizons: On  the horizon of the pode when $\lambda_1+\lambda_2\ge0$, and on the horizon of the antipode when $\lambda_1+\lambda_2\le0$. 
More specifically, the intersection points of the plane and the hyperboloid satisfy 
\be
(\lambda_1+\lambda_2\cos\Theta)^2+\lambda_2^2\sin^2\Theta =1,
\ee
which defines an ellipse in the plane $(\lambda_2,\lambda_1)$. For given $\lambda_2$,  the two roots for $\lambda_1$ are 
\be
\lambda_1^\pm(\lambda_2)=-\lambda_2\cos \Theta\pm \sqrt{1-\lambda_2^2\sin^2\Theta}, \quad \where\quad |\lambda_2|\le {1\over |\sin\Theta|}.
\ee 
The ellipse can be divided into two parts, both with endpoints $(\lambda_2,\lambda_1)=(0,1)$ and $(1,0)$. To describe them explicitly, it is useful to consider two cases:

\noindent $\bullet$ When $\cos\Theta\ge 0$ \ie $\Theta\in(0,\pi/2]\cup[3\pi/2,2\pi)$, the first part of the ellipse is the set of points
\be
\Big\{\big(\lambda_2,\lambda_1^+(\lambda_2),\big),~~ \lambda_2\in[0,1]\Big\},
\label{b+}
\ee 
which is shown in blue in Figure~\ref{ell+}.
%
%
Since an ellipse is convex, $\lambda_1^+(\lambda_2)+\lambda_2\ge 1$ is satisfied for all $\lambda_2\in[0,1]$. Hence, the geodesic parametrized as 
\be
\left\{\!\begin{array}{ll}
X^0&\!\!\!=T_1 \big(\lambda^+_1(\lambda_2)+\lambda_2\big)\\
X^1&\!\!\!=-|X^0|\esps\\
X^2&\!\!\!=\sqrt{1-\lambda_2^2\sin^2\Theta}\esps\\
X^3&\!\!\!=\lambda_2\sin\Theta\esps
\end{array}\right. 
\label{geoL}
\ee
lies on the horizon of the pode and satisfies 
\be
|X^0|\ge |T_1|, ~~ X^0\, T_1\ge 0
\qquad \ie\qquad |\sigma|\ge |\sigma_1|, ~~\sigma \sigma_1\ge0. 
\label{above}
\ee
Its length is 
\begin{align}
    L &=\int \sqrt{\dis \eta_{\mu\nu}\d X^\mu \d X^\nu}\nonumber\\
    &= \int_0^1 \d\lambda_2 \,\sqrt{\Big(\frac{\d X_2}{\d\lambda_2}\Big)^2 + \Big(\frac{\d X_3}{\d\lambda_2}\Big)^2}=\int_0^1{\d \lambda_2|\sin \Theta|\over \sqrt{1-\lambda_2^2\sin^2\Theta}}\nonumber\\
    &=\arcsin(|\sin\Theta|)\nonumber\\
    &=\left\{\!\begin{array}{ll}
    \Theta, &\mbox{if $\Theta\in(0,\pi/2]$}\\
    2\pi-\Theta,&\mbox{if $\Theta\in[3\pi/2,2\pi)$}\esps\end{array}\right.\nonumber\\
    &=\inf(\Theta,2\pi-\Theta).
\label{L1}
\end{align}
The second part of the ellipse corresponds to 
 \begin{align}
\Big\{\big(\lambda_2,\lambda_1^+(\lambda_2)\big),~~ \lambda_2\in\Big[{-1\over |\sin\Theta|},0\Big]\Big\}
&\cup\Big\{\big(\lambda_2,\lambda_1^-(\lambda_2)\big),~~ \lambda_2\in\Big[{-1\over |\sin\Theta|}, {1\over |\sin\Theta|}\Big]\Big\}\nonumber \\
&\cup\Big\{\big(\lambda_2,\lambda_1^+(\lambda_2)\big),~~ \lambda_2\in\Big[1, {1\over |\sin\Theta|}\Big]\Big\},
\label{b+'}
\end{align}
and is shown in red in Figure~\ref{ell+}.
Since $\lambda_1^+(\lambda_2)+\lambda_2$ is strictly positive at $\lambda_2=0$ and strictly negative at $\lambda_2=-1/|\sin\Theta|$, the geodesic associated to the second part of the ellipse extends on both horizons. Since 
\be
X^0=T_1(\lambda_1+\lambda_2),
\ee
$X^0$ and thus $\sigma=\arctan X^0$ changes sign along the curve.  
Its length is   
\begin{align}
    L' &=\left[\int_{-1\over |\sin\Theta|}^0+\int_{-1\over |\sin\Theta|}^{1\over |\sin\Theta|}+\int_1^{1\over |\sin\Theta|}\right]{\d \lambda_2|\sin \Theta|\over \sqrt{1-\lambda_2^2\sin^2\Theta}}\nonumber\\
    &=2\pi-\arcsin(|\sin\Theta|)\nonumber\\
    &=\left\{\!\begin{array}{ll}
    2\pi-\Theta, &\mbox{if $\Theta\in(0,\pi/2]$}\\
    \Theta,&\mbox{if $\Theta\in[3\pi/2,2\pi)$}\esps\end{array}\right.\nonumber\\
    &=\sup(\Theta,2\pi-\Theta).
    \label{L1'}
\end{align}

\noindent $\bullet$ When $\cos\Theta< 0$ \ie $\Theta\in(\pi/2,3\pi/2)$, the first part of the ellipse is the set 
 \be
\Big\{\big(\lambda_2,\lambda_1^+(\lambda_2)\big),~~ \lambda_2\in\Big[0,{1\over |\sin\Theta|}\Big]\Big\}\cup\Big\{\big(\lambda_2,\lambda_1^-(\lambda_2)\big),~~ \lambda_2\in\Big[1,{1\over |\sin\Theta|}\Big]\Big\},
\label{b-}
\ee
shown in blue in Figure~\ref{ell-}. Along its entire length, thanks to the convexity of the ellipse, $\lambda_1+\lambda_2\ge 1$ is true. This means that all points of the associated geodesic lie on the horizon of the pode and satisfy \Eq{above}. 
The length of the geodesic is 
\begin{align}
    L &=\left[\int_0^{1\over |\sin\Theta|}+\int_1^{1\over |\sin\Theta|}\right]{\d \lambda_2|\sin \Theta|\over \sqrt{1-\lambda_2^2\sin^2\Theta}}\nonumber\\
    &=\pi-\arcsin(|\sin\Theta|)\nonumber\\
    &=\left\{\!\begin{array}{ll}
    \Theta, &\mbox{if $\Theta\in(\pi/2,\pi]$}\\
    2\pi-\Theta,&\mbox{if $\Theta\in(\pi,3\pi/2)$}\esps\end{array}\right.\nonumber\\
    &=\inf(\Theta,2\pi-\Theta).
    \label{L2}
\end{align}
The second part of the ellipse is given by 
 \be
\Big\{\big(\lambda_2,\lambda_1^+(\lambda_2)\big),~~ \lambda_2\in\Big[{-1\over |\sin\Theta|},0\Big]\Big\}\cup\Big\{\big(\lambda_2,\lambda_1^-(\lambda_2)\big),~~ \lambda_2\in\Big[{-1\over |\sin\Theta|}, 1\Big]\Big\},
\label{b-'}
\ee
depicted in red in Figure~\ref{ell-}. Since $\lambda_1^+(\lambda_2)+\lambda_2$ is strictly positive at $\lambda_2=0$ and strictly negative at $\lambda_2=-1/|\sin\Theta|$, we conclude that the corresponding geodesic lies on both horizons and that $X^0$ and $\sigma$ change sign along it.  
Its length is given by
\begin{align}
    L' &=\left[\int_{-1\over |\sin\Theta|}^0+\int_{-1\over |\sin\Theta|}^1\right]{\d \lambda_2|\sin \Theta|\over \sqrt{1-\lambda_2^2\sin^2\Theta}}\nonumber\\
    &=\pi+\arcsin(|\sin\Theta|)\nonumber\\
    &=\left\{\!\begin{array}{ll}
    2\pi-\Theta, &\mbox{if $\Theta\in(\pi/2,\pi]$}\\
    \Theta,&\mbox{if $\Theta\in(\pi,3\pi/2)$}\esps\end{array}\right.\nonumber\\
    &=\sup(\Theta,2\pi-\Theta).
    \label{L2'}
\end{align}

\subsection{Curves of degenerate lengths, anchored on an arc of a screen}
\label{dege curves}

Let us reconsider the arc $A$ of angle $\Theta$ on the screen $H_1$ and the geodesics anchored on~$A$, which are described in Appendix~\ref{sec_embedding_one_arc}. Our aim is to show the existence of curves with lengths equal to those of the geodesics. 

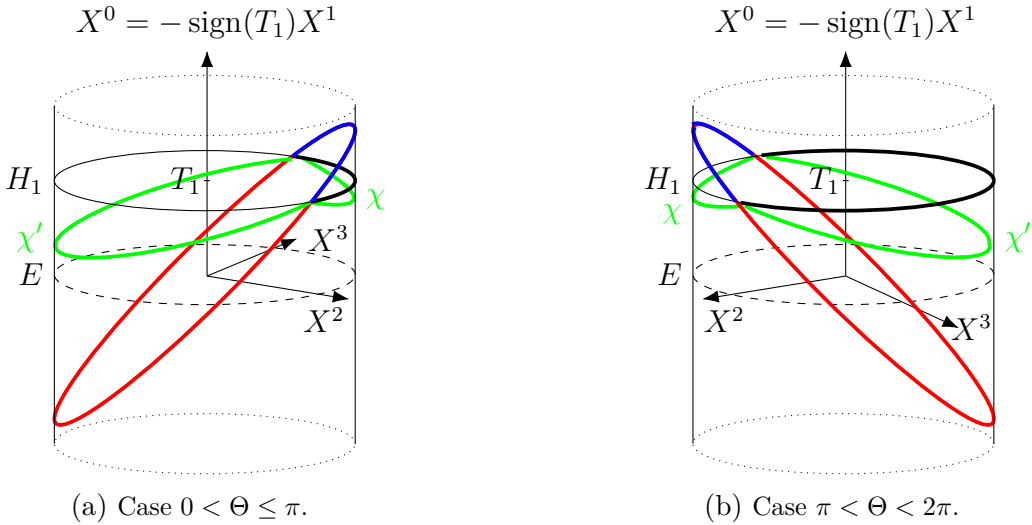
\begin{figure}[!h]
\begin{subfigure}[t]{0.48\linewidth}
\centering
\begin{tikzpicture}
	\pgfmathsetmacro{\persp}{0.2};
	\pgfmathsetmacro{\Ri}{1.5}
	\pgfmathsetmacro{\Re}{2}
	\pgfmathsetmacro{\H}{5.5+2}
	\pgfmathsetmacro{\e}{2.5}

    \draw[-{Latex[scale=1.3]}] (xyz cs:x=0.3,z=0.7,y=4) -- (xyz cs:x=1,z=-0.6,y=4) node[right] {$X^3$};
    \draw[-{Latex[scale=1.3]}] (xyz cs:x=0.3,z=0.7,y=4) -- (xyz cs:x=0.3,z=0.7,y=7) node[above] {$X^0=-\sign(T_1)X^1$};
    \draw[-{Latex[scale=1.3]}] (xyz cs:z=0.7,x=0.3,y=4) --        (xyz cs:z=1.5,x=2.5,y=4) node[left] {};
    \node[left] at (\Re, \H/2-0.6)  {${X^2}$};

    \tikzset{elliparc/.style args={#1:#2:#3}{%
insert path={++(#1:#3) arc (#1:#2:#3)}}}

	\draw[dotted] (0,6) ellipse ({\Re} and \Re*\persp);

    \draw[dashed] (0,\H/2) ellipse ({\Re} and \Re*\persp);

    \node[left] at (-\Re, \H/2)  {$E$};
    \node[right, green] at (\Re, \H/2+1)  {$\chi$}; 
    \node[left] at (-\Re, \H-\e)  {$H_1$}; 
    \node[left, green] at (-\Re, \H/2+0.5)  {$\chi'$};

    \draw[rotate around={45:(0,\H/2)}, red ,line width=0.5mm] (0,\H/2) ellipse ({1.4*\Re} and \Re*\persp);

    \draw[line width=0.5mm, green, rotate around={15:(\Re-0.58,\H-\e-0.28)}] (\Re-0.58,\H-\e-0.28) arc (-45:-360+49:{1.01*\Re} and 1.05*\Re*\persp);

    \draw[rotate around={-20:(\Re-0.02,\H-2.75)}, green,line width=0.5mm] (\Re-0.02,\H-2.75) arc (0:-35:{1.4*\Re} and \Re*\persp);
    \draw[rotate around={-20:(\Re-0.02,\H-2.75)}, green,line width=0.5mm] (\Re-0.02,\H-2.75) arc (0:45:{1.4*\Re} and \Re*\persp);

    \draw[line width=0.5mm] (\Re,\H-\e) arc (0:-45:{\Re} and \Re*\persp);
    \draw[line width=0.5mm] (\Re,\H-\e) arc (0:54:{\Re} and \Re*\persp);
    \draw (0,\H-\e) ellipse ({\Re} and \Re*\persp);

    \draw[rotate around={45:(\Re-0.02,\H/2+1.98)}, blue,line width=0.5mm] (\Re-0.02,\H/2+1.98) arc (0:-53:{1.4*\Re} and \Re*\persp);
    \draw[rotate around={45:(\Re-0.02,\H/2+1.98)}, blue,line width=0.5mm] (\Re-0.02,\H/2+1.98) arc (0:46:{1.4*\Re} and \Re*\persp);


    \draw (-0.01,\H-\e) -- (0.07,\H-\e);
    \node[left] at (0.1, \H-\e)  {$T_1$};
    
	\draw[dotted] (0,1.5) ellipse ({\Re} and \Re*\persp);
	
    \draw (-\Re,1.5)--++(0,4.5);
	\draw (\Re,1.5)--++(0,4.5);
\end{tikzpicture}
\caption{\footnotesize Case $0<\Theta\le \pi$. \label{cylL}}
\end{subfigure}
\quad \,
\begin{subfigure}[t]{0.48\linewidth}
\centering
\begin{tikzpicture}
	\pgfmathsetmacro{\persp}{0.2};
	\pgfmathsetmacro{\Ri}{1.5}
	\pgfmathsetmacro{\Re}{2}
	\pgfmathsetmacro{\H}{5.5+2}
	\pgfmathsetmacro{\e}{2.5}

    \draw[-{Latex[scale=1.3]}] (xyz cs:x=0.3,z=0.7,y=4) -- (xyz cs:x=2.5,z=2.5,y=4);
    \node[left] at (\Re+0.12, \H/2-0.7)  {${X^3}$};
    \draw[-{Latex[scale=1.3]}] (xyz cs:x=0.3,z=0.7,y=4) -- (xyz cs:x=0.3,z=0.7,y=7) node[above] {$X^0=-\sign(T_1)X^1$};
    \draw[-{Latex[scale=1.3]}] (xyz cs:z=0.7,x=0.3,y=4) --        (xyz cs:z=1.5,x=-1.3,y=4) node[left] {};
    \node[right] at (-\Re, \H/2-0.6)  {${X^2}$};

    \tikzset{elliparc/.style args={#1:#2:#3}{%
insert path={++(#1:#3) arc (#1:#2:#3)}}}

	\draw[dotted] (0,6) ellipse ({\Re} and \Re*\persp);

    \draw[dashed] (0,\H/2) ellipse ({\Re} and \Re*\persp);

    \node[left] at (-\Re, \H/2)  {$E$};
    \node[left, green] at (-\Re, \H/2+0.8)  {$\chi$}; 
    \node[left] at (-\Re, \H-\e)  {$H_1$}; 
    \node[right, green] at (\Re, \H/2+0.5)  {$\chi'$};

    \draw[rotate around={15:(0,\H-\e-0.3)}, line width=0.5mm, green] (\Re-3,\H-\e+0.6) arc (123:360-135:{\Re} and 0.95*\Re*\persp);

    \draw[rotate around={-45:(0,\H/2)}, red ,line width=0.5mm] (0,\H/2) ellipse ({1.4*\Re} and \Re*\persp);

     \draw[rotate around={-15:(0,\H-\e-0.3)}, green,line width=0.5mm] (\Re,\H-\e-0.3) arc (0:125:{\Re} and \Re*\persp);

     \draw[rotate around={-15:(0,\H-\e-0.3)}, green,line width=0.5mm] (\Re,\H-\e-0.3) arc (10:-130:{\Re} and \Re*\persp);
  
    \draw[line width=0.5mm] (\Re,\H-\e) arc (0:-133:{\Re} and \Re*\persp);
    \draw[line width=0.5mm] (\Re,\H-\e) arc (0:123:{\Re} and \Re*\persp);
    \draw (0,\H-\e) ellipse ({\Re} and \Re*\persp);
  
    \draw[rotate around={-45:(-\Re-0.02,\H/2+1.98)}, blue,line width=0.5mm] (-\Re-0.02,\H/2+2) arc (0:-46:{-1.4*\Re} and -\Re*\persp);
    \draw[rotate around={-45:(-\Re-0.02,\H/2+1.98)}, blue,line width=0.5mm] (-\Re+0.03,\H/2+2) arc (0:52:{-1.4*\Re} and -\Re*\persp);

    \draw (-0.01,\H-\e) -- (0.07,\H-\e);
    \node[left] at (0.1, \H-\e)  {$T_1$};
    
	\draw[dotted] (0,1.5) ellipse ({\Re} and \Re*\persp);
	
    \draw (-\Re,1.5)--++(0,4.5);
	\draw (\Re,1.5)--++(0,4.5);
\end{tikzpicture}
\caption{\footnotesize Case $\pi<\Theta< 2\pi$. \label{cylR}}
\end{subfigure}
    \caption{\footnotesize The submanifold $X^0=-\sign(T_1)X^1$ of dS$_3$ is a cylinder of radius 1. The arc of angle $\Theta$ lying on the screen $H_1$ is the black bold line. The two geodesics anchored to the arc lie on the cylinder. The shortest of the two (in blue) has length $L=\inf(\Theta,2\pi-\Theta)$, while the longest of the two (in red) has length $L'=\sup(\Theta,2\pi-\Theta)$.  The geodesic of length $L$ can be deformed in the cylinder into a curve $\chi$. That of length $L'$ can be deformed in the cylinder into a curve $\chi'$.  \label{cyl}}
\end{figure}

The results of Appendix~\ref{sec_embedding_one_arc} are illustrated in Figures~\ref{cylL} and~\ref{cylR}, 
%
%
which show  the submanifold of dS$_3$ satisfying $X^0=-\sign(T_1)X^1$, respectively  when $\Theta\in(0, \pi]$ and $\Theta\in(\pi, 2\pi)$. From \Eq{H12}, we see that it is a cylinder of radius 1, whose part $\sign(T_1)X^0\ge 0$ ($\sign(T_1)X^0\le 0$) is half of the horizon of the pode (antipode). The arc $A$ (in black bold) lies on $H_1$, which is the circle at $X^0=T_1$, while the bifurcate horizon $E$ is the circle at $X^0=0$. The section between them, $0\le \sign(T_1)X^0\le |T_1|$, corresponds to the boundary of the blue rectangle along the horizon of the pode in Figure~\ref{PL}. The geodesic of length $L$ is represented in blue, while that of length $L'$ is shown in red. In the following, we describe deformations of the geodesics for which  the length functional remains constant. The discussion can be divided in two cases: 

\noindent $\bullet$ When $\Theta\in(0, \pi]$, let us deform continuously along the cylinder the geodesic of length $L=\Theta$, while keeping fixed the endpoints.  We obtain an arbitrary curve $\chi$ represented in green in Figure~\ref{cylL}.\footnote{By continuous deformation along the cylinder, we mean that the geodesic of length $L=\Theta$ and the curve~$\chi$ are homotopic in the cylinder, \ie that all intermediate steps of the deformation are curves lying on the cylinder. Hence, $\chi$ cannot be the geodesic of length $L'=2\pi-\Theta$. Nor can it have a non-trivial winding number along the cylinder. } Choosing an arbitrary affine parameter $\lambda\in[0,1]$ along it,  the coordinates of its points are arbitrary functions  $\tilde X^\mu(\lambda)$ satisfying the  conditions
\be
\left\{\!\begin{array}{ll}
\tilde X^0=-\sign(T_1)\tilde X^1, \quad &\mbox{with}\quad  \tilde X^0(0)\!=\tilde X^0(1)=T_1\\
\tilde X^2+i\tilde X^3=e^{i\tilde \theta_2},\quad &\mbox{with}\quad \; \tilde \theta_2(0)=0, \quad \tilde \theta_2(1)=\Theta \esps
\end{array}\right.\!\!.
\label{top left + down right}
\ee
Since $-(\d \tilde X^0)^2+(\d \tilde X^1)^2\equiv 0$, when $\tilde \theta_2(\lambda)$ is monotonically  increasing, $\chi$ is still of length $L=\Theta$, even though it is not a geodesic. Indeed, its length is given by
\begin{align}
\int\sqrt{(\d \tilde X^2)^2 + (\d \tilde X^3)^2}&=\int \sqrt{(-\sin\tilde \theta_2)^2\d\tilde \theta_2^2+(\cos\tilde \theta_2)^2\d\tilde \theta_2^2}\nonumber\\
&=\int_0^\Theta\d\tilde \theta_2=\Theta.
\label{length}
\end{align}
On the contrary, if the function $\tilde \theta_2(\lambda)$ is not monotonic, extra positive integrals appear in the second line of the above equation and the result is strictly greater than~$\Theta$. 

When the geodesic of length $L=\Theta$ is deformed enough so that $\chi$ crosses the bifurcate horizon $E$, $\tilde X^0=-\sign(T_1)\tilde X^1$ changes sign at intersection points. In that case, we can construct a new curve $\check\chi$,  where the coordinates $(\tilde X^0,\tilde X^1)$ of all the points satisfying  $\sign(T_1)\tilde X^0=-\tilde X^1<0$ are changed to $(-\tilde X^0,\tilde X^1)$. The image points lie on the submanifold $X^0=+\sign(T_1)X^1$ of dS$_3$, which is a second cylinder of radius 1. The parametric coordinates $\tilde X^\mu(\lambda)$, $\lambda\in[0,1]$, of $\check\chi$ satisfy
\be
\left\{\!\begin{array}{ll}
\tilde X^0=\sign(T_1)|\tilde X^1|, \quad &\mbox{with}\quad  \tilde X^0(0)\!=\tilde X^0(1)=T_1\\
\tilde X^2+i\tilde X^3=e^{i\tilde \theta_2}, \quad &\mbox{with}\quad \; \tilde \theta_2(0)=0, \quad \tilde \theta_2(1)=\Theta\esps
\end{array}\right. \!\!.
\label{top left + top right}
\ee
Since $-(\d \tilde X^0)^2+(\d \tilde X^1)^2\equiv 0$ remains an identity, \Eq{length} applies  and the length of $\check\chi$ equals $L=\Theta$, when $\tilde \theta_2(\lambda)$ is monotonically  increasing. When $\tilde \theta_2(\lambda)$ is not monotonic, the length is strictly greater. One may think that starting from a differentiable curve $\chi$, our construction always yields a non-differentiable curve $\check\chi$. Indeed, the projection of $\check\chi$ on the Penrose diagram in Figure~\ref{PL} extends on the pode and antipode horizons satisfying $\sign(T_1)\,\sigma\ge 0$ and bifurcates brutally at $E$. However, the diagram is misleading, as all differentiable curves $\chi$ satisfying $\d \tilde X^0/\d \lambda=0$ when $\tilde X^0=0$ result in differentiable curves $\check\chi$.
However, the curves $\check \chi$ are not infinitely differentiable. An easy way to show it is to consider the field $\tilde X^1+\sign(T_1)\tilde X^0$, which is identically vanishing when $ \sign(T_1)\tilde X^0 \ge 0$ and varying non-trivially when $\sign(T_1)\tilde X^0 < 0$. As a result, it is not analytic and its derivative of high enough order cannot be continuous everywhere.

When $\tilde X^0=-\sign(T_1)\tilde X^1$ changes sign along the deformed curve $\chi$, we can construct an alternative curve $\hat\chi$  by flipping the sign of $\tilde X^1$ instead of $\tilde X^0$. To be specific, we change the coordinates $(\tilde X^0,\tilde X^1)$ of every point of $\chi$  satisfying  $\sign(T_1)\tilde X^0=-\tilde X^1<0$ to $(\tilde X^0,-\tilde X^1)$. The image points lie on the second cylinder, which satisfies $X^0=+\sign(T_1)X^1$ in dS$_3$. The curve $\hat\chi$ extends entirely on the horizon of the pode and its parametric coordinates $\tilde X^\mu(\lambda)$, $\lambda\in[0,1]$, satisfy
\be
\left\{\!\begin{array}{ll}
|\tilde X^0|=-\tilde X^1, \quad &\mbox{with}\quad  \tilde X^0(0)\!=\tilde X^0(1)=T_1\\
\tilde X^2+i\tilde X^3=e^{i\tilde \theta_2}, \quad &\mbox{with}\quad \; \tilde \theta_2(0)=0, \quad \tilde \theta_2(1)=\Theta \esps
\end{array}\right. \!\!.
\label{top left + down left}
\ee
The length of $\hat\chi$ is $L=\Theta$ when $\tilde \theta_2(\lambda)$ is monotonically increasing and it is longer otherwise. The curves $\hat \chi$ are not infinitely differentiable. 

Notice that the set of possible curves $\chi$, $\check \chi$ and $\hat\chi$ is also the set of curves lying on the horizons that are homologous to the arc in the exterior region. 

Similarly, the geodesic of length $L'=2\pi-\Theta$ can be deformed continuously  along the horizons, while remaining attached to the arc. One of them, $\chi'$, is shown in green in Figure~\ref{cylL}. The deformed curves, denoted $\chi'$, $\check\chi'$ or $\hat\chi'$, are parametrized by $\tilde X^\mu(\lambda)$, $\lambda\in[0,1]$, satisfying respectively \Eq{top left + down right},~(\ref{top left + top right}) and~(\ref{top left + down left}), except that the second conditions in these equations become
\be
\tilde X^2+i\tilde X^3=e^{i\tilde \theta_2}, \quad \mbox{with}\quad \; \tilde \theta_2(0)=\Theta, \quad \tilde \theta_2(1)=2\pi.
\label{2d}
\ee
 The lengths of the deformed curves  are $L'=2\pi-\Theta$ when $\tilde \theta_2(\lambda)$ is monotonically increasing and are greater otherwise. 

\noindent $\bullet$ We can proceed the same way in the case where $\Theta\in(\pi, 2\pi)$. The geodesic of length $L'=\Theta$ can be deformed continuously along  the horizons, with its ends anchored to the endpoints of the arc. One of them, $\chi'$, is shown in green in Figure~\ref{cylR}. The deformed curves, denoted $\chi'$, $\check\chi'$ or $\hat\chi'$, are  parametrized by  $\tilde X^\mu(\lambda)$, $\lambda\in[0,1]$, satisfying respectively \Eq{top left + down right},~(\ref{top left + top right}) and~(\ref{top left + down left}). They all have the same  length $L'=\Theta$ when $\tilde \theta_2(\lambda)$ is monotonically increasing, while they are longer when $\tilde \theta_2(\lambda)$ is not monotonic. The possible curves $\chi'$, $\check \chi'$ and $\hat\chi'$ are those lying on the horizons that are homologous to the arc in the exterior region. 

Finally, the geodesic of length $L=2\pi-\Theta$ can be deformed along the horizons, keeping its endpoints fixed to the arc. One of them, $\chi$, is shown in green in Figure~\ref{cylR}. The deformed curves, denoted $\chi$, $\check\chi$ or $\hat\chi$, are parametrized by $\tilde X^\mu(\lambda)$, $\lambda\in[0,1]$, which satisfy respectively \Eq{top left + down right},~(\ref{top left + top right}) and~(\ref{top left + down left}), except that the second conditions in these equations are \Eq{2d}. 
 Their lengths  are $L=2\pi-\Theta$ when $\tilde \theta_2(\lambda)$ is monotonically increasing but are greater otherwise.


\subsection{Length extremization on a domain with boundary}
\label{nu(lambda)}

In ${\rm d S}_3$, there is one geodesic curve anchored to an arc of angle $\Theta$  in the screen $H_1$ and lying in the causal diamond in the static patch of the pode. It is the geodesic of length $L=\inf(\Theta,2\pi-\Theta)$ described in Appendices~\ref{sec_embedding_one_arc} and~\ref{dege curves}. It actually lies on the portion of the horizon of the pode 
represented in Figure~\ref{fig:Penrose_dS_ext} as the line segment between $H_1$ and the corner $(\theta_1,\sigma)=(0,\sign(\sigma_1)\pi/2)$. 
The aim of this appendix is to show that a continuous family of curves extending on this boundary of the causal diamond of the left interior region are solutions of a length-extremization problem. 

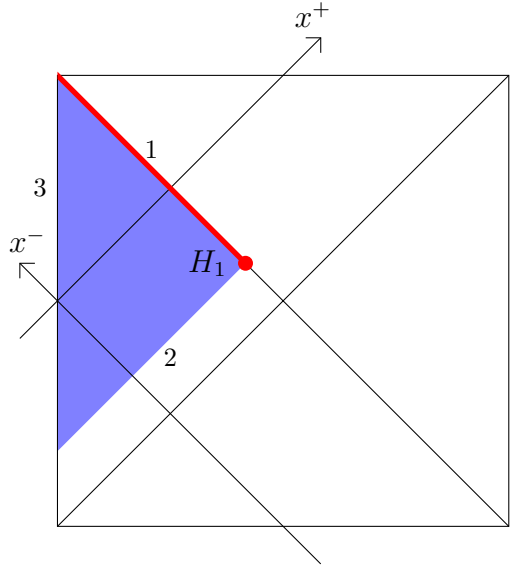
\begin{figure}[!h]
\centering

\begin{tikzpicture}

\path
       +(3,3)  coordinate (IItopright)
       +(-3,3) coordinate (IItopleft)
       +(3,-3) coordinate (IIbotright)
       +(-3,-3) coordinate(IIbotleft)
      
       ;

\fill[fill=blue!50] (-3,3) -- node[pos=0.5, above] {{\footnotesize $1$}} (-1/2,1/2) -- node[pos=0.4, below] {{\footnotesize $2$}} (-3,-2) -- node[pos=0.7, left] {{\footnotesize $3$}} cycle;
           
\draw (IItopleft) --
      (IItopright) --
      (IIbotright) -- 
      (IIbotleft) --
      (IItopleft) -- cycle;
      
\draw (IItopleft) -- (IIbotright)
              (IItopright) -- (IIbotleft) ;

\draw (-3.5,0.5) -- (1/2,-3.5);
\draw (-3.5,-0.5) -- (1/2,3.5);
\draw (-3.5,0.3) -- (-3.5,0.5) -- node[midway, above, sloped] {$x^-$} (-3.3,0.5) ;
\draw (0.5,3.3) -- (0.5,3.5) -- node[midway, above, sloped] {$x^+$} (0.3,3.5) ;
              

\fill[fill=red] (-1/2,1/2+0.05) -- (-3,3.05) -- (-3,3-0.05) -- (-1/2,1/2-0.05) -- cycle; 

\node at (-1/2,1/2) [circle,fill,inner sep=2pt, red, label = left:$H_1$]{};

\end{tikzpicture}
\caption{\footnotesize The causal diamond of all Cauchy slices $\Sigma_{\rm left}'$ has 3 boundaries on the complex plane $(x^+,x^-)$. Curves anchored to the arc of angle $\Theta$ (or $2\pi-\Theta$) in~$H_1$ and lying on the boundary 1 are the only spacelike solutions of the length-extremization problem. 
\label{tri}}
\end{figure}

In terms of the lightlike coordinates~(\ref{x+-}), the metric~(\ref{eq:dS3_metric})  of dS$_3$ can be written as 
\begin{equation}
\d s^2=\frac{1}{\cos^2\!\left(\frac{x^++x^-}{2}\right)}\left(-\d x^+ \d x^-+\sin^2\!\left(\frac{x^+-x^-}{2}\right)\d\theta_2^2\right)\!.
\label{m+-}
\end{equation}
For simplicity, we will assume that $H_1$ is at a positive generic conformal time $\sigma_1$, \ie satisfying $0<\sigma_1<\pi/2$. In that case the causal diamond in the left static patch is the domain\footnote{Notice that even if the line $\theta_1=(x^+-x^-)/2=0$ is a boundary of the domain in the complex plane $(x^+,x^-)$, it is not a boundary of the static patch of the pode. (See \Sect{dSgeo}.)}
\begin{equation}
\left\{\!\begin{array}{l}
\dis x^+\le \frac{\pi}{2}\\
\dis x^-\ge 2\sigma_1-{\pi\over 2}\\
\dis x^+-x^-\ge 0
\end{array}\right.\!\!.
\label{triangle}
\end{equation}
It has the shape of a triangle shown in blue in Figure~\ref{tri}. Its boundary is composed of 3 segments of line denoted  1, 2, 3. 
%
%
Let us consider the arc of angle $\Theta\in(0,2\pi)$ in $H_1$. Any curve in the diamond and  anchored to the arc can be parametrized by coordinates $(\tilde x^+(\lambda),\tilde x^-(\lambda), \tilde \theta_2(\lambda))$, where $\lambda\in[0,1]$ is an arbitrary affine parameter.  The length of the curve is 
\be
\ell=\int_0^1 d\lambda\,\mathcal{L},
\label{elldef}
\ee
where the Lagrangian $\mathcal{L}$ follows from \Eq{m+-},
\begin{equation}
\mathcal{L}\big(\tilde x^+,\tilde x^-,\dot {\tilde x}^+,\dot {\tilde x}^-,\dot{\tilde \theta}_2\big)=\frac{\sqrt{-\dot {\tilde x}^+\dot {\tilde x}^-+\sin^2(\frac{\tilde x^+-\tilde x^-}{2})\dot{\tilde \theta}_2^2}}{\cos(\frac{\tilde x^++\tilde x^-}{2})},
\end{equation}
and where the dots stand for derivatives with respect to $\lambda$.

Let us now relax the constraint that all curves should lie from the onset in the causal diamond or even in the square Penrose diagram. To be specific, let us allow $\tilde x^+$, $\tilde x^-$ and $\tilde \theta_2$ to be fields taking values in $\R$. The conditions that a curve be anchored to the arc become
\begin{align}
&\tilde x^+(0)=\tilde x^+(1)={\pi\over 2},\qquad \tilde x^-(0)=\tilde x^-(1)=2\sigma_1-{\pi\over 2},\nonumber \\ 
&\tilde \theta_2(0)=0,~~ \tilde \theta_2(1)=\Theta+2\pi w ~~\quad \mbox{or}~~\quad \tilde \theta_2(0)=\Theta,~~\tilde \theta_2(1)=2\pi+2\pi w,\quad ~~w\in\natural.
\label{bcx}
\end{align}
Notice that $\tilde \theta(\lambda)$ not being restricted to an interval allows a non-trivial winding number~$w$.\footnote{Negative winding numbers are not allowed because they produce the same set of curves with only their orientations reversed.} Our goal is to extremize the length for unconstrained coordinate fields $\tilde x_\pm$, $\tilde \theta_2$, while imposing the constraint of being located inside the causal diamond only on-shell. This is done by considering a generalized length-functional 
\begin{align}
\hat \ell[\tilde x^+,\tilde x^-,\tilde \theta_2, \nu_I,a_I]=\ell[\tilde x^+,\tilde x^-,\tilde \theta_2]&+\int_0^1 \d \lambda\, \Big\{\nu_1\left(\tilde x^+-\frac{\pi}{2}+a_1^2\right)\nonumber \\
&+ \nu_2\left(\tilde x^--2\sigma_1+\frac{\pi}{2}-a_2^2\right)+\nu_3\left(\tilde x^+-\tilde x^--a_3^2\right)\!\Big\},
\end{align}
where $\nu_I(\lambda)$, $I\in\{1,2,3\}$, are Lagrange multipliers and $a_I^2(\lambda)$ are the positive distances from the boundary segments of the diamond. The extra fields $\nu_I$, $a_I$, which are non-dynamical, take values in $\R$. 

The extrema of the ``action'' $\hat \ell$ are found by solving the ``equations of motion of the 9  fields.'' Setting to zero the variation of $\hat \ell$ with respect to $x^+$ and $x^-$, we obtain 
\begin{subequations}
\label{ex}
\begin{align}
\frac{\partial\mathcal{L}}{\partial \tilde x^+}-\frac{\d}{\d\lambda}\frac{\partial\mathcal{L}}{\partial \dot {\tilde x}^+}+\nu_1+\nu_3&=0,\label{ex+}\\
\frac{\partial\mathcal{L}}{\partial \tilde x^-}-\frac{\d}{\d\lambda}\frac{\partial\mathcal{L}}{\partial \dot {\tilde x}^-}+\nu_2-\nu_3&= 0,\label{ex-}
\end{align}
\end{subequations}
where we have 
\begin{align}
\frac{\partial\mathcal{L}}{\partial \tilde x^+}&=\phantom{-}\frac{1}{2}\frac{1}{\cos^2\tilde \sigma}\frac{\cos \tilde x^- \sin\tilde \theta_1\, \dot{\tilde \theta}_2^2-\sin\tilde \sigma\, \dot {\tilde x}^+ \dot {\tilde x}^-}{\sqrt{-\dot {\tilde x}^+\dot {\tilde x}^-+\sin^2\tilde \theta_1\, \dot{\tilde \theta}_2^2}},\nonumber \\
\frac{\partial\mathcal{L}}{\partial \tilde x^-}&=-\frac{1}{2}\frac{1}{\cos^2\tilde \sigma}\frac{\cos \tilde x^+ \sin\tilde \theta_1\, \dot{\tilde \theta}_2^2+\sin\tilde \sigma\, \dot {\tilde x}^+ \dot {\tilde x}^-}{\sqrt{-\dot {\tilde x}^+\dot {\tilde x}^-+\sin^2\tilde \theta_1\, \dot{\tilde \theta}_2^2}},\nonumber\\
\frac{\partial\mathcal{L}}{\partial\dot {\tilde x}^+}&=-\frac{1}{2}\frac{1}{\cos\tilde \sigma} \frac{\dot {\tilde x}^-}{\sqrt{-\dot {\tilde x}^+\dot {\tilde x}^-+\sin^2\tilde \theta_1\, \dot{\tilde \theta}_2^2}},\nonumber\\
\frac{\partial\mathcal{L}}{\partial\dot {\tilde x}^-}&=-\frac{1}{2}\frac{1}{\cos\tilde \sigma} \frac{\dot {\tilde x}^+}{\sqrt{-\dot {\tilde x}^+\dot {\tilde x}^-+\sin^2\tilde \theta_1\, \dot{\tilde \theta}_2^2}},
\end{align}
and where $\tilde \sigma=(\tilde x^++\tilde x^-)/2$ and  $\tilde \theta_1=(\tilde x^+-\tilde x^-)/2$. 
Since $\mathcal{L}$ is independent of $\tilde \theta_2$, the equation of motion $\delta \hat \ell/ \delta \tilde \theta_2=0$ can be trivially integrated once. This yields 
\begin{equation}
C\equiv\frac{\partial\mathcal{L}}{\partial\dot{\tilde \theta}_2}=\frac{1}{\cos\tilde \sigma}\frac{\sin^2\tilde \theta_1\, \dot{\tilde \theta}_2}{\sqrt{-\dot {\tilde x}^+\dot {\tilde x}^-+\sin^2\tilde \theta_1\, \dot{\tilde \theta}_2^2}},
\label{eT}
\end{equation}
where $C$ is a conserved quantity. 
Varying $\hat \ell$ with respect to the Lagrange multipliers leads to the equations 
\begin{subequations}
\label{en}
 \begin{align}
a_1^2&= -x^++\frac{\pi}{2},\label{en1}\\
a_2^2&=x^--2\sigma_1+\frac{\pi}{2},\label{en2}\\
a_3^2&=x^+-x^-,\label{en3}
\end{align}
\end{subequations}
which imply $\hat \ell=\ell$ when they are satisfied. Finally,  the variations with respect to $a_I$ give
\begin{subequations}
\label{ea}
\begin{align}
\nu_1a_1&=0,\label{ea1}\\
\nu_2a_2&=0,\label{ea2}\\
\nu_3a_3&=0.\label{ea3}
\end{align}
\end{subequations}

Thanks to \Eqs{en}, all solutions to the 9 equations satisfying the boundary conditions~(\ref{bcx}) have $(x^+(\lambda),x^-(\lambda))$, $\lambda\in[0,1]$, located inside the blue triangle in Figure~\ref{tri}. Therefore, we can organize our search of solutions from the locations of the associated curves in this triangle: 

- When $(x^+(\lambda),x^-(\lambda))$, $\lambda\in[0,1]$, evolves along the boundary segment~1, we have $x^+\equiv \pi/2$. In that case, \Eq{eT} reduces to $\sign(\dot{\tilde \theta}_2)\equiv C$. The function $\tilde \theta_2(\lambda)$ is thus monotonic and actually increasing due to the boundary conditions~(\ref{bcx}): $\sign(\dot{\tilde \theta}_2)\equiv 1$. Moreover, we have $a_1\equiv 0$ from \Eq{en1}, which implies that \Eq{ea1} is satisfied. \Eqs{en2},~(\ref{en3}) determine $a_{2,3}^2(\lambda)\ge 0$, $\lambda\in[0,1]$. 

When $\lambda$ is such that $a_{2,3}^2(\lambda)> 0$, we have $\nu_{2,3}(\lambda)=0$ from \Eq{ea2},~(\ref{ea3}). In that case, \Eq{ex-} is trivially solved, while \Eq{ex+} determines $\nu_1(\lambda)$. 

When $\lambda$ is such that $a_{2}(\lambda)= 0$, the point of the curve sits on $H_1$ and we have $a_3^2(\lambda)>0$. \Eq{ea2} is thus satisfied, while $\nu_3(\lambda)=0$ from  \Eq{ea3}. In that case, \Eq{ex-} imposes $\nu_2(\lambda)=0$, while \Eq{ex+} determines $\nu_1(\lambda)$. 

Similarly, when $a_{3}(\lambda)= 0$, the point of the curve is located at the upper left corner of the Penrose diagram and we have $a_2^2(\lambda)>0$. Reasoning as in the previous case, one can see that all equations are satisfied. 

As a result, we have found curves $(\tilde x^+,\tilde x^-, \tilde \theta_2)$ supplemented by the $\nu_I$ and $a_I$ functions that extremize $\hat \ell$. Applying \Eq{elldef}, their lengths are given by 
\be
\hat \ell=\ell=\int_0^1 \d \lambda\, \dot{\tilde \theta}_2=\left\{\!\begin{array}{ll}\dis \int_0^{\Theta+2\pi w}\d\tilde\theta_2&\!\!\!=\Theta+2\pi w\\ \mbox{or}\\\dis \int_\Theta^{2\pi+2\pi w}\d\tilde\theta_2&\!\!\!=2\pi-\Theta+2\pi w\esps\end{array}\right.\!\!.
\ee
In fact, the solutions with vanishing winding numbers are all curves homotopic, in the portion of the horizon corresponding to segment 1 in Figure~\ref{tri}, to the arc of angle $\Theta$ or to its complement of angle $2\pi-\Theta$ in $H_1$, and with coordinate $\theta_2$ monotonic on all their lengths.

- When $(x^+(\lambda),x^-(\lambda))$ partially evolves along the bulk of the boundary segment~2, \ie not on its endpoints,  we have $x^-(\lambda)= 2\sigma_1-\pi/2$ for all $\lambda\in K$, where $K$ is a subset of~$[0,1]$. For $\lambda \in K$, we also have $a_{1,3}^2(\lambda)>0$ from \Eqs{en1},~(\ref{en3})  and thus $\nu_{1,3}(\lambda)=0$ from \Eqs{ea1},~(\ref{ea3}). As a result, \Eq{ex+} reduces to $\dot {\tilde \theta}_2 \sin 2\sigma_1 =0$ in $K$, \ie $\dot {\tilde \theta}_2\equiv 0$. 
This means that $\lambda\in K$ corresponds to lightlike portions of the curve, since constant $x^-$ and~$\theta_2$ define lightlike directions. These portions of the curve are thus geodesics between their endpoints.  However, restricting to spacelike curves extremizing $\hat \ell$, none lies even partially along the bulk of the boundary segment~2.


- When $(x^+(\lambda),x^-(\lambda))$ partially evolves along the boundary segment~3, we have $\tilde \theta_1(\lambda)=0$ for all $\lambda\in K$, where $K$ is a subset of~$[0,1]$. The associated portions of the curve coincide with the worldline of the observer at the pode and are therefore timelike. Indeed, the metric reduces to $\d s^2=-\d \tilde \sigma^2/\cos^2 \tilde \sigma\le0$. Even if we are mostly interested in spacelike curves, we can however analyze these timelike portions.   

For all $\lambda\in K$, \Eqs{eT}, (\ref{ex+}), (\ref{ex-}) are simplified to 
\be
C=0, \qquad \nu_1+\nu_3=0, \qquad \nu_2-\nu_3=0,
\label{3e}
\ee
leaving the functions $\tilde \theta_2(\lambda)$ and $\tilde x^+(\lambda)=x^-(\lambda)$ free. Since $x^+(\lambda)= x^-(\lambda)$, \Eq{en3} yields $a_3(\lambda)=0$ and \Eq{ea3} is satisfied, while \Eqs{en1},~(\ref{en2}) determine  $a_{1,2}^2(\lambda)\ge 0$.

When $\lambda\in K$ is such that $a_{1,2}^2(\lambda)> 0$, we have $\nu_{1,2}(\lambda)=0$ from \Eq{ea1},~(\ref{ea2}), and $\nu_{3}(\lambda)=0$ from \Eqs{3e}. 

When $\lambda\in K$ is such that $a_2(\lambda)=0$ and $a_1^2(\lambda)>0$, \Eq{ea2} is satisfied and $\nu_{1}(\lambda)=0$ from \Eq{ea1}.  \Eqs{3e} then yields $\nu_2(\lambda)=\nu_3(\lambda)=0$. 

When $\lambda\in K$ is such that $a_1(\lambda)=0$ and $a_2^2(\lambda)>0$, \Eq{ea1} is satisfied and $\nu_{2}(\lambda)=0$ from \Eq{ea2}.  \Eqs{3e} then yields $\nu_1(\lambda)=\nu_3(\lambda)=0$. 

To summarize, the portions of the curves parametrized by $\lambda\in K$ satisfy all local conditions for extremizing $\hat \ell$, whatever the functions $\tilde \theta_2(\lambda)$ and  $\tilde \sigma(\lambda)$ are.  However, restricting to spacelike curves extremizing $\hat \ell$, none is allowed to lie even partially along the boundary segment~3. 

- Finally, let us consider any curve lying entirely in the bulk of the triangular domain (except at  its endpoints and other possible isolated points). At generic point,  we have $a_{1,2,3}^2(\lambda)>0$ from \Eqs{en} and $\nu_{1,2,3}(\lambda)=0$ from \Eqs{ea}. Hence, $\hat\ell\equiv \ell$ almost everywhere and extremizing $\hat \ell$ amounts to looking for geodesics lying in the bulk of the triangular domain, which we know do not exist, as shown in Appendix~\ref{sec_embedding_one_arc}.

To conclude this \Appendix{nu(lambda)}, all spacelike curves anchored to the arc of angle $\Theta$ in $H_1$ and lying inside the causal diamond located in the static patch of the pode are actually extending along the section of the horizon denoted 1 in Figure~\ref{tri}, with $\tilde \theta_2$ increasing and arbitrary winding number $w\in\natural$ around the horizon.\footnote{Recall that this section of the horizon is a cylinder, as seen in Appendix~\ref{dege curves}.}


\subsection{Geodesic connecting symmetric points on the screens}
\label{arm}

In this section, we consider the screens $H_1$ and $H_2$, respectively located in the Penrose diagram at $(\theta_{10},\sigma_0)$ and $(\pi-\theta_{10},\sigma_0)$, where $(\theta_{10},\sigma_0)$ satisfies \Eq{dH1}. Our goal is to compute the length of an ``arm along the barrel.'' What is meant by arm is the geodesic that connects in the exterior region two symmetric points on $H_1$ and $H_2$, \ie located on the screens at the same coordinate $\theta_2$. The result being independent of~$\theta_2$, we will take $\theta_2=0$. The derivation will be done by applying two methods. 

\subsubsection{Using embedding coordinates}
\label{armsem}

In the conventions of \Eq{H12},  the exterior region of dS$_3$ are the points whose embedding coordinates satisfy
\be
 |X^0|\ge |X^1|\quad \and\quad \mbox{\Eq{hyperboloid}}.
\label{er}
\ee
Using \Eqs{h1} and~(\ref{Ts}), the endpoints of the curves of extremal lengths we look for are located at
\begin{equation}
    (T,-|T|,1,0),  \qquad (T,|T|,1,0),
\end{equation}
where $T=\tan \sigma_0$, and the plane passing through them and the origin is the set
\be
\Big\{\lambda_1\big(T,-|T|,1,0\big)+\lambda_2\big(T,|T|,1,0\big), ~\where~\lambda_1,\lambda_2\in\R\Big\}. 
\label{pl2}
\ee
Indeed, defining 
\be u=\lambda_2 + \lambda_1,\qquad v=\lambda_2-\lambda_1,
\label{uv}
\ee
the endpoints are reached for $(v,u)=(-1,1)$ and $(1,1)$, while $(v,u)=(0,0)$ matches the origin. The intersection of the plane and the hyperboloid~(\ref{hyperboloid}) are the geodesics we are interested in. This intersection amounts to the set of points $(X^0,X^1,X^2,X^3)=(uT,v|T|,u,0)$, such that 
\be 
u^2(1-T^2)+v^2T^2=1.
\label{eh}
\ee
From \Eq{er}, those  located in the exterior region  satisfy
\be
|u|\ge |v|.
\label{u>v}
\ee

\noindent $\bullet$ When $|T|\le 1$, \Eq{eh} defines an ellipse\footnote{We include the limit cases $T=0$ and $T=\pm1$.} in the plane $(v, u)$ and the roots for $u$ are given by 
\be
u^\pm(v)=\pm\sqrt{1-v^2T^2\over 1-T^2},\quad \where\quad |v|\le {1\over |T|}.
\ee
The ellipse can be divided into two parts, both ending at the points $(v,u)=(-1,1)$ and~$(1,1)$. The first part corresponds to the set
 \be
\Big\{\big(v,u^+(v),\big),~~ v\in[-1,1]\Big\},
\label{eext}
\ee 
which gives rise to a geodesic fully contained in the exterior region. It is the arm we are looking for. Its length is 
\begin{align}
    L_{\rm arm} &= \int \sqrt{\dis \eta_{\mu\nu}\d X^\mu\d X^\nu}=\int_{-1}^1{\d v |T|\over \sqrt{1-v^2 T^2}}\nonumber\\
    &=2\arcsin(|T|)\nonumber \\
    &=2\arcsin(\sinh|\tau_0|),
\label{L1'arm}
\end{align}
where we use \Eq{Ts} in the last equality to define the global time $\tau_0$. Since $L_{\rm arm}$ is real positive, the arm is spacelike. 
The second part of the ellipse corresponds to the set 
\begin{align}
\Big\{\big(v,u^+(v)\big),~~ v\in\left[-{1\over |T|}, -1\right]\Big\}&\cup\Big\{\big(v,u^-(v)\big),~~ v\in\left[-{1\over |T|}, {1\over |T|}\right]\Big\}\nonumber \\ 
&\cup\Big\{\big(v,u^+(v)\big),~~ v\in\left[1,{1\over |T|}\right]\Big\}
\label{b-2}.
\end{align}
As opposed to the previous case, it leads to an  irrelevant geodesic lying in both interior regions and the exterior one.  

\noindent $\bullet$ When $|T|>1$, \Eq{eh} defines an hyperbola in the plane $(v, u)$ and the roots for $u$ are given by
\be
u^\pm(v)=\pm\sqrt{v^2T^2-1\over T^2-1},\quad \where\quad |v|> {1\over |T|}.
\ee
As before, the hyperbola can be divided into two parts, one of them giving rise to a geodesic, the arm,\footnote{ The denomination ``arm'' in this case is artificial since the geodesic is actually composed of two disconnected parts reaching points at infinity in dS$_3$.} lying entirely in the exterior region, while the second part yields a geodesic extending in both interior regions and the exterior region. However, there is no need to describe them in detail since the infinitesimal length along them is 
\begin{align}
\d s&=\sqrt{\dis \eta_{\mu\nu}\d X^\mu\d X^\nu}\nonumber \\
&={\d v |T|\over \sqrt{1-v^2 T^2}}\, ,
\end{align}
which is pure imaginary for $v>1/|T|$. In other words, they are timelike and thus irrelevant.


\subsubsection{Using equations of motion}
\label{two arcs:eq_of_motion}

Results of the previous subsection can also be derived using a more common method based on the geodesic equation. To be specific, we want to compute in another way the length of a spacelike arm, \ie the size of the geodesic connecting in the exterior region the points $(\theta_{10}, \sigma_0, 0)$ and $(\pi-\theta_{10},\sigma_0,0)$, where $(\theta_{10},\sigma_0)$ satisfy \Eq{dH1}. Since these points have the same coordinate $\theta_2=0$ and the metric in \Eq{eq:dS3_metric} is independent of $\theta_2$, we will search for geodesics at constant $\theta_2\equiv0$.

\begin{figure}[!h]
    \centering
\begin{tikzpicture}
\begin{scope}[transparency group]
\begin{scope}[blend mode=multiply]
\path
       +(3,3)  coordinate (IItopright)
       +(-3,3) coordinate (IItopleft)
       +(3,-3) coordinate (IIbotright)
       +(-3,-3) coordinate(IIbotleft)
      
       ;
\draw (IItopleft) --
      (IItopright) --
      (IIbotright) -- 
      (IIbotleft) --
      (IItopleft) -- cycle;

\draw (IItopleft) -- (IIbotright)
              (IItopright) -- (IIbotleft) ;

\draw[line width= 0.5 pt]plot[variable=\t,samples=1000,domain=-86.2:86.2] ({-0.1*tan(\t)},{-0.1*sec(\t)+3});

\draw[line width= 0.5 pt,dashed]plot[variable=\t,samples=1000,domain=-88.1:-86.2] ({-0.1*tan(\t)},{-0.1*sec(\t)+3});

\draw[line width= 0.5 pt,dashed]plot[variable=\t,samples=1000,domain=86.2:88.1] ({-0.1*tan(\t)},{-0.1*sec(\t)+3});

\draw (1.2,3) node[below] {$J=-\infty$};

\draw[dotted] (-3,1.3)--(3,1.3);   
\node at (-3,1.3) [label = left:$\sigma_0$]{};
\draw[dotted] (-1.3,0)--(-1.3,3);   
\node at (-1.3,-0.1) [label = below:$\theta_{10}$]{};

\draw[line width= 0.5 pt,violet]  plot[variable=\t,samples=1000,domain=-71:71] ({-0.5*tan(\t)},{-0.5*sec(\t)+3});

\draw[line width= 0.5 pt,violet,dashed]  plot[variable=\t,samples=1000,domain=-80.4:-71] ({-0.5*tan(\t)},{-0.5*sec(\t)+3});

\draw[line width= 0.5 pt,violet,dashed]  plot[variable=\t,samples=1000,domain=71:80.4] ({-0.5*tan(\t)},{-0.5*sec(\t)+3});

\draw[line width= 0.5 pt,blue]  plot[variable=\t,samples=1000,domain=-40:40] ({-1.5*tan(\t)},{-1.5*sec(\t)+3.3});

\draw[line width= 0.5 pt,blue,dashed]  plot[variable=\t,samples=1000,domain=-63:-40] ({-1.5*tan(\t)},{-1.5*sec(\t)+3.3});

\draw[line width= 0.5 pt,blue,dashed]  plot[variable=\t,samples=1000,domain=40:63] ({-1.5*tan(\t)},{-1.5*sec(\t)+3.3});

\draw[color=cyan,dashed] (-3,0)-- node[midway, above, sloped] {$J=-1$}(3,0);

\node at (-3,3) [label = left:$\pi/2$]{};
\node at (-3,0) [label = left:$0$]{};
\node at (-3,-3) [label = left:$-\pi/2$]{};

\node at (3.3,0) [label = below left:$\pi$]{};

\draw[->] (-3, 3) -- (-3, 3.5) node[above]{ $\sigma$};
\draw[->] (3,0) -- (3.5,0) node[right]{ $\theta_1$};

\end{scope}
\end{scope}
\end{tikzpicture}
    \caption{\footnotesize Spacelike geodesics between points $(\theta_{10},\sigma_0)$ and $(\pi-\theta_1,\sigma_0)$ at $\theta_2=0$,  located on the horizons of the pode and antipode. They approach a seemingly null geodesic as $J \rightarrow -\infty$. By continuing them into the causal patches, they reach the pode and antipode at $\sigma=0$ and their total lengths are $\pi$.}
    \label{fig:geodesics}
\end{figure}
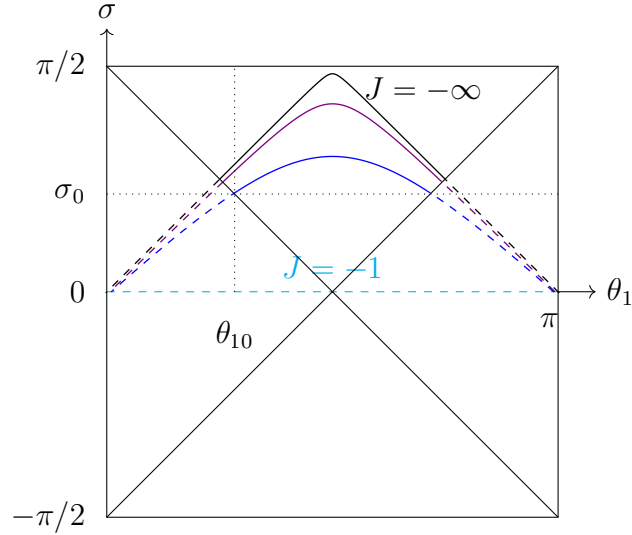

The length $L_{\rm arm}$ we are looking for is an extremum of the functional 
\begin{equation}
\ell=\int \d s=\int_{\theta_{10}}^{\pi-\theta_{10}} \d\theta_1\,\mathcal{L}(\sigma,\dot\sigma),
\end{equation}
where $\cal L$ is a Lagrangian whose form follows from the conformal metric~(\ref{eq:dS3_metric}), 
\begin{equation}
\mathcal{L}(\sigma,\dot\sigma)=\frac{\sqrt{1-\dot\sigma^2}}{\cos\sigma}.
\end{equation}
In this expression, the ``dots'' stand for derivatives with respect to $\theta_1$ and we have set $\d\theta_2\equiv 0$. Since $\mathcal{L}$ is independent of $\theta_1$, the on-shell Hamiltonian is  conserved,
\begin{equation}
J\equiv\frac{\partial\mathcal{L}}{\partial\dot\sigma}\dot\sigma-\mathcal{L}=\frac{-1}{\cos\sigma\sqrt{1-\dot\sigma^2}}.
\label{j}
\end{equation} 
In other words,  this expression, where $J<0$ is a constant, is nothing but the first integral of the equation of motion. Since $\theta_1\to \pi-\theta_1$ is a symmetry of the metric~(\ref{eq:dS3_metric}), which is also satisfied by the endpoints of the geodesic, the whole trajectory satisfies 

\be
\sigma(\theta_1)\equiv \sigma(\pi-\theta_1) \quad \Longrightarrow\quad \dot\sigma(\theta_1)\equiv-\dot \sigma(\pi-\theta_1).
\ee 
As a result, the fixed point $\theta_1=\pi/2$ is a turning point, $\dot\sigma(\pi/2)=0$, and we obtain from \Eq{j} that
\begin{equation}\label{eq:sigma_M(J)}
\sigma(\pi/2)=\epsilon \arccos\frac{1}{-J}\, ,\quad \mbox{for some }\epsilon\in\{+1,1\}. 
\end{equation}
This result tells us that we must have 
\be
-J\ge 1,
\label{-j}
\ee  
for the geodesic to exist. Assuming that $\dot\sigma$ does not change sign when $\theta_1<\pi/2$, \Eq{j} implies
\be
{J\cos \sigma\, \d\sigma \over \sqrt{J^2\cos^2\sigma-1}}=\eta\d \theta_1\, ,\quad \theta_1\le\pi/2\, , \quad \mbox{for some }\eta\in\{+1,1\}. 
\ee
Integrating the right-hand side from $\theta_1$ to $\pi/2$ and imposing the boundary condition~(\ref{eq:sigma_M(J)}), we find 
\be
\arctan{J\sin\sigma\over \sqrt{J^2\cos^2\sigma-1}}=-(\epsilon+\eta){\pi\over 2}+\eta\theta_1,\quad \theta_1\le{\pi\over 2}.
\ee
Since the left-hand side is in the range $[-\pi/2,\pi/2]$, we have $\epsilon+\eta=0$, which leads to the geodesic
\be
\arctan{J\sin\sigma\over \sqrt{J^2\cos^2\sigma-1}}=-\epsilon\theta_1,\quad \theta_1\le{\pi\over 2},
\label{geo}
\ee
which can be completed for $\theta_1>\pi/2$ thanks to the symmetry $\theta_1\to \pi-\theta_1$. Notice that the geodesic can be extended in the causal patches of the pode and antipode, \ie for $\theta_1\in[0,\theta_{10}]$ and $[\pi-\theta_{10},\pi]$. In particular, one sees that 
\begin{equation}\label{eq:sigma(0)=0}
\sigma(0)=\sigma(\pi)=0,
\end{equation}
which means that the geodesic passes through the pode and antipode at $\sigma=0$. \Fig{fig:geodesics} shows the geodesics arising for various values of $-J\ge 1$ and $\epsilon=+1$.
By applying \Eq{geo} at $(\theta_{10},\sigma_0)$ and using the definition of the horizon of the pode, \Eq{dH1}, one obtains
\begin{equation}\label{eq:sigma_0(J)}
\sigma_0=\epsilon\arccos\frac{-J}{\sqrt{2J^2-1}}.
\end{equation}
Unicity of the geodesic we have found in the exterior region justifies {\it a posteriori} the assumption we made below \Eq{-j}. 
Using  \Eq{geo}, one can express the extremal value of~$\ell$~as
\begin{align}
L_{\rm arm}&=2\epsilon\int_{\sigma_0}^{\sigma({\pi\over 2})}\frac{\d\sigma}{\cos\sigma}\frac{1}{\sqrt{J^2\cos^2\sigma-1}}\nonumber \\
\label{eq:length_integral}
&=2\epsilon \left[\arctan\frac{\sin\sigma}{\sqrt{J^2\cos^2\sigma-1}}\right]_{\sigma_0}^{\sigma({\pi\over 2})}\nonumber \\
&= \arctan\sqrt{J^2-1}.\esps
\end{align}
To write this result in term of the global time $\tau_0$, we use \Eq{eq:sigma_0(J)} and the relation~(\ref{tau-sigma}) to find 
\be
J^2={1\over 2-\cosh^2\tau_0}>1.
\ee
This yields
\begin{align}
L_{\rm arm}&=2\arctan{|\sinh\tau_0|\over \sqrt{2-\cosh^2\tau_0}}\nonumber\\
&=2\arcsin(\sinh|\tau_0|),
\end{align}
where the constraint $J^2>1$ translates into
\be
|\tau_0|\le\arcsinh 1,
\ee
thus reproducing \Eq{L1'arm}.

Before concluding this subsection, let us note that the length of the geodesic between the pode $(\theta_1,\sigma)=(0,0)$ and antipode $(\theta_1,\sigma)=(\pi,0)$ is 
\be
2\epsilon\int_{0}^{\sigma({\pi\over 2})}\frac{\d\sigma}{\cos\sigma}\frac{1}{\sqrt{J^2\cos^2\sigma-1}}=2\epsilon \arctan\frac{\sin\sigma({\pi\over 2})}{\sqrt{J^2\cos^2\sigma({\pi\over 2})-1}}= \pi,
\ee
which is independent of $J$. This may be surprising since the trajectory in \Eq{geo} approaches the null geodesic $\sigma=\epsilon \theta_1$, where $\theta_1\le\pi/2$, as $J\to -\infty$, and that null geodesics have zero proper lengths. The resolution of this puzzle is that for arbitrary $J\le -1$, most of the length is obtained from the neighborhood of the turning point $\theta_1=\pi/2$. In fact, we have for arbitrary $\zeta>0$
\begin{align}
2\epsilon\int_{\sigma({\pi\over 2})-\epsilon\zeta}^{\sigma({\pi\over 2})}\frac{\d\sigma}{\cos\sigma}\frac{1}{\sqrt{J^2\cos^2\sigma-1}}&=\pi-2 \arctan{\sin[\arccos({1\over -J})-\zeta)]\over \sqrt{J^2\cos^2[\arccos({1\over -J})-\zeta]-1}}\nonumber \\
&\underset{J\to-\infty}\longrightarrow\pi \underset{\zeta\to 0^+}\longrightarrow\pi,
\end{align}
which shows that the tip of the seemingly ``lightlike geodesic'' contributes to the entire length.


\subsection{Geodesic connecting points of the two horizons}
\label{appendix_non_sym}

This section generalizes the computation of the length of an ``arm along the barrel,'' which was presented in \Sect{armsem}. First, we consider $H_1$ and $H_2$ respectively at  different conformal times $\sigma_1$ and $\sigma_2$, satisfying \Eq{si}. 
Moreover, choosing the geodesic endpoint on $H_1$ to be at $\theta_2=0$, the second endpoint on $H_2$ can be twisted by an angle $\alpha$.  Our goal is to compute the length of the geodesic lying in the exterior region and connecting these two points. 

Using embedding coordinates, the plane that passes through the two endpoints and the origin of Minkowski spacetime in four dimensions is the set
\be
\Big\{\lambda_1\big(T_1,-|T_1|,1,0\big)+\lambda_2\big(T_2,|T_2|,\cos\alpha,\sin\alpha\big), ~\where~\lambda_1,\lambda_2\in\R\Big\}, 
\label{pl2'}
\ee
with $T_1$, $T_2$ related to $\sigma_1$, $\sigma_2$ as in \Eq{Ts}.
With the definitions of $u$, $v$ given in \Eq{uv}, the endpoints and the origin are reached at $(v,u)=(-1,1)$, $(1,1)$ and $(0,0)$. The intersection of this plane and the hyperboloid~(\ref{hyperboloid}) is the set of points 
\be
\left({u\over 2}\,(T_2+T_1)\!+\!{v\over 2}\, (T_2-T_1), {u\over 2}\,( |T_2|\!-\!|T_1|)+{v\over 2}\,( |T_2|\!+\!|T_1|),{u\over 2}\, ( \cos\alpha+1)\!+\!{v\over 2}\, ( \sin\alpha-1)\!\right)\!,
\ee
satisfying
\be 
u^2(1-\hat T^2)+v^2\hat T^2=1, \quad \where\quad \hat T^2=T_1T_2+\sin^2{\alpha\over 2}\, .
\ee
This result is obtained by using the fact that $T_1T_2\ge0$, which follows from \Eq{si}. It is identical to \Eq{eh}, under the replacement $T^2\to \hat T^2$. Moreover, the condition~(\ref{er}) of being located in the exterior region turns out to be again \Eq{u>v}. As a result, we can conclude the following:

\noindent $\bullet$ When $|\hat T|\le 1$ \ie $T_1T_2\le \cos^2{\alpha\over 2}$, there is a spacelike geodesic fully contained in the exterior region. Its length is 
\begin{align}
   \hat L_{\rm arm} &=2\arcsin(|\hat T|)\nonumber \\
    &=2\arcsin\sqrt{\sinh\tau_1\sinh\tau_2+\sin^2{\alpha\over 2}},\quad\where\quad \sinh\tau_1\sinh\tau_2\le \cos^2{\alpha\over 2}. 
\label{hL1'}
\end{align}
A second geodesic exists. It is spacelike but extends in the two interior and the exterior regions. 

\noindent $\bullet$ When $|\hat T|> 1$ \ie $T_1T_2> \cos^2{\alpha\over 2}$, the conclusions are the same except that the geodesics are timelike. 


\section{A phase transition in the two-arc subsystem in \bm dS$_3$?}
\label{ptran}

As in Appendix~\ref{Appendix:geodesics_dS}, we work  in dS$_3$ and choose a Cauchy slice $\Sigma$ such that the screens $H_1$ and~$H_2$ are circles. We consider the subsystem $A$ of $H_1\cup H_2$ comprising a pair of identical and perfectly aligned arcs respectively in $H_1$ and $H_2$. Throughout this appendix, we will use the notations of \Sect{misal}. The configuration corresponds to arcs parametrized by $\theta_2\in[0,\Theta]$, where $\Theta\in(0,2\pi)$, \ie with $\alpha=\alpha'=0$. Our goal is to show that phase transitions occur when the minimal extremal curve $\chi_{\rm ext}$ is not allowed to be lightlike. For simplicity, we will take the screens at identical global times $\tau_0\equiv \tau_1=\tau_2$~\cite{Shaghoulian:2021cef}. 

The minimal lengths of the pairs of curves described in cases~($i$) and~($ii$) of \Sect{misal} are
\be
L_{\rm arcs}(\Theta) =2\inf(\Theta,2\pi-\Theta),
\ee 
while that of the arms~($iii$) is given by 
\be
    L_{\rm arms}(\tau_0)= 4\arcsin({\sinh|\tau_0|}), \quad \when \quad |\tau_0|\le \arcsinh 1.
\ee
In the class of pairs of curves~($vi$), the spacelike ones have nonvanishing winding numbers $w_1$ and~$w_2$, which implies their lengths to be greater than or equal to $4\pi>L_{\rm arcs}$. As a result, the length of $\chi_{\rm ext}$ is the smallest between $L_{\rm arcs}(\Theta)$ and $L_{\rm arms}(\tau_0)$. 

\begin{figure}[!h]
    \begin{subfigure}[t]{0.48\linewidth}
        \centering
\begin{tikzpicture}
\begin{axis}[
    axis lines = middle,
    xlabel = $\tau_0$,
    ylabel = {Entropy},
    x label style={at={(axis description cs:1,0.15)},anchor=north},
    y label style={at={(axis description cs:0.5,1.2)},rotate=0,anchor=north},
    no markers,
    xmin=-0.5,
    xmax=0.5,
    ymax=8*(0.6+0.6^3/3+0.6^5/6)+1,
    xtick={-0.3,0, 0.3},
    xticklabels={{\footnotesize $-\tau_{\rm t}(\Theta)$},$0$, {\footnotesize $\tau_{\rm t}(\Theta)$}},
    ytick={0,4*(0.6+0.6^3/3+0.6^5/6),8*(0.6+0.6^3/3+0.6^5/6)},
    yticklabels={$0$,$\frac{L_{\rm arcs}(\Theta)}{4G\hbar}$,$\frac{L_{\rm arcs}(\Theta)}{2G\hbar}$},
    y post scale=0.7,
    x post scale=0.9,
    legend style={at={(1.2,0.8)},anchor=east, font=\small},
] 

\addplot [
    domain=-1:0, 
    samples=100, 
    color=blue,
    line width=0.6mm
    ]
    {(\x<-0.3)*(4*(0.6+0.6^3/3+0.6^5/6))+(\x>-0.3)*(-4*(2*x+8*x^3/3+2^5*x^5/6))};

\addplot [
    domain=0:1, 
    samples=100, 
    color=blue,
    line width=0.6mm
    ]
    {(\x<0.3)*(4*(2*x+8*x^3/3+2^5*x^5/6))+(\x>0.3)*(4*(0.6+0.6^3/3+0.6^5/6))};

\addplot [
    domain=0:1, 
    samples=100, 
    color=red,
    line width=0.6mm
    ]
    {(\x<0.3)*(4*(2*x+8*x^3/3+2^5*x^5/6)+4*(0.6+0.6^3/3+0.6^5/6))+(\x>0.3)*(8*(0.6+0.6^3/3+0.6^5/6))};

\addplot [
    domain=-1:0, 
    samples=100, 
    color=red,
    line width=0.6mm
    ]
    {(\x<-0.3)*(8*(0.6+0.6^3/3+0.6^5/6))+(\x>-0.3)*(-4*(2*x+8*x^3/3+2^5*x^5/6)+4*(0.6+0.6^3/3+0.6^5/6))};
    
\draw[dotted] (-0.3,0)--(-0.3,5.5);  
\draw[dotted] (0.3,0)--(0.3,5.5);
\end{axis}

\draw[-{Latex[scale=1.3]}] (6,0) -- (6.2,0); 
\draw[-{Latex[scale=1.3]}] (3.09,3.8) -- (3.09,4);

\end{tikzpicture}
\caption{\footnotesize At fixed arc angle $\Theta$ and as a function of global time $\tau_0$. \label{Stau}}
  \end{subfigure}
  \quad 
    \begin{subfigure}[t]{0.48\linewidth}
\begin{tikzpicture}
\begin{axis}[
    axis lines = left,
    xlabel = $\Theta$,
    ylabel = {Entropy},
    x label style={at={(axis description cs:1,0.15)},anchor=north},
    y label style={at={(axis description cs:0.14,1.1)},rotate=-90,anchor=east},
    no markers,
    xmin=0,
    xmax=1.1,
    ymax=2,
    xtick={0,0.35,0.65, 1},
    xticklabels={$~$,{\footnotesize $\Theta_{\rm t}(\tau_0)$},{\footnotesize $2\pi-\Theta_{\rm t}(\tau_0)$},{\footnotesize $2\pi$}},
    ytick={0,0.7,1.4},
    yticklabels={$~$,$\frac{L_{\rm arms}(\tau_{0})}{4G\hbar}$,$\frac{L_{\rm arms}(\tau_{0})}{2G\hbar}$},
    y post scale=0.7,
    x post scale=0.9,
    legend style={at={(1.2,0.8)},anchor=east, font=\small},
]

\addplot [
    domain=0:1, 
    samples=100, 
    color=blue,
    line width=0.6mm
    ]
    {(\x<0.35)*(2*x) + (0.35<\x)*(\x<0.65)*(0.7) + (\x>0.65)*(\x<1)*(-2*(x-1))};

\addplot [
    domain=0:1, 
    samples=100, 
    color=red,
    line width=0.6mm
    ]
    {(\x<0.35)*(4*x) + (0.35<\x)*(\x<0.5)*(1.4+2*(x-0.35)) + (0.5<\x)*(\x<0.65)*(1.7-2*(x-0.5)) + (\x>0.65)*(\x<1)*(-4*(x-1))};
\draw[dotted] (0.35,0)--(0.35,1.4);  
\draw[dotted] (0.65,0)--(0.65,1.4); 
\end{axis}

\draw[-{Latex[scale=1.3]}] (6,0) -- (6.2,0); 
\draw[-{Latex[scale=1.3]}] (0,3.8) -- (0,4);

\end{tikzpicture}
\caption{\footnotesize At fixed global time $\tau_0$ and as a function of arc angle $\Theta$. \label{SPhi}}
  \end{subfigure}

\caption{\footnotesize Geometrical contributions to the entropies of two identical and aligned arcs on both screens, by applying the monolayer (blue) or bilayer (red) prescriptions.}
    \label{fig:S_arcs}
\end{figure}
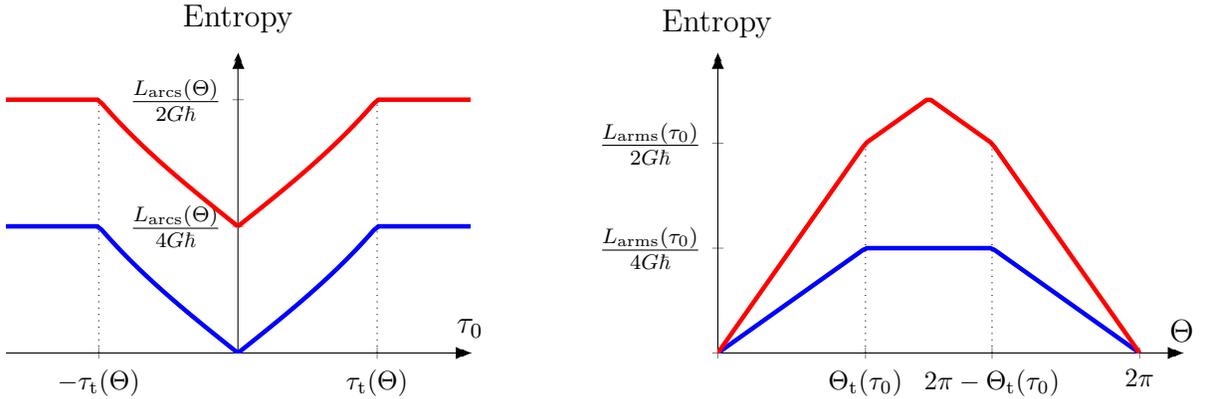

At given $\Theta$, since $L_{\rm arcs}(\Theta)$ is  independent of time and $L_{\rm arms}(\tau_0)$ increases from 0 with~$|\tau_0|$, there are transitions at some times $\pm \tau_{\rm t}(\Theta)$ satisfying $L_{\rm arms}=L_{\rm arcs}$, which yields 
\begin{equation}\label{eq:transition_time}
    \tau_{\rm t}(\Theta) = \arcsinh\!\left(\!\sin{\frac{\Theta}{2}}\right)\!.
\end{equation}
Hence, the monolayer entropy of the two-arc subsystem is given by
\begin{equation}\label{eq:molonayer_sym_arcs}
    S_{\rm mono}(\Theta,\tau_0)= \frac{1}{4G\hbar} \left\{\!\begin{array}{ll}
       L_{\rm arms}(\tau_0) \, , & \mbox{if }|\tau_0| \leq \tau_{\rm t}(\Theta)\espD\\
     L_{\rm arcs}(\Theta) \, , &\mbox{if }|\tau_0| \geq \tau_{\rm t}(\Theta)
    \end{array}\right.+\O(G\hbar)^0.
\end{equation}
Its classical contribution is plotted in blue in Figure~\ref{Stau}. 
%
%
At fixed $\tau_0$, allowing $\Theta$ to vary, transitions occur at the angles
\be
\Theta_{\rm t}(\tau_0)=2\arcsin(\sinh|\tau_0|)~~ \and ~~ 2\pi-\Theta_{\rm t}(\tau_0), \quad \mbox{if} \quad |\tau_0|\le \arcsinh 1.
\ee
The monolayer entropy can therefore  alternatively be written  as
\begin{equation}\label{eq:molonayer_sym_arcs2}
    S_{\rm mono}(\Theta,\tau_0)= \frac{1}{4G\hbar} \left\{\!\begin{array}{ll}
       L_{\rm arcs}(\Theta) \, , & \mbox{if }\Theta \in[0,\Theta_{\rm t}(\tau_0)]\cup [2\pi-\Theta_{\rm t}(\tau_0),2\pi]\espD\\
     L_{\rm arms}(\tau_0) \, , &\mbox{if }\Theta \in[\Theta_{\rm t}(\tau_0),2\pi-\Theta_{\rm t}(\tau_0)]
    \end{array}\right.+\O(G\hbar)^0.
\end{equation}
Its classical leading contribution is drawn in blue in Figure~\ref{SPhi}. 
Notice that in this result, we include the limit cases \mbox{$\Theta\to 0$} and $\Theta\to 2\pi$, since the formula produces correctly a vanishing geometrical contribution.   
 The transitions are analogous to the Hartman-Maldacena transition in the eternal AdS black hole examples \cite{Hartman:2013qma}. Indeed, in the higher dimensional analogue, the two arcs are replaced by caps 
 on the spherical screens, while the two arms are replaced by a connected surface joining the two screens and having the same boundaries as the caps.
 The transition is between a connected extremal surface and two disconnected extremal surfaces ending on the boundaries of the caps.

 In the bilayer proposal, we have to add to \Eqs{eq:molonayer_sym_arcs} and~(\ref{eq:molonayer_sym_arcs2}) the classical contributions arising from curves $\chi_{\rm left}$ and $\chi_{\rm right}$ in the two interior regions. As explained above \Eq{sb}, their total length being $L_{\rm arcs}$, we obtain  
\begin{equation}\label{bilayer_definition}
    S_{\rm bil}(\Theta,\tau_0) = S_{\rm mono}(\Theta,\tau_0) +\frac{L_{\rm arcs}(\Theta)}{4G\hbar},
\end{equation}
where $\Theta\in[0,2\pi]$. Again, we include the cases $\Theta\to0$ and $\Theta\to 2\pi$ since the extra contribution $L_{\rm arcs}/(4G\hbar)$ vanishes in these limits and does not spoil vanishing of the entropy for a pure state. The variations of $S_{\rm bil}$ as a function of $\tau_0$ and $\Theta$ are shown in red in Figures~\ref{Stau} and~\ref{SPhi}, respectively. 

Note that when $|\tau_0|\ge \tau_{\rm t}$, the leading contribution to the entropy saturates to a constant value equal to $L_{\rm arcs}(\Theta)/(4G\hbar)$ in the monolayer case and $L_{\rm arcs}(\Theta)/(2G\hbar)$ in the bilayer case. Therefore, the holographic entanglement entropy obeys a volume law, rather than an area law, suggesting as in the case of a single-arc subsystem that the holographic theory on the screens is non-local~\cite{Shaghoulian:2021cef}. 


\end{appendices} 
 

\bibliographystyle{jhep}
\bibliography{Main}

\end{document}